\documentclass[lettersize,journal]{IEEEtran}
\IEEEoverridecommandlockouts

\def\BibTeX{{\rm B\kern-.05em{\sc i\kern-.025em b}\kern-.08em
    T\kern-.1667em\lower.7ex\hbox{E}\kern-.125emX}}
\ifCLASSINFOpdf
\fi
\usepackage{verbatim}
\usepackage{algorithm}
\usepackage{graphicx}
\usepackage{cite}
\usepackage{bm}
\usepackage{amssymb,amsfonts}
\usepackage{algorithmicx}
\usepackage[noend]{algpseudocode}
\usepackage{setspace}
\usepackage{float}
\usepackage{subfigure}
\usepackage{epstopdf}
\usepackage{amsthm}
\newtheorem{lemma}{Lemma}

\usepackage{stfloats}
\usepackage{enumerate}
\usepackage{multirow}
\usepackage{latexsym}
\usepackage{multicol,lipsum}
\usepackage{cuted} 

\usepackage{tikz}
\usepackage{textcomp}
\usepackage{xcolor}
\usepackage{cancel}

\usepackage[cmex10]{amsmath}
\usepackage[cmex10]{mathtools}

\algdef{SE}[DOWHILE]{Do}{doWhile}{\algorithmicdo}[1]{\algorithmicwhile\ #1}%
\allowdisplaybreaks
\makeatletter 
\def\@eqnnum{{\normalsize \normalcolor (\theequation)}} 

\makeatother 
\begin{document}
\title{Robust and Secure Blockage-Aware Pinching Antenna-assisted Wireless Communication}
\author{Ruotong Zhao,~\IEEEmembership{Graduate Student Member,~IEEE,} Shaokang Hu,~\IEEEmembership{Member,~IEEE,}\\  Deepak Mishra,~\IEEEmembership{Senior Member,~IEEE,} and Derrick Wing Kwan Ng,~\IEEEmembership{Fellow,~IEEE} 
\thanks{R. Zhao, S. Hu, D. Mishra, and D. W. K. Ng are with the School of Electrical Engineering and
Telecommunications, University of New South Wales, Sydney, NSW 2052,
Australia (e-mails: ruotong.zhao@student.unsw.edu.au, \{shaokang.hu, d.mishra, w.k.ng\}@unsw.edu.au).}
\thanks{The work of R. Zhao was supported in part by the Commonwealth through an Australian Government Research Training Program Scholarship [DOI: https://doi.org/10.82133/C42F-K220] and the Australian Research Council's Discovery Projects (DP240101019). The work of S. Hu and D. W. K. Ng was supported by the Australian Research Council's Discovery Projects (DP240101019). The work of D. Mishra was supported by the Australian Research Council's Discovery Early Career Researcher Award (DE230101391).}
\thanks{This article was presented in part at the 2025 IEEE Globecom~\cite{conference}.}
\vspace{-1em}}

\maketitle

\begin{abstract}
In this work, we investigate a blockage-aware pinching antenna (PA) system designed for secure and robust wireless communication. The considered system comprises a base station equipped with multiple waveguides, each hosting multiple PAs, and serves multiple single-antenna legitimate users in the presence of multi-antenna eavesdroppers under imperfect channel state information (CSI). To safeguard confidential transmissions, artificial noise (AN) is deliberately injected to degrade the eavesdropping channels. 
Recognizing that conventional linear CSI-error bounds become overly conservative for spatially distributed PA architectures, we develop new geometry-aware uncertainty sets that jointly characterize eavesdropper position and array-orientation errors. Building upon these sets, we formulate a robust joint optimization problem that determines per-waveguide beamforming and AN covariance, individual PA power-ratio allocation, and PA positions to maximize the system sum rate subject to secrecy constraints. The highly non-convex design problem is efficiently 
addressed via a low computational complexity iterative algorithm that capitalizes on block coordinate descent, penalty-based methods, majorization–minimization, the S-procedure, and Lipschitz-based surrogate functions. Simulation results demonstrate that the sum rate achieved by the proposed algorithm outperforms conventional fixed-antenna systems by $4.7$ dB, offering substantially improved rate and secrecy performance. In particular, (i) adaptive PA positioning preserves LoS to legitimate users while effectively exploiting waveguide geometry to disrupt eavesdropper channels, and (ii) neglecting blockage effects in the PA system significantly impacts the system design, leading to performance degradation and inadequate secrecy guarantees.
\end{abstract}


\section{Introduction}
To satisfy the rapidly escalating data traffic demands anticipated in the upcoming sixth-generation (6G) wireless networks, multiple-input multiple-output (MIMO) technology incorporating flexible antenna architectures has been widely envisioned as a critical enabler for effectively exploiting the additional spatial degrees of freedom (DoF) through dynamically reconfiguring the propagation environment and reshaping wireless channels~\cite{wu2023movable}.
Depending on the specific implementations, a flexible antenna can be realized by leveraging mechanical actuators~\cite{wu2023movable} or liquid-metal designs~\cite{wong2020fluid}, with the latter known as fluid antennas; however, their repositioning range is typically limited to only a few wavelengths, often rendering them insufficient to maintain reliable line-of-sight (LoS) connectivity~\cite{ding2024flexible}. 
Considering that emerging communication systems are migrating toward higher carrier frequencies, pinching antenna (PA) technology has emerged as a novel and promising flexible antenna solution that operates by utilizing dielectric waveguides in which radiating elements are activated by selectively pinching different dielectric particles to simultaneously regulate both path loss and phase shift~\cite{xu2025pinching}. This unique mechanism provides stronger LoS connectivity and lower propagation loss compared with alternative flexible antenna approaches~\cite{ding2024flexible}.
Additionally, PAs facilitate adaptive antenna configurations by simply activating or deactivating elements along the waveguide via pinching operations, offering high flexibility in both the number and placement of radiating elements.
This spatial capability enables precise control over complex communication environments and supports effective transmission in dynamic or obstructed scenarios.
Moreover, the physics-based hardware model in~\cite{wang2025modeling} has demonstrated that the power of PAs can be actively adapted by adjusting their coupling lengths along the waveguide, reinforcing their practical feasibility for dynamic amplitude and phase control.

Despite these advancements, existing studies on PA-assisted systems predominantly rely on overly optimistic assumptions, especially the persistent assumptions of idealized LoS propagation and perfect channel state information (CSI), e.g.,~\cite{conference, li2025mimo, tegos2024minimum, xu2025joint}. Indeed, these assumptions inadequately capture the complexities of practical propagation environments, typically assuming a dominant LoS link between PAs and users without explicitly accounting for inevitable blockage effects. However, realistic urban deployments of PAs are inherently subject to numerous blockages, such as buildings and vegetation, which can severely disrupt LoS connectivity~\cite{guo2021multiple, yi2023trajectory}. Although~\cite{ding2025los} demonstrated that PA-assisted systems can achieve significantly lower outage probabilities of data rates compared to conventional antenna systems under probabilistic LoS assumptions, it did not directly investigate the implications of blockages on resource allocation strategies. Similarly, a recent study in~\cite{wang2025pin} analyzed PA performance under blockage effects, exploiting a waveguide architecture with a simplified PA-selection model and reported performance gains. 
However, its restricted assumptions, e.g., fixed PA positions, single active PA per waveguide, and serving only one user per waveguide, limit spatial flexibility, prevent coherent multi-waveguide beamforming, and ultimately constrain system performance. Furthermore, since existing simplified LoS channel models deviate significantly from real-world conditions, especially at high frequencies such as the millimeter-wave bands~\cite{yi2022joint}, a more realistic blockage-aware channel model becomes imperative for better performance characterization and accurate resource allocation.
Consequently, a comprehensive and rigorous investigation into blockage-aware resource allocation strategies in PA-assisted systems remains underexplored.

On the other hand, since wireless communication is inherently broadcast in nature, confidential signals remain vulnerable to eavesdropping~\cite{yu2020robust, ren2025robust}. In this context, the combined effect of blockage and the PA system's pronounced deployment flexibility creates both additional vulnerabilities and new degrees of freedom in system design for physical layer security (PLS). In fact, rather than viewing blockage solely as an inherent impairment, it can be regarded as a design that acts as a spatial barrier and can be strategically exploited to suppress unauthorized signal reception. This makes a blockage-aware PA system model essential for characterizing both new challenges and design opportunities in PLS.
While blockages inherently restrict feasible PA deployment regions and complicate the optimization landscape for maintaining reliable LoS connectivity with legitimate users, they can also be strategically harnessed to obstruct LoS paths toward potential eavesdroppers (EAs) through carefully designed PA positions, thereby effectively reducing information leakage and mitigating multi-user interference. However, previous works generally neglected the presence of EAs within PA-assisted systems, and studies such as~\cite{conference, ding2025los}, may therefore underestimate security risks, especially in scenarios where both EAs and PAs are located in close proximity to legitimate users.
Indeed, recent studies have highlighted the capability of PA positioning flexibility to enhance PLS by placing antennas closer to legitimate users while distancing them from potential EAs~\cite{kaidi2025physical, sun2025physical}. Nonetheless, these existing investigations addressing PLS in PA-assisted systems remain generally limited to oversimplified configurations, such as considering a single-antenna eavesdropper, neglecting waveguide propagation loss and eavesdropper data-rate leakage tolerance. Additionally, although artificial noise (AN) injection is an effective method for mitigating eavesdropping, it has not yet been incorporated into PA-based PLS frameworks~\cite {sun2025physical, kaidi2025physical, badarneh2025physical}.

Existing studies often over-optimistically assume the availability of perfect CSI, e.g.~\cite{conference, xu2025pinching, wang2025modeling, li2025mimo, tegos2024minimum}, which is generally impractical. Although pilot-based estimation and geometric propagation modeling have improved CSI acquisition accuracy in modern networks~\cite{su2023sensing,zhou2025channel}, obtaining reliable CSI remains inherently challenging under blockages, especially with constrained training resources~\cite{yao2023robust,xiu2024robust}. This challenge is indeed further exacerbated in the presence of passive eavesdroppers, whose CSI is extremely difficult to estimate due to their silent and non-cooperative nature~\cite{ren2025robust,yu2020robust,hu2021robust}. To date, to the best of our knowledge, the only study that considers imperfect CSI in PA systems is~\cite{zeng2025robust}, which minimizes the transmit power subject to a data rate outage probability constraint. However, its applicability is limited to a single waveguide with a single PA and does not provide a generalized characterization applicable to multiple PAs or multi-waveguide configurations, making the model unsuitable for distributed PA networks. Moreover, unlike conventional antenna arrays, e.g.,~\cite{hu2021robust, xiu2024robust,yu2020robust}, PA systems feature multiple PAs that can be flexibly distributed along different spatial positions, resulting in geometry-dependent phase variations across propagation paths that may partially cancel or reinforce each other.
Consequently, existing near-field CSI-error bounds~\cite{ren2025robust, xiu2024robust}, which rely on co-located antennas and linear approximations, become overly loose and inapplicable for practical distributed PA architectures. Despite the critical importance of capturing these practical effects, current literature lacks a geometry-aware robust CSI-uncertainty framework tailored to PA systems that jointly accounts for blockage conditions, antenna positional displacement, and orientation uncertainties. This significant gap motivates the development of a new uncertainty model capable of accurately characterizing the CSI mismatch in distributed PA networks.
It is also important to note that prior studies have often overlooked the intrinsic characteristics of dielectric waveguides, instead treating the power distribution across multiple PAs as uniform~\cite{xu2025pinching, ding2024flexible, wang2025pin}. 

Although several works have examined the impact of dielectric waveguide propagation~\cite{hu2025, wang2025modeling}, some are limited to single-PA-per-waveguide configurations, whereas others employ a propagation model that ignores frequency dependence and characterizes attenuation solely by PA location. Unfortunately, such simplified treatment leads to inaccurate modeling of signal attenuation, thereby hindering the efficiency of resource allocation~\cite{balanis2024balanis}.
Additionally, although the authors in~\cite{conference} highlighted that PA systems are notably suitable for multicasting scenarios, it considered only a specific broadcasting setting and did not account for the complex practical environment involving both PLS threats and blockage effects, which jointly restrict feasible PA placements and significantly increase the complexity of resource allocation.

Motivated by these identified research gaps and associated practical challenges, this paper aims to develop a robust and secure resource allocation framework for a blockage‑aware PA‑assisted system, where a base station is equipped with multiple waveguides, each carrying multiple PAs, to serve multiple single‑antenna users in the presence of passive multi‑antenna EAs under imperfect CSI. In this regard, we construct a three‑dimensional (3D) blockage‑aware channel model that jointly captures waveguide attenuation and free‑space propagation, which explicitly characterizes the geometric conditions for LoS availability. We adopt a bounded CSI uncertainty for legitimate users, while deriving a new geometry‑aware uncertainty bound for passive EAs that captures position- and orientation-induced errors inherent to distributed near-field deployments. Specifically, the main contributions of this paper are summarized as follows:
\begin{enumerate}
    \item We formulate a robust joint design as an optimization problem to maximize the system sum rate subject to stringent secrecy constraints that limit information leakage. The proposed framework jointly optimizes the transmit signal beamforming matrix, the artificial noise (AN) covariance, the PA power-ratio allocation, and the continuous PA positions, while proactively accounting for the deleterious effects of imperfect CSI.
    \item The formulated problem is inherently non-convex and mathematically intractable, exacerbated by infinitely many inequality constraints induced by CSI uncertainty and the intricate coupling among the optimization variables. To address this, we develop a computationally efficient block coordinate descent (BCD)-based algorithm. By leveraging the S-procedure and majorization–minimization (MM) techniques, the proposed algorithm circumvents the intractable constraints and alternately optimizes the system variables, thereby guaranteeing convergence to a high-quality stationary solution. 
    \item We reveal that the continuous PA placement simultaneously dictates three PA-dependent channel functions operating at vastly different spatial scales: large-scale path loss and blockage-induced attenuation (on the order of meters), and phase shifts (on the order of a few wavelengths). To address this challenging nonconvexity arising from the strong coupling among these effects, we devise a tailored two-stage optimization strategy that first establishes a coarse PA placement based on macroscopic blockage geometry and path loss, followed by a fine-grained position refinement to achieve accurate phase alignment.
    \item Simulation results verify the tightness of the proposed uncertainty bound for distributed multi‑PA configurations and demonstrate significant gains in system sum rate over conventional fixed‑antenna baselines, especially in a blockage-rich environment. They further demonstrate that ignoring blockage can lead to performance degradation and even infeasible secrecy constraints, whereas adaptive PA positioning establishes strong LoS links for legitimate users while suppressing eavesdropping channels, thereby reducing the required AN power to satisfy secrecy targets.
\end{enumerate}

\begin{figure}[t]
\centerline{\includegraphics[width=3.2in]{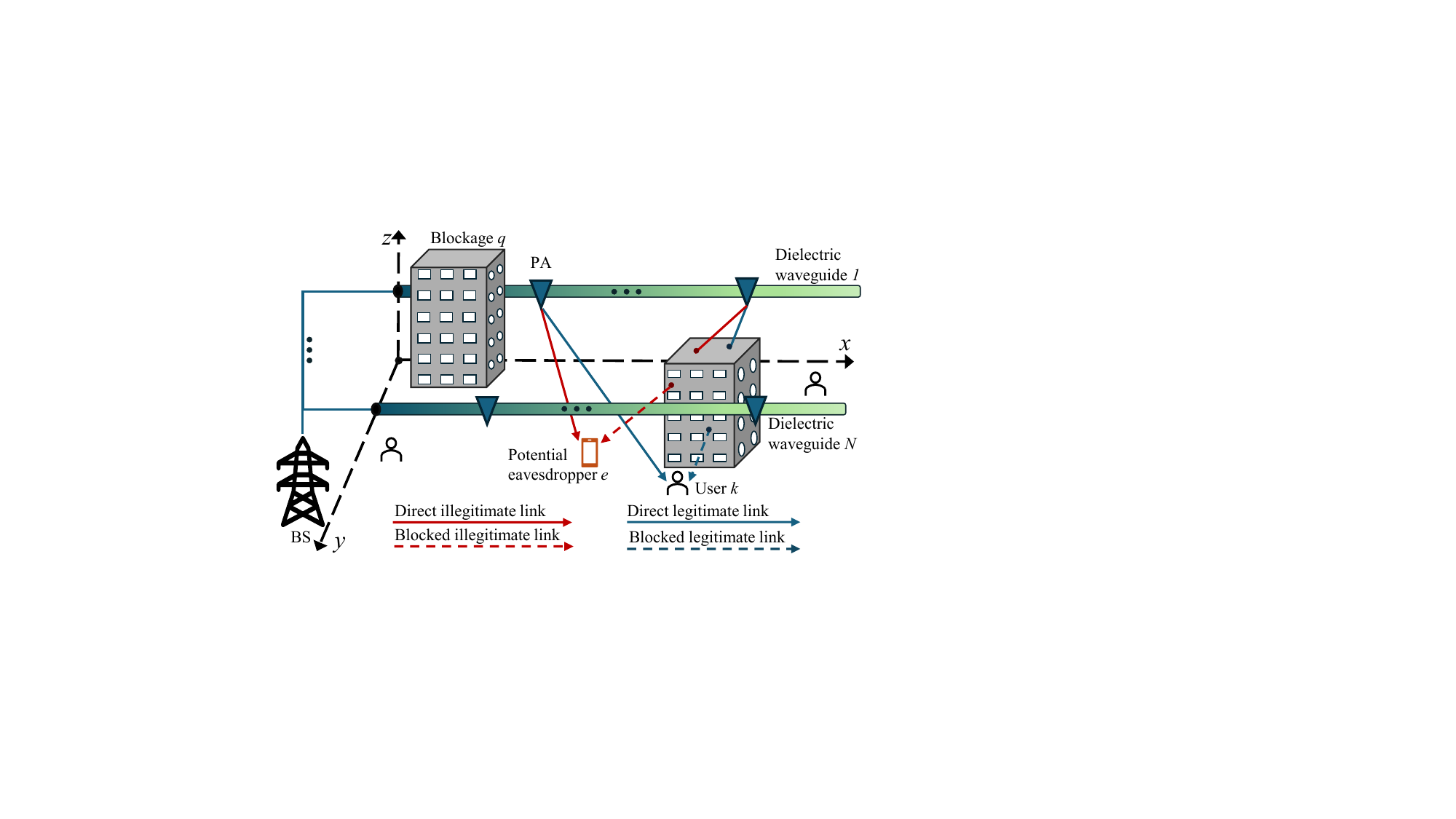}}
\vspace{-1em}
\caption{A downlink PA-assisted communication system.}
\vspace{-1em}
\label{fig:SYSTEM}
\end{figure}
\textbf{Notation:} 
Boldface capital and lowercase letters denote matrices and vectors, respectively. $\textbf{A}^\top$, $\textbf{A}^H$, $\text{Tr}(\textbf{A})$, and $\text{Rank}(\textbf{A})$ represent the transpose, the Hermitian transpose, the trace, and the rank of  $\textbf{A}$, respectively. $\textbf{A} \succ \bm{0}$ and $\textbf{A} \bm{\succeq 0}$ indicate that $\textbf{A}$ is a positive definite and positive semi-definite matrix, respectively. The operator $\mathrm{diag}(\textbf{a})$ forms a diagonal matrix from vector $\textbf{a}$,  $\mathrm{Diag}(\textbf{A})$ extracts the main diagonal of matrix $\textbf{A}$,  and $\text{blkdiag}\left(\textbf{A}_1, ..., \textbf{A}_n\right)$ denotes a block diagonal matrix whose diagonal components are $\textbf{A}_1, ..., \textbf{A}_n$. 
$\textbf{A} \odot \textbf{B}$ represents the Hadamard product of two matrices $\textbf{A} \ \text{and} \ \textbf{B}$.
${I}_N$ is the $N \times N$ identity matrix; 
we adopt $\mathbb{C}^{N \times M}$ ($\mathbb{R}^{N \times M}$) and $\mathbb{H}^{N}$ to denote the spaces of $N \times M$ complex (real) and $N \times N$ Hermitian matrices, respectively. The largest eigenvalue of matrix $\textbf{A}$ and its associated eigenvector are denoted by $\lambda_{\rm max}\hspace{-1mm}\left(\hspace{-0.2mm}\textbf{A}\hspace{-0.2mm}\right)$ and $\bm{\lambda}_{\rm max}\hspace{-1mm}\left(\hspace{-0.2mm}\textbf{A}\hspace{-0.2mm}\right)$, respectively.
$|\cdot|$, $\left(\cdot\right)^*$, $\mathbb{R}^+$, and $\mathbb{E}\{\cdot\}$ denote the absolute value, complex conjugate, set of positive real numbers, and statistical expectation, respectively. Moreover, $\|\cdot\|$, $\|\cdot\|_2$, and $\|\cdot\|_*$ denote the norm, spectral norm, and nuclear norm, respectively. $\text{vec}([\textbf{a}_1, ..., \textbf{a}_n])$ and $\text{vec}\{a_{n,m}\}_{n,m}$ represent the vectorization of the $n$ vectors and $nm$ scalars, respectively. We utilize $j = \sqrt{-1}$,  $\Re(a)$, and $\Im(a)$ to represent the imaginary unit of a complex number, the real and imaginary parts of the complex number $a$, respectively. The circularly symmetric complex Gaussian (CSCG) distribution is denoted by $\mathcal{CN}(\mu, \sigma^2)$, with mean $\mu$ and variance $\sigma^2$, and $\sim$ stands for ``distributed as".

\begin{table}[t!]
\renewcommand{\arraystretch}{1.1} 
\small
\caption{Key System Parameters}
\label{tab:nomenclature}
\centering
\begin{tabular}{@{}l p{0.76\columnwidth}@{}}
\hline\hline
\textbf{Symbol} & \textbf{Definition} \\
\hline
$\mathcal{N}, \mathcal{M}$ & Sets of waveguides and PAs per waveguide \\
$\mathcal{K}, \mathcal{G}, \mathcal{Q}$ & Sets of legitimate users, EAs, and blockages \\
$K, G, Q$ & Numbers of users, EAs, and blockages \\
$L, d$ & Waveguide length and mounting height \\
$\gamma, \alpha$ & Minimum PA spacing and attenuation constant \\
$\epsilon_r, \text{tan}\left(\delta_e\right)$ & Dielectric constant and effective electric loss tangent \\
$c, f_c, \lambda$ & Speed of light, carrier frequency, and wavelength \\
$y_n^{ \rm FP}$ & $y$-coordinate of waveguide $n$'s feed point\\
$x^{\mathrm{Pin}}_{n,m}$ & $x$-coordinate of PA $m$ on the waveguide $n$\\
$\bm{\phi}_{k}, \mathbf{e}_{g}$ & 3D location of user $k$ and reference point of EA $g$ \\
$u_k$, $x_{k}$, $y_{k}$ & User $k$ and its location along the $x$-axis and $y$-axis\\
$p_{n,m}, p_{n,m}^{\max}$ & Power ratio for PA $m$ and its maximum limit \\
$\textbf{E}_n$ & Diagonal selection matrix with $[\textbf{E}_n]_{n,n} = 1$\\
$\mathcal{I}_{k,q}$ & Hyperplane index set of blocked region $\mathcal{D}_{k,q}$ \\
$(\textbf{a}_{k,q}^i)^{\hspace{-0.6mm}\top}\hspace{-0.8mm}, b_{k,q}^i$ & Normal vector and offset of hyperplane $i$ \\
$\hat{\textbf{h}}_{k}, \bm{\triangle}\textbf{h}_{k}$ & Estimated CSI vector and its estimation error for $u_k$ \\
$\bm{\Psi}_{k}$ & Continuous uncertainty region for $u_k$ \\
$\wp_{k}>0$ & A known bound parameter for $u_k$ \\
$\hat{\theta}_g, \triangle \theta_g$ & Estimated angle and corresponding error of EA $g$ \\
\hline\hline
\end{tabular}
\end{table}
\section{System Model}\label{sec:System}
\subsection{Blockage-Aware System Setup}\label{sec:network}
The key parameters are summarized in Table I and the proposed system is illustrated in Fig. 1, where a base station (BS) feeds signals to $N$ dielectric waveguides and each of them can accommodate at most $M$ activated PAs\footnote{One practical implementation of a PA system is to mount PAs on movable platforms along a pre-installed track parallel to the waveguide~\cite{ding2024flexible}, thereby enabling low-cost deployment and facilitating dense installations with fast and flexible repositioning~\cite{hu2025}. Moreover, since the positioning resolution achievable by modern actuation mechanisms is typically significantly smaller than the wavelength, the PA locations are modeled as continuous variables for modeling and optimization purposes~\cite{wang2025modeling, xu2025joint}. An explicit discrete-position formulation is reserved for future study.}.
Moreover, each single-antenna user $k$ requires distinct information $s_k \in \mathbb{C}, \forall k \in \mathcal{K}$, satisfying $\mathbb{E}\{|s_k|^2\} = 1$ and $\mathbb{E}\{s_{j}^*s_k\} = 0, \forall j \neq k$.
Since multiple users are served simultaneously, the transmitted signal $\textbf{s} \in \mathbb{C}^{N \times 1}$ is formed as a superposition of independent signals intended for each user $k$, along with an AN vector, $\textbf{v} \in \mathbb{C}^{N \times 1}$, generated by the BS to combat potential EAs, as
$\textbf{s} = \sum_{k=1}^{K}\textbf{w}_k s_k + \textbf{v}$.
Note that $\textbf{w}_k=[w_{k,1},\dots,w_{k,N}]^\top\in\mathbb{C}^{N\times1}$ denotes the beamforming vector for user $k$ with elements $w_{k,n} \in \mathbb{C}, \forall n$, indicating the beamforming coefficient on the waveguide $n$. The AN vector $\textbf{v}\sim\mathcal{CN}(\bm{0},\textbf{V})$ is modeled as a complex Gaussian vector with covariance matrix $\textbf{V}\in\mathbb{H}^N,\textbf{V} \bm{\succeq 0}$ and  $\sum_{k=1}^{K}|w_{k,n}|^2 + \text{Tr}(\textbf{VE}_n)\leq P_n^{\max},\forall n\in\mathcal{N}$.

\begin{figure}[t]
\centerline{\includegraphics[width=3.3in]{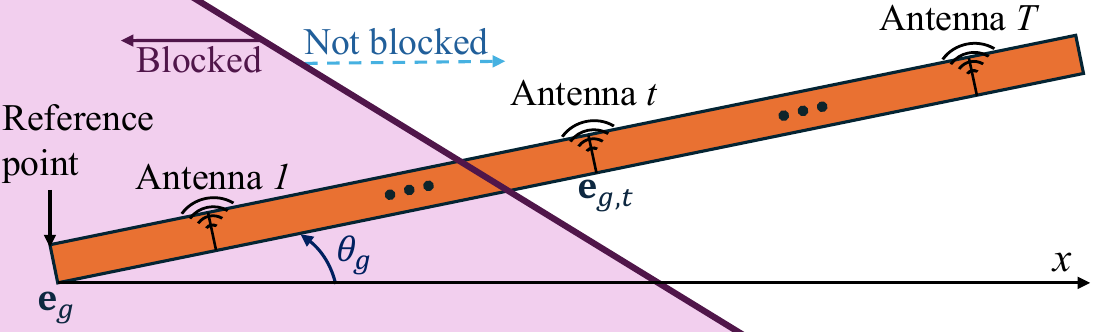}}
\vspace{-0.5em}
\caption{Illustration of the ULA of EA $g$, depicting the antenna indexing, blockage regions, and the orientation angle $\theta_g$ with respect to the $x$-axis.}
\vspace{-1em}
\label{fig:EavAntenna}
\end{figure}

This work employs a 3D Cartesian coordinate system. Specifically, the positions of the feed point\footnote{Since the waveguide coordinates $y_n^{\rm FP}$ are predetermined hardware parameters, we assume a sufficiently large separation between adjacent dielectric waveguides such that inter-waveguide evanescent coupling (crosstalk) is negligible~\cite{wang2025modeling, gu2026gap}.} and the PA $m$ of waveguide $n$ are denoted by $\bm{\psi}_n^{\mathrm{FP}} = \left[0,y_n^{ \rm FP},d\right]^\top$ and  $\bm{\psi}_{n,m}^{\mathrm{Pin}} = \left[x^{\mathrm{Pin}}_{n,m}, y_n^{ \rm FP}, d\right]^\top$, respectively. 
Furthermore, each non-cooperative potential EA is equipped with a uniform linear array (ULA) comprising $T > 1$ antenna elements, subject to $K + GT \leq N$. Without loss of generality, we assume that the reference point of EA $g$ is located at one end of its ULA~\cite{li2025mimo}, denoted as $\textbf{e}_{g} = \left[x^{\rm E}_{g}, y^{\rm E}_{g}, 0\right]^\top$, as shown in Fig.~\ref{fig:EavAntenna}. Specifically, antenna $t$ of EA $g$ is the $t$-th antenna, counting from the reference point, with its position denoted as $\textbf{e}_{g,t} = \left[x^{\rm E}_{g,t}, y^{\rm E}_{g,t}, 0\right]^\top$, $\forall t \in \{1, ..., T\}$. The spacing between the reference point and antenna $1$, as well as between any two adjacent antennas\footnote{Antenna arrays typically maintain a half-wavelength spacing because this configuration strikes an effective balance between avoiding grating lobes in unintended directions and reducing mutual coupling between antenna elements. It also simplifies the antenna array design by yielding more predictable and easily controlled radiation patterns~\cite{mailloux2017phased, cui2022channel}.} of EA $g$, is set to $\frac{\lambda}{2}$~\cite{ren2025robust, li2025mimo}.

\begin{figure}[t] 
	\centering
	\subfigure[Blockage areas affecting user $u_{k}$.]{
		\includegraphics[width=1.8in]{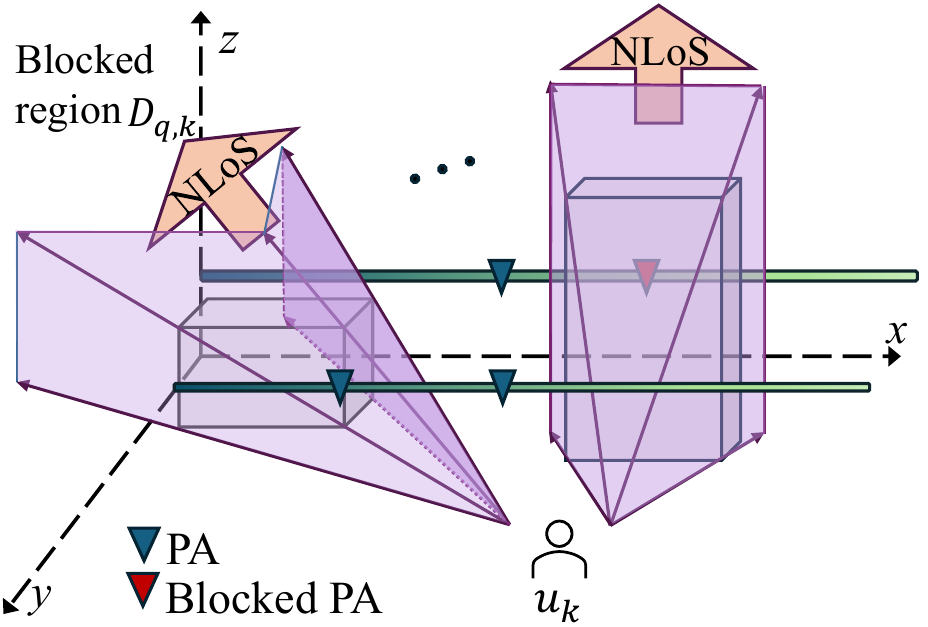}
	}\hspace{-3mm}
	\subfigure[Detailed view of blockage $q$.]{
		\includegraphics[width=1.47in]{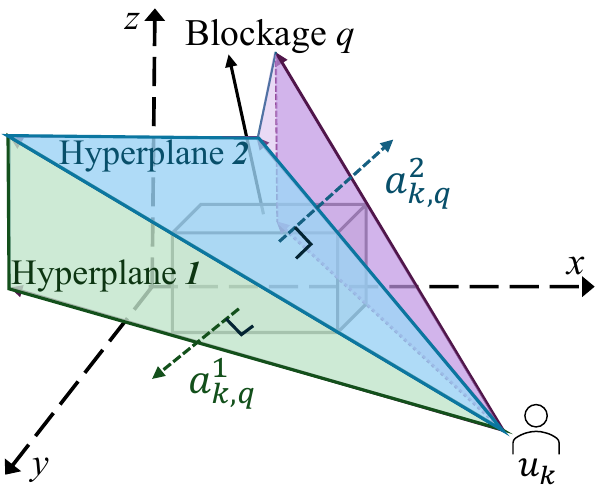}
	}
	\caption{Geometric illustration of a 3D blockage scenario.}
    \vspace{-1.3em}
	\label{fig:blockage}	
\end{figure}

Existing works on PA-assisted systems frequently adopt an overly optimistic assumption of ideal LoS propagation conditions~\cite{wang2025modeling, liu2025pinching}. However, practical deployment environments contain many physical obstructions, such as buildings and vegetation, and this blockage effect becomes more pronounced at higher frequencies with shorter wavelengths~\cite{hu2025}. Therefore, explicitly incorporating blockage effects into system modeling is essential for improving both modeling accuracy and the effectiveness of subsequent resource allocation and beamforming design.
As illustrated in Fig.~\ref{fig:blockage}(a), the $3$D coordinates and dimensions of the $Q$ randomly positioned cuboid-shaped blockages\footnote{The blockages are conservatively modeled as enclosing cuboids, which is a commonly adopted approximation~\cite{yi2022joint,yi2023trajectory}. Over the considered optimization timescale, static environmental blockers, such as buildings and dense vegetation, can be treated as quasi-static, and their geometric information, e.g., contours and height profiles, can be obtained from open-source geographic databases~\cite{yi2022joint,yi2023trajectory}. Therefore, the proposed framework is developed under a nominal blockage map, while explicit blockage uncertainty is beyond the scope of this paper.} are assumed to be known a priori.
These blockages obstruct the LoS links between a specific user $u_{k}$ and certain potential PAs located in the corresponding regions, thereby resulting in user-specific blocked areas. 
Specifically, the blocked region attributable to blockage $q$ with respect to user $u_{k}$ as shown in~ Fig.~\ref{fig:blockage}(b) can be represented as~\cite{yi2022joint}:
\begin{equation}
\mathcal{D}_{k,q} = \{\textbf{x} \in \mathbb{R}^3 \mid (\textbf{a}_{k,q}^i)^\top \textbf{x} - b_{k,q}^i \leq 0,\, i \in \mathcal{I}_{k,q} \}, \forall k, q, 
\end{equation}
where $\textbf{x}$ represents all possible locations of PAs. 

We define $d_{k,q}^{n,m} = \underset{i \in \mathcal{I}_{k,q}}{\max}\left\{(\textbf{a}_{k,q}^i)^\top\bm{\psi}_{n,m}^{\rm Pin} - b_{k,q}^i\right\}$ to quantify LoS availability. A strictly positive minimum value, i.e., $\underset{q \in \mathcal{Q}}{\min}\left\{d_{k,q}^{n,m}\right\} > 0$, indicates that the location of this particular PA, $\bm{\psi}_{n,m}^{\rm Pin}$, is unblocked by all $Q$ blockages defined with respect to user $u_{k}$. In contrast, a non-positive minimum value indicates that this specific PA is blocked by at least one blockage from the perspective of $u_{k}$.
To maximize system performance, PA placements should ideally maintain LoS conditions with legitimate users while opportunistically exploiting blocked conditions with respect to EAs. Note that the evaluation of LoS/blockage conditions for EAs follows the same methodology and procedures outlined above for legitimate users and is omitted for brevity.

\subsection{Channel Model}
Considering the impact of blockages, the wireless channel between the waveguide $n$ and $u_{k}$ is modeled by exploiting the LoS\footnote{In near-field configurations, NLoS components typically experience more than 20 dB of additional path loss relative to the LoS path~\cite{chen2025robust}. Indeed, any residual NLoS contribution can be modeled as a norm-bounded perturbation and absorbed into the imperfect CSI uncertainty set~\cite{ren2025robust, xing2023location}, without affecting algorithmic structure. We therefore omit explicit NLoS terms.} spherical wave near-field channel model~\cite{xu2025rate, suzuki2022pinching}, denoted as $\Tilde{\textbf{h}}_{n,u_{k}} = [\Tilde{h}_{n,u_{k}}^1, ..., \Tilde{h}_{n,u_{k}}^M]^\top \in \mathbb{C}^{M \times 1} $, where $\Tilde{h}_{n,u_{k}}^m = \tfrac{(\eta_{k}^{n,m})^{1/2} e^{-j\frac{2\pi}{\lambda}\left\| \bm{\phi}_{k} - \bm{\psi}^{\rm{Pin}}_{n,m}\right\|} }{\left\|\bm{\phi}_{k} - \bm{\psi}^{\rm{Pin}}_{n,m}\right\|}$, with
\begin{equation}\label{eq: step}
\eta_{k}^{n,m} = \begin{cases}
\frac{c^2}{16\pi^2f_c^2}, & \min\limits_{q \in \mathcal{Q}} \{ d_{k,q}^{n,m} \} > 0, \\
0, & \min\limits_{q \in \mathcal{Q}} \{ d_{k,q}^{n,m} \} \le 0.
\end{cases}
\end{equation}
Thus, a non-zero path gain is assigned only when PA $m$ on waveguide $n$ maintains an unobstructed LoS path to user $u_k$.
Note that $\textbf{x}^{\rm{Pin}}_{n} \triangleq \{x_{n,1}^{\rm{Pin}}, ..., x_{n,M}^{\rm{Pin}}\}$ is defined as the set of all PA locations associated with waveguide $n$. 

Since the channel coefficient described in~\eqref{eq: step} is a discontinuous step function, we approximate it with a continuous and differentiable function as in~\cite{yi2023trajectory}:
\begin{equation}
\eta_{k}^{n,m} = \hat{\eta} \zeta_{k}^{n,m}, \ \hat{\eta} =  \tfrac{c^2}{16\pi^2f_c^2}, \ \forall n,m,k,
\end{equation}
where $\zeta_{k}^{n,m}$ employs a sigmoid function given by:
\begin{equation}\label{eq: sigmoid}
\zeta_{k}^{n,m} = \tfrac{1}{1+ {\rm exp}\left(-\vartheta \frac{{\rm min}_{q \in \mathcal{Q}} \{d_{k,q}^{n,m}\}}{\left\|\bm{\phi}_{k} - \bm{\psi}^{\rm{Pin}}_{n,m}\right\|} \right)},
\end{equation}
and $\vartheta$ is a smoothing parameter that governs the transition rate from LoS to NLoS conditions in~\eqref{eq: sigmoid}. The normalization by $\left\|\bm{\phi}_{k} - \bm{\psi}^{\mathrm{Pin}}_{n,m}\right\|$ ensures consistent channel blockage evaluation along the same direction with respect to user $u_k$~\cite{yi2023trajectory}. Furthermore, to extend the model to more practical conditions, we consider partial shadowing, where only a subset of antenna elements may be blocked by obstacles, as depicted in Fig.~\ref{fig:EavAntenna}, while the remaining elements retain LoS links to the PAs. Similarly, the wireless channel between the $M$ PAs on waveguide $n$ and antenna $t$ of EA $g$, denoted by $\tilde{\textbf{h}}_{n,g,t}^{\mathrm{E}} \in \mathbb{C}^{M \times 1}$, follows the same mathematical structure as that for a legitimate user and is therefore omitted for brevity.
The blockage-related metric is defined as $d_{g,t,q}^{n,m} = \underset{i \in \mathcal{I}_{g,t,q}}{\max}\left\{(\textbf{a}_{g,t,q}^i)^\top\bm{\psi}_{n,m}^{\rm Pin} - b_{g,t,q}^i\right\}$, in which $(\textbf{a}_{g,t,q}^i)^\top$ and $b_{g,t,q}^i$ follow the same geometric definitions and structure as legitimate users.

Since there are $M$ PAs positioned along the same waveguide, the signals transmitted by different PAs differ primarily in their relative phase shifts~\cite{ding2024flexible}. 
Accordingly, the phase-shift vector associated with the $M$ PAs positioned on waveguide $n, \forall n$, excluding propagation losses, is defined as~\cite{tegos2024minimum}:
\begin{align}\label{eq:channel_prop}
\bar{\textbf{h}}_n \left(\textbf{x}^{\rm{Pin}}_{n}\right) \hspace{-1mm}=\hspace{-1mm} \left[e^{-j\frac{2\pi \left\|\bm{\psi}_n^{\rm{FP}} - \bm{\psi}_{n,1}^{\rm{Pin}}\right\|}{\hat{\lambda}}}\hspace{-0.2em}, ..., e^{-j\frac{2\pi \left\|\bm{\psi}_n^{\rm{FP}} - \bm{\psi}_{n,M}^{\rm{Pin}}\right\|}{\hat{\lambda}}}\right]^\top \hspace{-0.5em},
\end{align}
where the guided wavelength is $\hat{\lambda} = \frac{\lambda}{\eta_{\mathrm{eff}}}$ with $\eta_{\mathrm{eff}} > 1$ represents the effective refractive index of the dielectric waveguide~\cite{ding2024flexible, liu2025pinching}. 
Therefore, we define the combined phase-shift vector, and the wireless channel vectors of user $u_{k}$ and antenna $t$ of EA $g$ as 
{\small$\bar{\textbf{h}} = \left[\bar{\textbf{h}}_{1}^\top, ..., \bar{\textbf{h}}_{N}^\top\right]^\top \in \mathbb{C}^{MN \times 1}$, 
$\Tilde{\textbf{h}}_{k} = \left[\Tilde{\textbf{h}}_{1, u_{k}}^\top, ..., \Tilde{\textbf{h}}_{N, u_{k}}^\top\right]^\top \in \mathbb{C}^{MN \times 1}$}, and 
{\small$\Tilde{\textbf{h}}_{g,t}^{\rm E} = \left[(\Tilde{\textbf{h}}_{1, g, t}^{\rm E})^\top, ..., (\Tilde{\textbf{h}}_{N, g, t}^{\rm E})^\top\right]^\top \in \mathbb{C}^{MN \times 1}$},
respectively. Consequently, the effective wireless channel vector towards legitimate user $u_{k}$, incorporating both phase-shift and wireless propagation components, is expressed as:
$\textbf{h}_{k} = \Tilde{\textbf{h}}_{k} \odot \bar{\textbf{h}}, \forall k$.
Similarly, the wireless channel matrix toward EA $g$, which comprises $T$ antennas, is defined as $\Tilde{\textbf{H}}_{{\rm E}_g} = [\Tilde{\textbf{h}}_{g,1}^{\rm E}, ..., \Tilde{\textbf{h}}_{g,T}^{\rm E}] \in \mathbb{C}^{MN \times T}$,
where the effective channel matrix corresponding to EA $g$ with the phase matrix $\bar{\textbf{H}} = \bar{\textbf{h}} [1, ..., 1]^\top \in \mathbb{C}^{MN \times T}$ is 
$\textbf{H}_{{\rm E}_g} = \Tilde{\textbf{H}}_{{\rm E}_g} \odot \bar{\textbf{H}}, \forall  g$.

\subsection{PA System CSI Uncertainty Models}
In the proposed system, limited pilot resources and the degraded channel estimation accuracy inherent to blocked channels render perfect CSI unattainable, even for legitimate users~\cite{zhou2025channel}. Consequently, despite utilizing pilot-based channel estimation techniques, the obtained CSI is inevitably imperfect. Furthermore, acquiring accurate CSI for EAs is even more challenging. Since EAs typically remain silent to conceal their existence and do not proactively cooperate with the BS~\cite{ren2025robust, hu2021robust}, precise CSI acquisition for such nodes is practically infeasible. To explicitly characterize these uncertainties, we assume that legitimate users leverage the pilot-based channel estimation method~\cite{zhou2025channel} and adopt a deterministic CSI uncertainty model~\cite{chen2025robust}. Therefore, the CSI uncertainty range from the $MN$ PAs to legitimate user $u_{k}$ is defined as:
\begin{align}
\bm{\Psi}_{k} = \{\textbf{h}_{k} | \hat{\textbf{h}}_{k} + \bm{\triangle}\textbf{h}_{k}, || \bm{\triangle}\textbf{h}_{k} ||_2 \leq \wp_{k}, \forall k \}.
\end{align}

On the other hand, although advanced localization techniques, such as various sensing technologies, can yield coarse estimates of an EA's position and orientation~\cite{xiu2024robust}, passive EAs may still reveal partial CSI information through local-oscillator power leakage at their front ends, even without interacting with the BS~\cite{yu2020robust,hu2021robust}. 
However, unlike in the far field, where a planar wavefront illuminates every element with practically identical amplitude, allowing orientation variations of a ULA can be absorbed into small perturbations of the effective direction of arrival. By contrast, since the impinging wavefront is spherical in the radiative near field, the element-dependent amplitudes and nonlinear phase curvature are highly sensitive to array orientation, thereby elevating orientation to a critical channel parameter for precise focal beamforming\cite{ramezani2025localization,xiu2024robust}.
Hence, this work explicitly captures the impacts of imperfect CSI arising from uncertainties in both positions and orientations of EAs.
Specifically, we assume that the BS possesses only coarse location information of each EA, denoted as $\{\hat{\textbf{e}}_{g} | \hat{\textbf{e}}_{g} = [\hat{x}_{g}^{\rm E}, \hat{y}_{g}^{\rm E}, 0]^\top, \forall g \in \mathcal{G}\}$. Given that the antenna spacing of each EA's ULA is $\frac{\lambda}{2}$~\cite{cui2022channel}, we can further estimate the positions of individual antenna elements as $\hat{\textbf{e}}_{g,t} = \hat{\textbf{e}}_{g} + t \frac{\lambda}{2} \textbf{u}_g, \forall g, t$, where $\textbf{u}_g = \textbf{R}_g(\triangle \theta_g) \hat{\textbf{u}}_g$, $\textbf{R}_g(\triangle \theta_g)$ and $\hat{\textbf{u}}_g$ are the rotation matrix and nominal direction vector for EA $g$, respectively:
\begin{align}\label{eq: orientaion}
\textbf{R}_g(\triangle \theta_g) \hat{\textbf{u}}_g  = \begin{bmatrix}
\cos(\hat{\theta_g} + \triangle \theta_g)\\
\sin(\hat{\theta_g} + \triangle \theta_g)\\
0
\end{bmatrix}, \forall g.
\end{align}
Hence, the position of antenna $t$ at EA $g$ can be modeled by the estimated reference location $\hat{\textbf{e}}_{g,t}$, a common\footnote{Recent studies have demonstrated that in near-field scenarios operating at $28$ GHz, the achievable estimation accuracy for 3D position is below $1$ cm, while rotation-matrix errors correspond to angular resolutions better than one degree~\cite{zhang2025array}. Such centimeter-level positioning accuracy and sub-degree orientation estimation justify employing a first-order linear approximation for the orientation perturbation $\triangle \theta_g$ in this work.} location perturbation, the orientation induced $\bm{\triangle} \textbf{e}_{g}$ and $t \frac{\lambda}{2} \textbf{u}_g$, i.e.,
\begin{align}\label{eq: location}
\textbf{e}_{g,t} \hspace{-0.8mm}=\hspace{-0.5mm} \hat{\textbf{e}}_{g} \hspace{-0.8mm}+\hspace{-0.5mm} \bm{\triangle} \textbf{e}_{g} \hspace{-0.8mm}+\hspace{-0.5mm} \tfrac{t\lambda}{2} \textbf{u}_g, \hspace{-0.5mm}|| \bm{\triangle} \textbf{e}_{g} || \hspace{-0.5mm}\leq\hspace{-0.5mm} \wp_{g}^{\rm E}\hspace{-0.5mm}, |\triangle \theta_g| \hspace{-0.5mm}\leq\hspace{-0.5mm} \wp^{\rm arc}_{g}, \forall g, t,
\end{align}
where the known bounds $\wp^{\rm arc}_{g}$, $\wp_{g}^{\mathrm{E}} > 0$ reflect the achievable localization and orientation accuracies~\cite{xiu2024robust,ren2025robust,cheng2021direction, schmidt1986multiple}.

Considering the above location and orientation-induced uncertainties, the resultant CSI error matrix for EA $g$, aggregated across all $MN$ PAs, $\bm{\triangle}\textbf{H}_{{\rm E}_g} = \textbf{H}_{{\rm E}_g} - \hat{\textbf{H}}_{{\rm E}_g} \forall g$, is given by:
\begin{align}\label{eq: CSIBound}
\bm{\triangle}\textbf{H}_{{\rm E}_g}  = \Tilde{\textbf{H}}_{{\rm E}_g} (\bm{E}) \odot \bar{\textbf{H}} - \Tilde{\textbf{H}}_{{\rm E}_g} (\hat{\bm{E}}) \odot \bar{\textbf{H}}, 
\end{align}
where $\hat{\textbf{H}}_{{\rm E}_g}$ represents the estimated effective channel matrix constructed from the estimated positions of $G$ EAs, each with $T$ antennas and accounting for orientation uncertainties, denoted collectively as $\hat{\bm{E}}$. Meanwhile, $\bm{E}$ represents the collection of the actual antenna positions of all EAs.
Consequently, the CSI uncertainty region for EA $g$ is expressed as:
\begin{align}\label{eq: Psi_eav}
\Psi_{g}^{\rm E} \hspace{-1mm}= &\big\{\hspace{-0.5mm}\textbf{H}_{{\rm E}_g} \hspace{-0.5mm}|\hat{\textbf{H}}_{{\rm E}_g} \hspace{-1.4mm}+\hspace{-1mm} \bm{\triangle} \textbf{H}_{{\rm E}_g}\hspace{-0.3mm},\hspace{-0.3mm} ||\hspace{-0.2mm}\triangle \textbf{e}_{g} \hspace{-0.2mm}|| \hspace{-0.3mm}\leq\hspace{-0.5mm} \wp_{g}^{\rm E}\hspace{-0.3mm},\hspace{-0.3mm} |\hspace{-0.2mm}\triangle \theta_g\hspace{-0.2mm}| \hspace{-0.3mm}\leq\hspace{-0.5mm} \wp^{\rm arc}_{g}\hspace{-0.5mm}, \hspace{-0.3mm}\forall g \hspace{-0.2mm}\big\}\hspace{-0.2mm}.
\end{align}
Note that the coupled CSI uncertainties originate from EAs in both location and orientation, severely impacting both large-scale path loss and element-wise phase alignment in the near-field. Moreover, the distributed architecture of PA arrays introduces path-specific channel perturbations that may partially offset one another under EA position perturbations, rendering existing linear uncertainty bounds~\cite{ren2025robust} inaccurate for PA systems. 
Consequently, deriving a tractable yet appropriately geometry-aware uncertainty bound for EA CSI constitutes one of the key technical challenges addressed in Section III-B.
\\ \textit{Remark 1: The uncertainty set} $\Psi_g^E$ \textit{in (10) also accommodates bounded EA mobility within one CSI and resource-allocation update interval. Assume that EA} $g$ \textit{is updated every} $\Delta t$ \textit{seconds, with bounded translational speed} $v_{g,\max}$ \textit{and angular rate} $\omega_{g,\max}$, \textit{then selecting} 
$\wp_g^E \ge \wp_{g,\mathrm{loc}} + v_{g,\max}\Delta t$ and 
$\wp_g^{\mathrm{arc}} \ge \wp_{g,\mathrm{ori}} + \omega_{g,\max}\Delta t$
\textit{guarantees that the resulting channel variation remains inside }$\Psi_g^E$~\cite{Wei2023}. \textit{Therefore, the worst-case secrecy constraint in (C6) continues to hold over the update interval, which reflects robustness to bounded EA mobility.}

\subsection{Waveguide Propagation Model}
\begin{figure}[t]
\centerline{\includegraphics[width=3.45in]{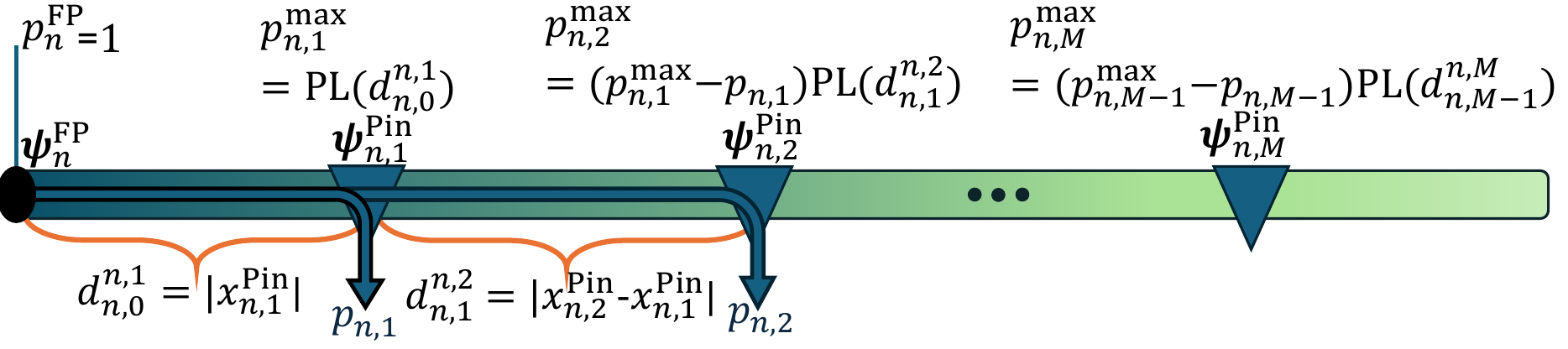}}
\caption{Illustration of the available power allocation among PAs along the $n$-th dielectric waveguide considering propagation attenuation.}
\label{fig:power}
\vspace{-1em}
\end{figure}

Noting that existing studies often oversimplify PA systems, limiting the accuracy of resource allocation strategies, this work adopts a more general model~\cite{ding2024flexible, hu2025, xu2025pinching}. In the proposed model, each PA can independently and dynamically adjust both its spatial position and allocated transmit-power ratio, while accounting for waveguide propagation losses. As depicted in Fig.~\ref{fig:power},
the propagation loss between two consecutive PAs, e.g., the $m$-th and ($m-1$)-th PA, in the $n$-th waveguide is denoted as ``PL($d^{n,m}_{n,m-1}$)'' with their spatial separation $d_{n,m-1}^{n,m} = x_{n,m}^{\rm Pin} - x_{n,m-1}^{\rm Pin}$.
The power-ratio allocated to the $m$-th PA in the $n$-th waveguide is denoted by $p_{n,m}$, which can be dynamically managed\footnote{In practice, the power-ratio extracted by each PA depends on its coupling length along the dielectric waveguide~\cite{wang2025modeling, okamoto2021fundamentals}. To abstract from specific hardware implementations, we assume that PAs dynamically adjust their coupling lengths via an auto-scaling mechanism, enabling precise control of the extracted power-ratio.} as well as its spatial position\footnote{Given the small physical length of PAs relative to practical waveguides~\cite{okamoto2021fundamentals, hu2025}, their impact on propagation loss is neglected.}, subject to constraints $0 \leq x_{n,m}^{\rm{Pin}} \leq L, d_{n,m-1}^{n,m} \geq \gamma$, the maximum available power-ratio at the $m$-th PA of the $n$-th waveguide, $\forall n,m$, can be mathematically modeled as follows:
\begin{align}\label{eq:power}
p^{\rm{max}}_{n,m} \hspace{-1mm}= e^{-2 \alpha \hspace{-0.2mm} \left\|\hspace{-0.3mm}\bm{\psi}_n^{\rm{FP}}\hspace{-0.7mm}-\bm{\psi}_{n,m}^{\rm{Pin}}\hspace{-0.6mm} \right\|} \hspace{-1mm}-\hspace{-1.8mm} \sum_{t=1}^{m-1} \hspace{-0.7mm} p_{n,t} e^{-2\alpha\hspace{-0.3mm}\left\|\hspace{-0.4mm}\bm{\psi}_{n,m}^{\rm{Pin}} \hspace{-0.5mm}- \bm{\psi}_{n,t}^{\rm{Pin}}\hspace{-0.5mm}\right\|}, \hspace{-0.5mm}
\end{align}
where $\alpha = \hat{\lambda} \epsilon_r \pi f_c^2c^{-2} \text{tan}\left(\delta_e\right)\in \mathbb{R}^+$~\cite{hu2025, balanis2024balanis}.

Furthermore, propagation losses within each waveguide are modeled leveraging a power-ratio vector for each waveguide, $\textbf{p}_n \in \mathbb{R}^{M \times 1}$, whose entries $\sqrt{p_{n,m}}, \forall n,m$, represent the amplitude-scaling factor applied to the corresponding PA, such that the power constraint $p_{n,m} \leq p^{\max}_{n,m}$ holds as shown in~\eqref{eq:power}.
Consequently, the introduced model inherently captures critical trade-offs among waveguide propagation losses, wireless path losses, and power allocation among the PAs, thereby significantly enhancing the practical relevance and effectiveness of subsequent resource allocation optimization.


\section{Problem Statement}
\subsection{Performance Metric}
To simplify notation and without loss of generality, an equivalent diagonal power-ratio allocation matrix for all PAs is defined as 
$\textbf{P} = \text{diag}\left[\textbf{p}_{1}, ..., \textbf{p}_{N}\right]^\top \in \mathbb{R}^{MN \times N}$.
Consequently, the received signals at user $k$ and EA $g$ are expressed as:
\begin{align}\label{rec_signal}
r_{k} &\hspace{-0.6mm}=\hspace{-0.6mm}  \underbrace{\textbf{h}_{k}^H \textbf{P}\textbf{w}_ks_k}_{\text{Desired signal}} + \underbrace{\underset{j \neq k}{\sum}\textbf{h}_{k}^H \textbf{P} \textbf{w}_{j}s_{j}}_{\text{Multi-user interference}} + \underbrace{\textbf{h}_{k}^H \textbf{P}\textbf{v}}_{\text{AN injection}} + \epsilon_{k},
\forall k, \nonumber \\
\textbf{r}_{g}^{\rm E} & \hspace{-0.5mm}=\hspace{-0.5mm} \textbf{H}_{{\rm E}_g}^H \hspace{-0.5mm}\textbf{P}\textbf{w}_ks_k \hspace{-0.5mm}+\hspace{-1.4mm} \cancel{\underset{j \neq k}{\sum} \hspace{-0.5mm}\textbf{H}_{{\rm E}_g}^H \hspace{-0.5mm} \textbf{P} \textbf{w}_{j}s_{j}} \hspace{-0.5mm}+\hspace{-0.5mm}  \textbf{H}_{{\rm E}_g}^H \hspace{-0.5mm}\textbf{P}\textbf{v} \hspace{-0.5mm}+\hspace{-0.5mm}  \bm{\epsilon}_{g},
\forall k, g,
\end{align}
respectively, where $\epsilon_{k} \sim \mathcal{CN}\left(0, \sigma_{k}^2\right)$ and $\bm{\epsilon}_{g} \sim \mathcal{CN}\left(\textbf{0}, \sigma_{g}^2 \textbf{I}_{T}\right)$ represent the additive white Gaussian noise (AWGN) at user $u_{k}$ and EA $g$, with $\sigma_{k}^2$ and $\sigma_{g}^2 \textbf{I}_{T}$ denoting the corresponding noise power and noise covariance matrix, respectively.
Thus, the achievable data rate for user $u_{k}$ can be written as:
\begin{equation}
R_{k} = \log_2\Big(1 + \tfrac{\left|\textbf{h}_{k}^H \textbf{P}\textbf{w}_k\right|^2}{\underset{j \neq k}{\sum}\Bigr|\textbf{h}_{k}^H \textbf{P} \textbf{w}_{j}\Bigr|^2 + \textbf{h}_{k}^H \textbf{P} \textbf{V}\textbf{P}^H \textbf{h}_{k} + \sigma_{k}^2}\Big), \forall k.
\end{equation}
Moreover, from a physical layer security perspective, we consider the worst-case scenario in which potential EAs possess advanced decoding capabilities, allowing them to perfectly remove multi-user interference and attempt to eavesdrop on the signal intended for user $k$~\cite{ng2016multiple, yu2020robust}. Thus, the eavesdropping capacity for decoding signal $s_k$ at EA $g$ is given by:
\begin{equation}
\label{eq: EAV_SINR}
    R^{\rm E}_{k,g} \hspace{-1mm}= \log_2 \text{det} \left(\textbf{I}_{T} + \textbf{J}_g^{-1}\textbf{H}_{{\rm E}_g}^H \textbf{P} \textbf{w}_k  \textbf{w}_k^H \textbf{P}^H\textbf{H}_{{\rm E}_g} \right), \forall k, g, 
\end{equation}
where
$\textbf{J}_g = \textbf{H}_{{\rm E}_g}^H \textbf{P} \textbf{VP}^H\textbf{H}_{{\rm E}_g} + \sigma_{g}^2 \textbf{I}_{T} \succ \textbf{0}$
represents the interference-plus-noise covariance matrix at EA $g$, accounting for AN and noise in the worst-case eavesdropping scenario.

\textit{\textbf{Remark 2}}: \textit{While the considered model assumes non-cooperative EAs, the proposed framework is not restricted to this assumption and can be readily extended to accommodate colluding EAs. In particular, if the $G$ EAs jointly process their intercepted signals, they can be equivalently modeled as a single distributed multi-antenna EA with a concatenated channel matrix} $\textbf{H}_\text{E,col}=[\textbf{H}_{E_1},\dots,\textbf{H}_{E_G}]\in\mathbb{C}^{MN\times GT}$. \textit{Consequently, the joint interference-plus-noise covariance becomes} $\textbf{J}_{\mathrm{col}}=\textbf{H}_\text{E,col}^{H}\textbf{P}\textbf{V}\textbf{P}^{H}\textbf{H}_\text{E,col} +\mathrm{blkdiag}(\sigma_1^2\textbf{I}_T,\dots,\sigma_G^2\textbf{I}_T)$~\cite{pinto2012collusion}, \textit{and the corresponding joint interception rate for user} $k$ \textit{is} $R^{E,\mathrm{col}}_{k}=\log_2\det\!\Big(\textbf{I}_{GT} +\textbf{J}_{\mathrm{col}}^{-1}\textbf{H}_\text{E,col}^{H}\textbf{P}\textbf{w}_k\textbf{w}_k^{H}\textbf{P}^{H}\textbf{H}_\text{E,col}\Big)$. \textit{In this context, the proposed robust design remains entirely applicable after replacing the individual leakage constraints with the corresponding joint leakage constraint}, $R^{E,\mathrm{col}}_{k}$, \textit{at the expense of increased matrix dimensionality in the subsequent transformations and optimization steps.}

\subsection{CSI Error Bound Determination}
Here, we derive a geometry‑aware Frobenius‑norm deterministic upper bound on the CSI error of EA $g$ in~\eqref{eq: CSIBound}, $\bm{\triangle} \textbf{H}_{{\rm E}_g} =\textbf{H}_{{\rm E}_g} -\hat{\textbf{H}}_{{\rm E}_g}$, induced by its position and orientation uncertainties.
Noting {\small$\left\|(\Tilde{\textbf{H}}_{{\rm E}_g}\hspace{-0.3mm} (\hspace{-0.3mm}\bm{E}_g\hspace{-0.3mm}) \hspace{-0.6mm}-\hspace{-0.6mm} \Tilde{\textbf{H}}_{{\rm E}_g}\hspace{-0.3mm} (\hspace{-0.3mm}\hat{\bm{E}}_g\hspace{-0.3mm}) 
) \hspace{-0.3mm}\odot \hspace{-0.3mm}\bar{\textbf{H}}\right\|  \hspace{-1mm}\leq\hspace{-1mm}  || \bar{\textbf{H}}||_\infty \hspace{-0.7mm} \left\|\Tilde{\textbf{H}}_{{\rm E}_g}\hspace{-0.3mm} (\hspace{-0.3mm}\bm{E}_g\hspace{-0.3mm}) \hspace{-0.6mm}-\hspace{-0.6mm} \Tilde{\textbf{H}}_{{\rm E}_g} \hspace{-0.3mm}(\hspace{-0.3mm}\hat{\bm{E}}_g\hspace{-0.3mm}) \right\|_F$}, and since {\small$||\bar{\textbf{H}}||_\infty = 1$} due to the unit-modulus nature, we establish the following proposition: \\
\textbf{Proposition 1:} \textit{Under the EA uncertainty model in}~\eqref{eq: orientaion} \textit{and}~\eqref{eq: location}, \textit{the CSI error of EA} $g$ \textit{satisfies }$\left\|\bm{\triangle}\textbf{H}_{{\rm E}_g} \right\|_F \leq \wp_g^{\rm tot}$, \textit{where:}
\begin{equation}\label{eq: CHannelBouns}
\wp_g^{\rm tot} \hspace{-1mm}= \hspace{-1mm}\tfrac{\lambda}{4 \pi} \sqrt{\hspace{-1mm}\sum^{N,M,T}_{n,m,t}\hspace{-2mm}\left(\hspace{-1mm}\left[\tfrac{\bm{\triangle}\textbf{R}_g^{(n,m,t)}}{(r_{g, {(\rm LB}}^{(n,m,t)}- \wp_g^{\rm E})^2}\right]^2 \hspace{-2.4mm}+\hspace{-0.6mm} \tfrac{2 \Omega (\tfrac{2\pi}{\lambda}\bm{\triangle}\textbf{R}_g^{(n,m,t)})}{(r_{g, {(\rm LB}}^{(n,m,t)}- \wp_g^{\rm E})^2}\hspace{-1mm}\right)},
\end{equation}
$\Omega(x) \triangleq 1 - \cos x$ \textit{if} $x < \pi$ \textit{and}  $\Omega(x) \triangleq 2$ \textit{otherwise. We denote} $\bm{\triangle} \textbf{R}_g^{(n,m,t)} \triangleq \wp_g^{\rm E} + t\lambda \sin (\tfrac{\wp^{\rm arc}_g}{2})\left| \hat{(\textbf{d}}_g^{(n,m,t)})^\top \textbf{R}_g (\pi/2)\hat{\textbf{u}}_g\right|$ \textit{is the geometry‑induced upper bound on} $|d-\hat{d}|^2$ \textit{obtained by projecting the orientation perturbation onto the LoS direction with} $\hat{\textbf{d}}_g^{(n,m,t)} \triangleq \tfrac{\hat{\textbf e}_g+\tfrac{t\lambda}{2}\hat{\textbf u}_g-\boldsymbol{\psi}^{\rm Pin}_{n,m}}{\hat{r}} \in \mathbb{R}^2$, $\hat{d}_g^{\hspace{-0.02mm}(\hspace{-0.5mm}n\hspace{-0.3mm},\hspace{-0.2mm}m\hspace{-0.2mm},\hspace{-0.2mm}t\hspace{-0.4mm})} = \left\|\hat{\textbf{e}}_{g,t} - \bm{\psi}^{\rm{Pin}}_{n,m}\right\|$, \textit{and} $d_g^{\hspace{-0.12mm}(\hspace{-0.5mm}n\hspace{-0.3mm},\hspace{-0.2mm}m\hspace{-0.2mm},\hspace{-0.2mm}t\hspace{-0.4mm})} = \left\| \textbf{e}_{g,t} - \bm{\psi}^{\rm{Pin}}_{n,m}\right\|$. \textit{The distance lower bound is:}
\begin{align}
&r_{g, {\rm LB}}^{(\hspace{-0.3mm}n\hspace{-0.1mm},\hspace{-0.1mm}m\hspace{-0.1mm},\hspace{-0.1mm}t\hspace{-0.2mm})}
\hspace{-1.5mm} = \hspace{-1.2mm}\Bigg[\hspace{-0.8mm}
\Big(\hspace{-0.8mm} \Big\{\hspace{-0.8mm} \big| \hat{\textbf u}_g^{\hspace{-0.8mm}\top}\!\big(\hat{\textbf e}_g \hspace{-0.8mm}+ \hspace{-0.8mm}\tfrac{t\lambda}{2}\hat{\textbf u}_g \hspace{-0.8mm}- \hspace{-0.8mm}\boldsymbol{\psi}^{\rm Pin}_{n,m}\big)
\big| \hspace{-0.8mm}- \hspace{-0.8mm} \tfrac{t\lambda}{2}\hspace{-0.9mm}\left(\hspace{-0.8mm}1 \hspace{-0.8mm}-\hspace{-0.8mm}\cos\!\Big(\hspace{-0.8mm}\wp_g^{\rm arc}\hspace{-0.8mm}\Big) \hspace{-0.5mm}\right)\hspace{-0.9mm}\Big\}_{\hspace{-0.7mm}+} \hspace{-0.3mm}\Big)^{\hspace{-0.7mm}2} \nonumber \\
& \; +\hspace{-1mm}\Big(\hspace{-0.6mm} \Big\{\hspace{-0.6mm} \big|
\big(\hspace{-0.4mm}\textbf R_g \hspace{-0.2mm}(\hspace{-0.2mm}\tfrac{\pi}{2}\hspace{-0.2mm})\hat{\textbf u}_g\hspace{-0.2mm}\big)^{\hspace{-1mm}\top}
\hspace{-0.7mm}\big(\hspace{-0.2mm}\hat{\textbf e}_g \hspace{-0.8mm}+\hspace{-0.7mm} \tfrac{t\lambda}{2}\hat{\textbf u}_g \hspace{-0.9mm}- \hspace{-0.6mm}\boldsymbol{\psi}^{\rm Pin}_{n,m} \hspace{-0.2mm}\big) \big|
\hspace{-0.7mm}-\hspace{-0.7mm} \tfrac{t\lambda \wp_g^{\rm arc}}{2}\hspace{-0.7mm}
\Big\}_{\hspace{-0.5mm}+}
\Big)^{\hspace{-1mm}2}
\hspace{-0.7mm}\Bigg]^{\hspace{-1mm}\frac{1}{2}}\hspace{-1mm}.
\end{align}
\vspace{-2em}
\begin{proof}
A straightforward expansion of the uncertainty of each link from a PA to an EA antenna yields:
\begin{align}\label{eq: profGamma}
&\Gamma_g^{(n,m,t)} = \left|\tfrac{e^{-j\frac{2\pi}{\lambda}d_g^{\hspace{-0.12mm}(\hspace{-0.5mm}n\hspace{-0.3mm},\hspace{-0.2mm}m\hspace{-0.2mm},\hspace{-0.2mm}t\hspace{-0.4mm})}} }{d_g^{\hspace{-0.02mm}(\hspace{-0.5mm}n\hspace{-0.3mm},\hspace{-0.2mm}m\hspace{-0.2mm},\hspace{-0.2mm}t\hspace{-0.4mm})}} - \tfrac{ e^{-j\frac{2\pi}{\lambda}\hat{d}_g^{\hspace{-0.02mm}(\hspace{-0.5mm}n\hspace{-0.3mm},\hspace{-0.2mm}m\hspace{-0.2mm},\hspace{-0.2mm}t\hspace{-0.4mm})}} }{\hat{d}_g^{\hspace{-0.12mm}(\hspace{-0.5mm}n\hspace{-0.3mm},\hspace{-0.2mm}m\hspace{-0.2mm},\hspace{-0.2mm}t\hspace{-0.4mm})}}\right|^2 \nonumber \\
=& \hspace{-1mm}\Big(\hspace{-1mm}\tfrac{1}{d_g^{\hspace{-0.1mm}(\hspace{-0.5mm}n\hspace{-0.3mm},\hspace{-0.2mm}m\hspace{-0.2mm},\hspace{-0.2mm}t\hspace{-0.4mm})}} \hspace{-0.3mm}- \hspace{-0.2mm}\tfrac{1}{\hat{d}_g^{\hspace{-0.02mm}(\hspace{-0.5mm}n\hspace{-0.3mm},\hspace{-0.2mm}m\hspace{-0.2mm},\hspace{-0.2mm}t\hspace{-0.4mm})}}\hspace{-1.4mm}\Big)^{\hspace{-1.4mm}2}  \hspace{-0.5mm}+\hspace{-0.3mm}
\tfrac{2\left(1-\cos \left(\frac{2\pi}{\lambda} ( d_g^{\hspace{-0.12mm}(\hspace{-0.5mm}n\hspace{-0.3mm},\hspace{-0.2mm}m\hspace{-0.2mm},\hspace{-0.2mm}t\hspace{-0.4mm})} - \hat{d}_g^{\hspace{-0mm}(\hspace{-0.5mm}n\hspace{-0.3mm},\hspace{-0.2mm}m\hspace{-0.2mm},\hspace{-0.2mm}t\hspace{-0.4mm})}) \right)   \right) }{d_g^{\hspace{-0.12mm}(\hspace{-0.5mm}n\hspace{-0.3mm},\hspace{-0.2mm}m\hspace{-0.2mm},\hspace{-0.2mm}t\hspace{-0.4mm})} \hat{d}_g^{\hspace{-0.0mm}(\hspace{-0.5mm}n\hspace{-0.3mm},\hspace{-0.2mm}m\hspace{-0.2mm},\hspace{-0.2mm}t\hspace{-0.4mm})}},
\end{align}
therefore 
\begin{equation}\label{eq: realBounds}
\left\|\Tilde{\textbf{H}}_{{\rm E}_g} (\bm{E}_g) - \Tilde{\textbf{H}}_{{\rm E}_g} (\hat{\bm{E}}_g) \right\|_F=\hspace{-0.5mm}\tfrac{\lambda \sqrt{\underset{n,m,t}{\sum}\Gamma_g^{(n,m,t)}}}{4\pi}.
\end{equation}
From the geometry in Fig. 2 and the uncertainty bounds in~\eqref{eq: orientaion} and~\eqref{eq: location}, the link distances satisfy $\left|d_g^{\hspace{-0.12mm}(\hspace{-0.5mm}n\hspace{-0.3mm},\hspace{-0.2mm}m\hspace{-0.2mm},\hspace{-0.2mm}t\hspace{-0.4mm})} - \hat{d}_g^{\hspace{-0.12mm}(\hspace{-0.5mm}n\hspace{-0.3mm},\hspace{-0.2mm}m\hspace{-0.2mm},\hspace{-0.2mm}t\hspace{-0.4mm})}\right| \leq \bm{\triangle}\hspace{-0.4mm}\textbf{R}_g^{\hspace{-0.12mm}(\hspace{-0.5mm}n\hspace{-0.3mm},\hspace{-0.2mm}m\hspace{-0.2mm},\hspace{-0.2mm}t\hspace{-0.4mm})}$ and $d_g^{\hspace{-0.12mm}(\hspace{-0.5mm}n\hspace{-0.3mm},\hspace{-0.2mm}m\hspace{-0.2mm},\hspace{-0.2mm}t\hspace{-0.4mm})}, \hat{d}_g^{\hspace{-0.12mm}(\hspace{-0.5mm}n\hspace{-0.3mm},\hspace{-0.2mm}m\hspace{-0.2mm},\hspace{-0.2mm}t\hspace{-0.4mm})} \geq r_{g, {\rm LB}}^{(\hspace{-0.3mm}n\hspace{-0.1mm},\hspace{-0.1mm}m\hspace{-0.1mm},\hspace{-0.1mm}t\hspace{-0.2mm})} - \wp_g^{\rm E}$. Substituting these bounds into~\eqref{eq: profGamma} and applying the definition of the function $\Omega(x)$, then summing over all $(n,m,t)$ yields the channel uncertainty upper bounds in~\eqref{eq: CHannelBouns}. 
\end{proof}
\vspace{-0.6em}
Thus, Proposition 1 converts the coupled EA position and orientation uncertainties into a single tractable Frobenius‑norm bound in~\eqref{eq: CHannelBouns}, which will be directly employed later.

\subsection{Problem Formulation}
We aim to maximize the sum of achievable rates for legitimate users\footnote{The proposed formulation incorporates information leakage constraints to provide higher resource allocation flexibility for heterogeneous applications compared to direct secrecy rate maximization~\cite{yu2020robust, hu2021robust}. Notably, the leakage threshold $R^{\rm{th}}_{k,g}$ enables the system operator to balance sum and secrecy rates.} while explicitly bounding the information decoded by EAs. Hence, the joint optimization of information and AN beamforming, the PA positions, and the individual PA power-ratio allocations is formulated as:
\begin{align}
&\underset{\textbf{V}\bm{\succeq 0} \in \mathbb{H}^{N}, \textbf{w}_k, \ \textbf{x}^{\rm{Pin}}_n, \ \textbf{P}} {\rm{maximize}} \hspace{2mm} \underset{k \in \mathcal{K}}{\sum} \hspace{2mm} \underset{\bm{\triangle}\textbf{h}_{k} \in \Psi_{k}}{\rm{min}} \{R_{k}\} \nonumber \\
\rm{s.t.}\ &({\rm C1}): 0 \leq x_{n,m}^{\rm{Pin}} \leq L, \ \forall n, m, \nonumber \\
&({\rm C2}): x_{n,m+1}^{\rm{Pin}} - x_{n,m}^{\rm{Pin}} \geq \gamma, \ \forall n,m\in\{1,...,M-1\}, \nonumber \\
&({\rm C3}): (\hspace{-0.4mm}\textbf{s}^{\rm C3}_{n,m}\hspace{-0.3mm})\hspace{-0.4mm}^\top \hspace{-0.5mm} \textbf{p}_n \hspace{-0.7mm} \leq \hspace{-0.7mm} e\hspace{-0.5mm}^{-2 \alpha \hspace{-0.3mm}\left\|\hspace{-0.3mm}\bm{\psi}_n^{\rm{FP}} \hspace{-1mm}-\bm{\psi}_{n,m}^{\rm{Pin}}\hspace{-0.5mm} \right\|},  \forall n,m, \nonumber \\
&({\rm C4}): \sum_{k=1}^{K}|w_{k,n}|^2 + \text{Tr}(\textbf{VE}_n)\leq P_n^{\max}, \forall n, \nonumber \\
&({\rm C5}): \sum_{n=1}^{N}P_n^{\max} \leq P_{\max}, \nonumber \\
&({\rm C6}): \underset{\bm{\triangle}\textbf{H}_{{\rm E}_g} \in \bm{\Psi}_g^{\rm E}}{\max} \ R_{k,g}^{\rm E} \leq R^{\rm{th}}_{k,g}, \ \forall k,g, 
\label{eq:formulation}
\end{align}
where $\textbf{s}^{\rm C3}_{n,m} \in \mathbb{R}^{M \times 1}$ is a predefined power-selection vector for the PA $m$ on the waveguide $n$. Specifically, $[\textbf{s}^{\rm C3}_{n,m}]_i = 1, \forall i = m$, $[\textbf{s}^{\rm C3}_{n,m}]_i = e^{-2 \alpha \left\|\bm{\psi}_{n,m}^{\rm{Pin}}-\bm{\psi}_{n,i}^{\rm{Pin}} \right\|}$, $ \forall 1 \leq i < m$, and $[\textbf{s}^{\rm C3}_{n,m}]_i = 0, \forall i > m$. Constraints $({\rm C1})$ and $({\rm C2})$ ensure that PA positions remain within the physical waveguide and maintain the minimum spacing $\gamma$ between adjacent PAs, respectively, for mitigating potential mutual coupling effects~\cite{zhu2025pinching}. Besides, constraint $({\rm C3})$ limits the allocated power-ratio $p_{n,m}$ of each PA to its maximum allowable value determined by the propagation attenuation and power-splitting model described previously. Furthermore, constraints $({\rm C4})$ and $({\rm C5})$ ensure that the total allocated power for each waveguide and across all waveguides does not exceed the available power budget for each waveguide $P_n^{\rm{max}}$ and the overall system power budget $P_{\rm{max}}$, respectively. Finally, constant $R^{\rm{th}}_{k,g}$ in constraint $({\rm C6})$ imposes an upper bound on the information leakage to EAs under worst-case CSI uncertainty, explicitly incorporating estimation imperfections.
Indeed, problem \eqref{eq:formulation} is inherently nonconvex due to the strong coupling among variables. Hence, we adopt a BCD scheme that alternately optimizes the blocks $\{\textbf{W}_k, \textbf{V}, \textbf{P}\}$ and $\{\textbf{x}_{n}^{\rm Pin}\}$ to obtain a high-quality suboptimal solution.

\section{Solution for Joint Optimization of $\textbf{w}_k, \textbf{V}$ and $\textbf{P}$}
In this section, we address the first subproblem of our proposed BCD framework, which can be formulated as:
\begin{align}
\underset{\textbf{V} \in \mathbb{H}^{N}, \ \textbf{w}_k, \ \textbf{P}} {\rm{maximize}} \hspace{2mm} \underset{k \in \mathcal{K}}{\sum} \hspace{2mm} \underset{\bm{\triangle}\textbf{h}_{k} \in \Psi_{k}}{\rm{min}} \{R_{k}\} \; \;
\rm{s.t.}\ ({\rm C3}) - ({\rm C6}).
\label{eq:formulation111}
\end{align}
The problem in~\eqref{eq:formulation111} remains nonconvex due to the quadratic coupling among the optimization variables, together with the infinitely many uncertainty-induced constraints, originating from the fractional rate expression and constraint $(\mathrm{C6})$.

\subsection{Problem Transformation}
To address the nonconvexity in the objective function of~\eqref{eq:formulation111}, we first introduce $\textbf{W}_k = \textbf{w}_k\textbf{w}_k^H, \forall k$, and then establish a commonly adopted performance lower bound~\cite{hu2021robust}, as:
\begin{align}
\label{eq: lower}
&\underset{||\bm{\triangle}\textbf{h}_{k} ||\leq \wp_{k}}{\rm{min}} \hspace{-2mm} \left\{\hspace{-1mm} \log_2 \hspace{-1mm}\left(\hspace{-1.8mm} 1 \hspace{-1mm} + \hspace{-1mm}\tfrac{\textbf{h}_{k}^H \textbf{P}\textbf{W}_k \textbf{P}^H \textbf{h}_{k}}{\underset{j \neq k}{\sum}\textbf{h}_{k}^H \textbf{P} \textbf{W}_{j} \textbf{P}^H \textbf{h}_{k} + \textbf{h}_{k}^H \textbf{P} \textbf{V} \textbf{P}^H \textbf{h}_{k} + \sigma_{k}^2}\right)\hspace{-1.7mm}\right\} \hspace{-1mm} \nonumber \\
\geq \hspace{-0.6mm} &  \log_2 \hspace{-1mm} \left(\hspace{-1.4mm} 1 \hspace{-0.7mm}+ \hspace{-0.7mm} \tfrac{ \underset{||\bm{\triangle}\textbf{h}_{k} || \leq \wp_{k}}{\rm{min}} \bigr\{\textbf{h}_{k}^H \textbf{P}\textbf{W}_k \textbf{P}^H \textbf{h}_{k}\bigr\}}{ \underset{||\bm{\triangle}\textbf{h}_{k} ||\leq \wp_{k}}{\rm{max}} \hspace{-0.7mm} \left\{\hspace{-0.3mm} \underset{j \neq k}{\sum}\textbf{h}_{k}^H \textbf{P} \textbf{W}\hspace{-0.6mm}_{j} \textbf{P}\hspace{-0.2mm}^H \textbf{h}_{k}\hspace{-0.4mm} \hspace{-0.4mm} + \textbf{h}_{k}^H \textbf{P} \textbf{V} \textbf{P}^H \textbf{h}_{k} \hspace{-0.2mm} \hspace{-0.2mm} + \sigma_{k}^2\hspace{-0.5mm}\right\}}\hspace{-0.6mm}\right)\hspace{-0.8mm}.
\end{align}
Leveraging logarithmic identities and introducing slack variables $\iota_k^{\mathrm N}$ and $\iota_k^{\mathrm D}$, the objective lower bound in~\eqref{eq: lower} can be equivalently rewritten in a difference-of-convex (DC) form as:
\begin{align}\label{eq: bounds}
&\underset{k \in \mathcal{K}}{\sum} \log_2 (\iota_k^{\rm N} + \iota_k^{\rm D} + \sigma_k^2) - \underset{k \in \mathcal{K}}{\sum} \log_2 (\iota_k^{\rm D} + \sigma_k^2) \; \; \text{s.t.}   \nonumber \\
&({\rm C7})\hspace{-1mm}: \hspace{-0.5mm}\iota_k^{\rm 
N} \hspace{-1mm} + \hspace{-1.8mm} \underset{||\bm{\triangle}\textbf{h}_{k} ||\leq \wp_{k}}{\rm{max}} \hspace{-1.7mm} \left\{\hspace{-0.8mm} - \hspace{-0.7mm}(\hspace{-0.25mm}\hat{\textbf{h}}_{k} \hspace{-0.8mm}+ \hspace{-0.8mm} \bm{\triangle}\textbf{h}_{k}\hspace{-0.6mm})\hspace{-0.4mm}^H \hspace{-0.6mm} \textbf{P}\textbf{W}\hspace{-0.7mm}_k \textbf{P}^H \hspace{-0.9mm}(\hat{\textbf{h}}_{k} \hspace{-1.0mm} + \hspace{-1.0mm} \bm{\triangle}\textbf{h}_{k} \hspace{-0.5mm})\hspace{-0.5mm}\hspace{-0.5mm} \right\} \hspace{-1mm} \leq \hspace{-1mm} 0,\hspace{-0.5mm} \forall k, \nonumber \\
&({\rm C8})\hspace{-1mm}: \hspace{-1mm} \underset{||\bm{\triangle}\textbf{h}_{k} ||\leq \wp_{k}}{\rm{max}} \hspace{-1mm} \left\{\hspace{-0.5mm}\underset{j \neq k}{\sum}\hspace{-0.5mm}(\hat{\textbf{h}}_{k} \hspace{-1mm}+\hspace{-1mm}\bm{\triangle}\textbf{h}_{k})^H \hspace{-0.5mm} \textbf{P} \textbf{W}_{j} \textbf{P}^H \hspace{-1mm}(\hat{\textbf{h}}_{k} \hspace{-1mm}+ \hspace{-1mm}\bm{\triangle}\textbf{h}_{k}) \right. \nonumber \\
&\left. \hspace{20mm} + (\hat{\textbf{h}}_{k} \hspace{-1mm}+ \hspace{-1mm}\bm{\triangle}\textbf{h}_{k}\hspace{-0.2mm})\hspace{-0.2mm}^H \hspace{-0.5mm}\textbf{P} \textbf{V} \textbf{P}^H \hspace{-0.8mm}(\hat{\textbf{h}}_{k} \hspace{-1mm}+\hspace{-1mm}\bm{\triangle}\textbf{h}_{k}\hspace{-0.4mm})\hspace{-0.3mm} \hspace{-0.3mm}\right\} \hspace{-0.7mm} \leq \hspace{-0.7mm} \iota_k^{\rm 
D}, \forall k, \nonumber \\
&({\rm C9}): \textbf{W}_k \bm{\succeq 0}, \forall k, 
\hspace{5mm} ({\rm C10}): \textbf{W}_k = \textbf{w}_k\textbf{w}_k^H, \forall k.
\end{align} 

Moreover, constraint $({\rm C6})$ is highly nonconvex, owing to the quadratic coupling of optimization variables and the presence of infinitely many inequality constraints arising from the corresponding CSI uncertainty sets. To overcome this difficulty, we convert it into a tractable form through linear matrix inequalities (LMIs) by defining $\hat{R}^{\rm th}_{k,g}=2^{R^{\rm th}_{k,g}-1}$ and introducing the following proposition and equivalent transformation:

\textbf{Proposition 2:} \textit{Since} ${\rm \textbf{J}_g^{-\frac{1}{2}}\textbf{H}_{{\rm E}_g}^H \textbf{P} \textbf{W}_k \textbf{P}^H\textbf{H}_{{\rm E}_g} \textbf{J}_g^{-\frac{1}{2}}}$ \textit{is a rank-one matrix, constraint} $({\rm C6})$ \textit{in}~\eqref{eq:formulation} \textit{is equivalent to:}
\begin{align}\label{eq: Pro2}
({\rm C6}) \Leftrightarrow &\hspace{-2mm}\underset{||\bm{\triangle} \textbf{H}_{{\rm E}_g}||_F \leq \wp_g^{\rm tot}}{\max}  \hspace{-1mm}\textbf{H}_{{\rm E}_g}^H \textbf{P} \textbf{W}_k \textbf{P}^H\textbf{H}_{{\rm E}_g} \nonumber \\ 
&- \hat{R}^{\rm th}_{k,g}\left( \textbf{H}_{{\rm E}_g}^H \textbf{P} \textbf{VP}^H\textbf{H}_{{\rm E}_g} \hspace{-0.7mm}+\hspace{-0.7mm} \sigma_{g}^2 \textbf{I}_{T} \right) \bm{\preceq 0}, \forall k,g.
\end{align}
\begin{proof}
According to Sylvester’s determinant identity $\det(\textbf{I} + \textbf{AB}) = \det(\textbf{I} + \textbf{BA})$~\cite{yu2020robust, hu2021robust}, we have
\begin{align}\label{eq: prof}
&\max_{\bm{\triangle} \textbf{H}_{E_g} \in \bm{\Psi}_g^{\rm E}} \log_2 \det\Big( \textbf{I}_T + \textbf{J}_g^{-1} \textbf{H}_{E_g}^H \textbf{P} \textbf{w}_k \textbf{w}_k^H \textbf{P}^H \textbf{H}_{E_g} \Big) \le R_{k,g}^{\rm th} \nonumber \\
&\Leftrightarrow \max_{\bm{\triangle} \textbf{H}_{E_g} \in \bm{\Psi}_g^{\rm E}} \log_2\Big(1 + \textbf{w}_k^H \textbf{P}^H \textbf{H}_{E_g} \textbf{J}_g^{-1} \textbf{H}_{E_g}^H \textbf{P w}_k\Big) \le R_{k,g}^{\rm th} \nonumber \\
&\Leftrightarrow \max_{\bm{\triangle} \textbf{H}_{E_g} \in \bm{\Psi}_g^{\rm E}} \text{Tr}\Big( \textbf{J}_g^{-1} \textbf{H}_{E_g}^H\textbf{ P w}_k \textbf{w}_k^H \textbf{P}^H \textbf{H}_{E_g} \Big) \le \hat{R}_{k,g}^{\text{th}} \nonumber \\
&\overset{(a)}{\Leftrightarrow} \hspace{-3mm} \max_{\bm{\triangle} \textbf{H}_{E_g} \in \bm{\Psi}_g^{\rm E}} \hspace{-1mm} \lambda_{\max}\Big( \textbf{J}_g^{-1/2} \textbf{H}_{E_g}^H \textbf{P w}_k \textbf{w}_k^H \textbf{P}^H \textbf{H}_{E_g} \textbf{J}_g^{-1/2} \Big) \hspace{-1mm}\le\hspace{-1mm} \hat{R}_{k,g}^{\text{th}} \nonumber \\
&\Leftrightarrow \hspace{-3mm} \max_{\bm{\triangle} \textbf{H}_{E_g} \in \bm{\Psi}_g^{\rm E}} \hspace{-1.2mm} \Big( \hspace{-1mm}\textbf{H}_{E_g}^H \textbf{P} \textbf{w}_k \textbf{w}_k^H \textbf{P}^H \textbf{H}_{E_g} \hspace{-1.5mm}-\hspace{-0.8mm} (2^{R_{k,g}^{\text{th}}} \hspace{-0.8mm}-\hspace{-0.8mm} 1) \textbf{J}_g \hspace{-0.8mm}\Big) \bm{\preceq 0},
\end{align}
where $(a)$ holds since the matrix $\textbf{J}_g^{-1} \textbf{H}_{E_g}^H \textbf{P} \textbf{w}_k \textbf{w}_k^H \textbf{P}^H \textbf{H}_{E_g}$ is rank one, and thus its largest eigenvalue equals its sole nonzero eigenvalue. Substituting the definition of $\textbf{J}_g$ into the last matrix inequality of~\eqref{eq: prof} yields Proposition~2.
\end{proof}
Although $(\mathrm{C6})$ still comprises infinitely many constraints, its LMI reformulation in~\eqref{eq: Pro2} substantially improves tractability. To further handle the quadratic coupling between $\{\textbf{P},\textbf{W}_k\}$ and $\{\textbf{P},\textbf{V}\}$, we recast $(\mathrm{C6})$ as:
\begin{align}\label{eq: Non_Convex}
&(\overline{{\rm C6}})\hspace{-1mm}: \hspace{-0.7mm}\underset{||\bm{\triangle}\textbf{H}_{{\rm E}_g}||_{F} \leq \wp_g^{\rm tot}}{\max} \hspace{-1mm} \textbf{H}_{{\rm E}_g}^H  \textbf{Y}_k \hspace{-0.3mm} \textbf{H}_{{\rm E}_g} \hspace{-0.7mm} \nonumber \\
& \hspace{20mm}- \hat{R}^{\rm th}_{k,g} \hspace{-0.2mm} \left(\textbf{H}_{{\rm E}_g}^H  \textbf{Z}\textbf{H}_{{\rm E}_g} \hspace{-0.5mm} +  \sigma_{g}^2 \textbf{I}_{T}\right) \bm{\preceq}  \bm{0}, \forall k,g, \nonumber \\ 
&({\rm C11})\hspace{-1mm}: \textbf{y}_k = \textbf{Pw}_k, \  \textbf{Y}_k = \textbf{y}_k \textbf{y}_k^H, \forall k, \nonumber \\
&({\rm C12})\hspace{-1mm}: \textbf{z} = \textbf{Pv}, \  \textbf{Z} = \textbf{z} \textbf{z}^H,\; \;  ({\rm C13})\hspace{-1mm}: \textbf{V} = \textbf{v} \textbf{v}^H.
\end{align}
Noting that new equality constraints induce nonconvexity, we adopt the Schur complement following~\cite{hu2021robust, wu2025globally}. Taking $\textbf{y}_k=\textbf{P}\textbf{w}_k$ as an example, it can be equivalently represented as:
\begin{align}
&({\rm C11a})\hspace{-1mm}:\hspace{-0.5mm}
\left[
  \begin{array}{ccc}
    \textbf{U}_k & \textbf{y}_k      & \textbf{P}         \\
    \textbf{y}_k^{H} & s_k  & \textbf{w}_{k}^{H} \\
    \textbf{P}^{H} & \textbf{w}_k  & \textbf{I}_{N}
  \end{array}
\right]
\bm{\succeq} \bm{0}, \forall k, \nonumber \\
&({\rm C11b})\hspace{-1mm}:\hspace{-1mm} \text{Tr}(\textbf{U}_k \hspace{-1mm}-\hspace{-0.8mm} \textbf{PP}^H\hspace{-0.5mm}) \hspace{-0.8mm} \leq \hspace{-0.5mm}0,  \hspace{-1mm}\forall k, 
({\rm C11c})\hspace{-1mm}:\hspace{-0.67mm}\textbf{U}_k\hspace{-0.65mm} \bm{\succeq} \hspace{-0.65mm} \bm{0}, s_k \hspace{-0.65mm}\ge \hspace{-0.65mm}0, \hspace{-1mm} \forall k,     
\end{align}
where $\textbf{U}_k \in \mathbb{C}^{MN \times MN}$ and $s_k$ are slack optimization variables. The other nonconvex equality in $({\rm C11})$, $\textbf{Y}_k = \textbf{y}_k \textbf{y}_k^H$, can be replaced by the following two constraints~\cite{hu2021robust}, as:
\begin{equation}\label{eq: shur}
\hspace{-2mm}({\rm C11d})\hspace{-1mm}:
\hspace{-0.5mm}\bm{\mathcal{Y}}_k \hspace{-0.8mm}=\hspace{-1.1mm} \left[\hspace{-1.4mm}
\begin{array}{cc}
    \textbf{Y}_k      \hspace{-1.5mm}& \textbf{y}_k         \\
    \textbf{y}_k^{H} \hspace{-1.5mm}& 1 
  \end{array}
      \hspace{-2mm} \right] \hspace{-1mm} \bm{\succeq} \hspace{-0.7mm} \bm{0}, \hspace{-0.5mm}\forall k, 
      ({\rm C11e})\hspace{-1mm}: \text{Rank} \hspace{-0.2mm}(\hspace{-0.25mm}\bm{\mathcal{Y}}_k\hspace{-0.25mm}) \hspace{-1mm}=\hspace{-1mm} 1, \hspace{-0.5mm}\forall k\hspace{-0.3mm}.
\end{equation}

Following the same structure, constraints $({\rm C10})$, $({\rm C12})$, and $({\rm C13})$ can be equivalently written as:
\begin{align}
& ({\rm C10a})\hspace{-1mm}:\hspace{-0.5mm} 
\bm{\mathcal{W}}_k \hspace{-1mm}=\hspace{-1mm} \left[\hspace{-1.7mm}
\begin{array}{cc}
    \textbf{W}_k      \hspace{-1.7mm}& \textbf{w}_k         \\
    \textbf{w}_k^{H} \hspace{-1.3mm}& 1 
  \end{array}
      \hspace{-2mm}\right] \hspace{-1.5mm} \bm{\succeq} \hspace{-1mm} \bm{0}, \hspace{-1mm}\forall k,   
      ({\rm C10b})\hspace{-1mm}:\hspace{-0.5mm} \text{Rank} (\hspace{-0.2mm}\bm{\mathcal{W}}_k\hspace{-0.5mm}) \hspace{-1mm}=\hspace{-1mm} 1, \forall k, \nonumber \\
&({\rm C12a})\hspace{-1mm}:\hspace{-0.5mm}
\left[\hspace{-1mm}
  \begin{array}{ccc}
    \bar{\textbf{U}} & \textbf{z}      & \textbf{P}         \\
    \textbf{z}^{H} & \bar{s}  & \textbf{v}^{H} \\
    \textbf{P}^{H} & \textbf{v}  & \textbf{I}_{N}
  \end{array}
\hspace{-1mm}\right]
\bm{\succeq} \bm{0}, \; \;
({\rm C12b})\hspace{-1mm}:\hspace{-0.5mm} \text{Tr}(\bar{\textbf{U}} - \textbf{PP}^H) \leq 0,  \nonumber \\
& ({\rm C12c})\hspace{-1mm}:\hspace{-0.5mm}  \bar{\textbf{U}} \bm{\succeq 0}, \bar{s} \ge 0, \;\;
({\rm C12d})\hspace{-1mm}:\hspace{-0.5mm}
\bm{\mathcal{Z}} = \left[
\begin{array}{cc}
    \textbf{Z}      & \textbf{z}         \\
    \textbf{z}^{H} & 1 
  \end{array}
      \right] \bm{\succeq} \bm{0}, \nonumber \\
&({\rm C12e})\hspace{-1mm}:\hspace{-0.5mm} \text{Rank} (\bm{\mathcal{Z}}) = 1, \nonumber \\
& ({\rm C13a})\hspace{-1mm}:\hspace{-0.5mm} 
\bm{\mathcal{V}} = \left[
\begin{array}{cc}
    \textbf{V}      & \textbf{v}         \\
    \textbf{v}^{H} & 1 
  \end{array}
      \right] \bm{\succeq} \bm{0},\;   
      ({\rm C13b})\hspace{-1mm}:\hspace{-0.5mm} \text{Rank} (\bm{\mathcal{V}}) = 1.
\end{align}
With the transformation in (26)-(28), the constraint sets $({\rm C11a})$ through $({\rm C11d})$ and $({\rm C12a})$ through $({\rm C12d})$ enforce the relations $\textbf{Y}_k=\textbf{P}\textbf{W}_k\textbf{P}^H, \forall k$, and $\textbf{Z}=\textbf{P}\textbf{V}\textbf{P}^H$, respectively. These permit the direct substitution of $\textbf{Y}_k$ and $\textbf{Z}$ for $\textbf{P}\textbf{W}_k\textbf{P}^H$ and $\textbf{P}\textbf{V}\textbf{P}^H$ in constraints $({\rm C7})$ and $({\rm C8})$ in~\eqref{eq: bounds}, thereby eliminating the explicit quadratic coupling in those constraints.

Note that $(\mathrm{C9})$ can be omitted since $\textbf{W}_k \succeq \textbf{0}$ is guaranteed by $(\mathrm{C10a})$. The nonconvexity due to coupling has been transformed into a series of convex constraints. Hence, a performance lower bound of problem~\eqref{eq:formulation111} can be written as:
\begin{align}
\label{eq: formu_sub1}
&
\operatorname*{maximize}_{\substack{\textbf{W}_k, \textbf{V} \in \mathbb{H}^{N},\; \mathcal{V} \in \mathbb{H}^{N+1}, \\  \mathcal{Y}_k,  \mathcal{W}_k, \mathcal{Z}\in \mathbb{H}^{MN+1}, \; 
\textbf{P}, \; \textbf{w}_k,\\ \iota_k^{\rm 
N},\; \iota_k^{\rm D}, \;
\textbf{U}_k, \;
\bar{\textbf{U}},\; s_k,\; \bar{s},\; \textbf{v}, \;\textbf{Y}_k, \; \textbf{y}_k, \; \textbf{Z}, \; \textbf{z}}}  \hspace{-7mm} \underset{k \in \mathcal{K}}{\sum} \hspace{-1.5mm} \left(\log_2 \hspace{-0.5mm} (\hspace{-0.5mm}\iota_k^{\rm N} \hspace{-1mm}+\hspace{-1mm} \iota_k^{\rm D} \hspace{-1.1mm}+\hspace{-0.8mm} \sigma_k^2) \hspace{-0.7mm}-\hspace{-0.7mm} \log_2 \hspace{-0.2mm} (\hspace{-0.2mm} \iota_k^{\rm D} \hspace{-0.99mm}+\hspace{-0.7mm} \sigma_k^2) \right) \nonumber \\
&\rm{s.t.}\ ({\rm C3})- ({\rm C5}), \overline{({\rm C6})}, ({\rm C7}), ({\rm C8}), ({\rm C10a}), ({\rm C10b}), \nonumber \\
&\hspace{6mm}({\rm C11a}) - ({\rm C12e}), ({\rm C13a}), \text{and} \ ({\rm C13b}).\end{align}

\subsection{S-Procedure}
To handle the infinitely many possibilities arising in constraints $\overline{({\rm C6})}$, $({\rm C7})$, and $({\rm C8})$, we introduce the following two Lemmas~\cite{hu2021robust, yu2020robust}:
\begin{lemma}(Standard S-Procedure~\cite{hu2021robust}): Let a function $f_m\left(\mathrm{\textbf{x}}\right), m\in \{1, 2\}$, be defined as:
\end{lemma}
\vspace{-1em}
\begin{equation}
    f_m \left(\textbf{x}\right) = \textbf{x}^H \textbf{C}_m \textbf{x} + 2\text{Re}\{\textbf{c}_m^H\textbf{x}\} + c_m,
\end{equation}
\textit{where} $\textbf{C}_m \in \mathbb{H}^N$, $\textbf{c}_m \in \mathbb{C}^{N \times 1}$, \textit{and} $c_m \in \mathbb{R}^{1 \times 1}$. \textit{Then, the implication} $f_1\left(\textbf{x}\right) \leq 0 \Rightarrow f_2\left(\textbf{x}\right) \ge 0$\textit{ holds if and only if there exists a} $\delta \geq 0$\textit{ such that}
\begin{equation}
\delta 
\begin{bmatrix} 
\textbf{C}_1 & \textbf{c}_1 \\ 
\textbf{c}_1^H & c_1 
\end{bmatrix} 
- 
\begin{bmatrix} 
\textbf{C}_2 & \textbf{c}_2 \\ 
\textbf{c}_2^H & c_2 
\end{bmatrix} 
\bm{\succeq 0}. 
\end{equation}
\begin{lemma}(Generalized S-Procedure~\cite{yu2020robust}): Let a function $f\left(\mathrm{\textbf{X}}\right)$ defined as:
\end{lemma}
\begin{equation}
    f \left(\textbf{X}\right) = \textbf{X}^H\hspace{-0.9mm}\textbf{AX} + \textbf{X}^H\hspace{-0.3mm}\textbf{B}+ \textbf{B}^H\hspace{-0.3mm}\textbf{X} + \textbf{C},
\end{equation}
\textit{where} $\textbf{X},\textbf{B} \in \mathbb{C}^{MN \times MM}$, $\textbf{A} \in \mathbb{H}^{MN}$, $\textbf{C} \in \mathbb{C}^{M \times M}$, \textit{and} $\textbf{D} \bm{\succeq 0} \in \mathbb{C}^{MN \times MN}$.  \textit{There exists
a} $\delta \geq 0 $ \textit{such that} $f\left(\mathrm{\textbf{X}}\right) \succeq 0, \forall \textbf{X} \in \{\textbf{X}| \text{Tr} (\textbf{DXX}^H) \leq 1\}$ \textit{is equivalent to}
\begin{equation}
\begin{bmatrix} 
\textbf{C} & \textbf{B}^H \\ 
\textbf{B} & \textbf{A} 
\end{bmatrix} 
- \delta
\begin{bmatrix} 
\textbf{I}_M & \bm{0} \\ 
\bm{0} & -\textbf{D} 
\end{bmatrix} 
\bm{\succeq} \bm{0}.
\end{equation}

By defining $\bar{\textbf{Z}}_k = \sum_{j \neq k} \textbf{Y}_j + \textbf{Z}$, constraints $({\rm C7})$ and $({\rm C8})$ can be equivalently recast by adopting Lemma 1, as:
\begin{align}\label{eq: C7_8}
&(\overline{{\rm C7a}})\hspace{-1.2mm}: \hspace{-1mm}
\begin{bmatrix}\hspace{-0.3mm}
\textbf{Y}_k \hspace{-0.4mm}+ \hspace{-0.2mm} \delta_{k}^{\rm C7} \textbf{I}_{MN} 
     \hspace{-4mm}& \textbf{Y}_k \hat{\textbf{h}}_k \\[.4em]
\hat{\textbf{h}}_k^H \textbf{Y}_k
     \hspace{-4mm}& \hat{\textbf{h}}_k^H \textbf{Y}_k \hat{\textbf{h}}_k \hspace{-0.3mm}-\hspace{-0.3mm} \delta_{k}^{\rm C7} \wp_k^{2}  \hspace{-0.3mm}-\hspace{-0.3mm} \iota_k^{\rm N}
\end{bmatrix} \bm{\succeq}  \hspace{-0.8mm} \bm{0},\forall k,  \nonumber \\
&\overline{({\rm C8a})}\hspace{-1mm}: \hspace{-1mm}
\begin{bmatrix} \hspace{-0.3mm}
\delta_{k}^{\rm C8} \textbf{I}_{MN} \hspace{-0.8mm}-\hspace{-0.4mm} \bar{\textbf{Z}}_k
    \hspace{-3mm}& -\bar{\textbf{Z}}_k \hat{\textbf{h}}_k \\[.4em]
-\hat{\textbf{h}}_k^H\bar{\textbf{Z}}_k
    \hspace{-3mm}&  \iota^{\rm D}_k- \delta_{k}^{\rm C8}\wp_k^{2}  \hspace{-0.2mm}-\hspace{-0.5mm} \hat{\textbf{h}}_k^H \hspace{-0.3mm} \bar{Z}_k \hspace{-0.3mm} \hat{\textbf{h}}_k 
\end{bmatrix} \hspace{-0.3mm}
\bm{\succeq} \hspace{-0.3mm} \bm{0}, \forall k, \nonumber \\
&(\overline{{\rm C7b}})\hspace{-1mm}: \hspace{-1mm} \delta_{k}^{\rm C7} \ge 0, \forall k,\; (\overline{{\rm C8b}})\hspace{-1mm}: \hspace{-1mm} \delta_{k}^{\rm C8} \ge 0, \forall k.
\end{align}
Similarly, we adopt Lemma $2$~\cite{yu2020robust} to handle constraint $\overline{({\rm C6})}$, which can be equivalently written in an LMI format, as:
\begin{align}
\overline{\overline{({\rm C6a})}}\hspace{-1mm}:&\hspace{-1mm} 
\begin{bmatrix}
\hspace{-0.7mm}(\hspace{-0.65mm}\hat{R}^{\rm th}_{k,g} \hspace{-0.5mm} \sigma_g^2 \hspace{-1.1mm}-\hspace{-0.7mm} \delta_{k,g}^{\rm E}\hspace{-0.45mm})\hspace{-0.3mm}\textbf{I}_{\hspace{-0.3mm}T} \hspace{-0.8mm}+\hspace{-0.8mm} \hat{\textbf{H}}_{{\rm E}_g}^H \hspace{-1mm} \textbf{M}_{\hspace{-0.3mm}k\hspace{-0.4mm},g}^{\rm E} \hspace{-0.3mm} \hat{\textbf{H}}_{{\rm E}_g} \hspace{-2mm} \hspace{-3mm}& \hat{\textbf{H}}_{{\rm E}_g}^H \textbf{M}_{k,g}^{\rm E} \\ 
\textbf{M}_{k,g}^{\rm E} \hat{\textbf{H}}_{{\rm E}_g} \hspace{-6mm}& \textbf{M}_{k,g}^{\rm E} \hspace{-1.5mm}+\hspace{-0.7mm} \tfrac{\delta_{k,g}^{\rm E} \textbf{I}_{MN}}{(\wp_g^{\rm tot})^2}  \hspace{-0.7mm}
\end{bmatrix} 
\hspace{-1.5mm} \bm{\succeq} \hspace{-1mm} \bm{0}\hspace{-0.5mm}, \forall k\hspace{-0.4mm}, \hspace{-0.4mm}g\hspace{-0.4mm}, \nonumber \\
\overline{\overline{({\rm C6b})}}\hspace{-1mm}:&  \delta_{k,g}^{\rm E} \ge 0, \forall k, g,
\end{align}
where $\delta_{k,g}^{\rm E}$ are slack variables, $\textbf{M}^{\rm E}_{k,g}=\hat{R}^{\rm th}_{k,g}\textbf{Z}-\textbf{Y}_k$, and $\overline{\overline{({\rm C6a})}}$ involves only $KG$ LMI constraints, which is significantly more tractable than the original semi-infinite form.

\subsection{MM Procedure}
Since the rank-one constraints $(\mathrm{C10b})$, $(\mathrm{C11e})$, $(\mathrm{C12e})$, and $(\mathrm{C13b})$ are nonconvex, we employ a DC reformulation and handle them by adopting an MM procedure. We first present the following lemma, taking $(\mathrm{C11e})$ as an example.
\begin{lemma}
The rank-one constraints $({\rm C11e})$ are equivalent to constraint $(\overline{{\rm C11e}})$ in a DC format, given by:
\end{lemma}
\vspace{-1em}
\begin{equation}\label{eq: RankOne}
(\overline{{\rm C11e}}): \ \left\|\bm{\mathcal{Y}}_k\right\|_* - \left\|\bm{\mathcal{Y}}_k\right\|_2 \leq 0, \ \forall k.
\end{equation}
\vspace{-2em}
\begin{proof}
For any $\textbf{X} \in \mathbb{H}^n \succeq \textbf{0}$ , the inequality $\left\|\textbf{X}\right\|_* = \sum_ix_i \geq \left\|\textbf{X}\right\|_2 = \text{max}\{x_i\}$ holds, where $x_i$ is the $i$-th singular value of $\textbf{X}$. Equality holds if and only if $\textbf{X}$ is rank-one.
\end{proof}
Following the same approach, the rank-one constraints $(\overline{\mathrm{C10b}})$, $(\overline{\mathrm{C12e}})$, and $(\overline{\mathrm{C13b}})$ can also be expressed in DC form.
With these transformations, all remaining sources of nonconvexity of the constraints and objective arise solely from their DC structure. Hence, we adopt an MM scheme based on first-order Taylor linearization and introduce the following propositions.
\\ \textbf{Proposition 3.} \textit{For the convex function} \( -\underset{k \in \mathcal{K}} {\sum} \log_2(\iota_k^{\mathrm{D}} + \sigma_k^2) \), \textit{an affine lower bound} \( \bar{R}(\iota_k^{\mathrm{D}(i_1)}) \) \textit{can be established at any given feasible} \( \iota_k^{\mathrm{D}(i_1)} \) \textit{in the} $(i_1\hspace{-1mm}+\hspace{-1mm}1)$-\textit{th iteration of the MM, as:}
\begin{align}
&-\underset{k \in \mathcal{K}}{\sum} \log_2(\iota_k^{\rm D}+\sigma_k^2) \geq \bar{R}({\iota_k^{{\rm D}(i_1)}}) \nonumber \\
\triangleq& -\underset{k \in \mathcal{K}}{\sum} \log_2(\iota_k^{{\rm D}(i_1)}+\sigma_k^2) - \underset{k \in \mathcal{K}}{\sum} \frac{1}{\ln 2} \frac{\iota_k^{\rm D} - \iota_k^{{\rm D}(i_1)}}{\iota_k^{{\rm D}(i_1)}+\sigma_k^2}.
\end{align}
\vspace{-1em}
\\ \textbf{Proposition 4.} \textit{For the convex quadratic term} $\text{Tr}(\textbf{X}^H\textbf{X})$, \textit{an affine global lower bound} \textit{can be established at any given feasible point} $\textbf{X}^{(i_1)}$ \textit{in the} $(i_1+1)$-th \textit{iteration of the MM, as:}
\begin{align}
\text{Tr}(\textbf{X}^H\textbf{X}) \hspace{-0.5mm}\geq \hspace{-0.5mm}\bar{L}(\textbf{X}^{(i_1)}) \triangleq - \left\|\textbf{X}^{(i_1)}\right\|^2_F \hspace{-1.5mm} + \hspace{-0.5mm}2\text{Tr}\left(\hspace{-0.5mm}(\textbf{X}^{(i_1)})^H \textbf{X}\hspace{-0.5mm}\right).
\end{align}
\vspace{-1em}
\\ \textbf{Proposition 5.}\textit{ For the DC rank-one surrogate} \(\|\textbf{X}\|_{*}-\|\textbf{X}\|_{2}\le 0\) \textit{with} \(\textbf{X}\in\mathbb{H}^{n} \bm{\succeq 0}\), \textit{a convex subset can be established at any given feasible local point} \(\textbf{X}^{(i_{1})}\) \textit{in the} $(i_1+1)$-th \textit{iteration of the MM, denoted as:}
\begin{equation}\label{eq: RankOneSCA}
\left\|\textbf{X}\right\|_* \hspace{-1.5mm} -\hspace{-1mm} \text{Tr}\hspace{-1mm}\left[\hspace{-0.5mm}\bm{\lambda}_{\rm max}\hspace{-1mm}\left(\hspace{-0.8mm}\textbf{X}^{(\hspace{-0.4mm}i_{\hspace{-0.2mm}1}\hspace{-0.4mm})\hspace{-0.8mm}}\right) \hspace{-0.8mm}\bm{\lambda}^H_{\rm max} \hspace{-0.8mm}\left(\hspace{-0.8mm}\textbf{X}^{(\hspace{-0.4mm}i_{\hspace{-0.2mm}1}\hspace{-0.4mm})} \hspace{-0.8mm}\right) \hspace{-0.8mm}\left(\hspace{-0.8mm}\textbf{X} \hspace{-0.8mm}-\hspace{-0.8mm} \textbf{X}^{(\hspace{-0.4mm}i_{\hspace{-0.2mm}1}\hspace{-0.4mm})}\hspace{-0.8mm}\right) \right] \hspace{-0.8mm}-\hspace{-0.8mm}  \left\|\textbf{X}^{(\hspace{-0.4mm}i_{\hspace{-0.2mm}1}\hspace{-0.4mm})}\right\|_{2} \hspace{-1.3mm}\leq \hspace{-0.8mm}0.
\end{equation}

The subsets of the related non-convex constraint sets can be obtained by adopting Proposition~4, as:
\begin{align}
(\overline{{\rm C11b}}):& \text{Tr}(\textbf{U}_k) + \left\|\textbf{P}^{(i_1)}\right\|^2_F - 2\text{Tr}\left((\textbf{P}^{(i_1)})^H\textbf{P}\right) \leq 0, \forall k, \nonumber \\
(\overline{{\rm C12b}}):& \text{Tr}(\bar{\textbf{U}}) + \left\|\textbf{P}^{(i_1)}\right\|^2_F - 2\text{Tr}\left((\textbf{P}^{(i_1)})^H\textbf{P}\right) \leq 0,  
\end{align}
respectively. All remaining DC-form rank-one constraints $(\overline{{\rm C10b}})$, $(\overline{{\rm C12e}})$, and $(\overline{{\rm C13b}})$ follow directly from Proposition 5; details are omitted due to page limits. Here, 
$(\overline{\overline{{\rm C10b}}})\Rightarrow(\overline{{\rm C10b}}), 
(\overline{{\rm C11b}})\Rightarrow({\rm C11b}), 
(\overline{\overline{{\rm C11e}}})\Rightarrow(\overline{{\rm C11e}}),
(\overline{{\rm C12b}})\Rightarrow({\rm C12b}), 
(\overline{\overline{{\rm C12e}}})\Rightarrow(\overline{{\rm C12e}}), $ and $
(\overline{\overline{{\rm C13b}}})\Rightarrow(\overline{{\rm C13b}})$.

For notational simplicity, we define
$\iota = \{\iota_k^{\rm N}, \iota_k^{\rm D}\}$ and 
$\delta = \{\delta_k^{\rm C7}, \delta_k^{\rm C8}, \delta_{k,g}^{\rm E}\}$.
Thus, a lower bound of~\eqref{eq:formulation111} is obtained by solving the following problem in the $(i_1 + 1)$-th MM iteration:
\begin{align}
\label{eq: formu_subFin}
&\operatorname*{maximize}_{\substack{\textbf{W}_k, \textbf{V} \in \mathbb{H}^{N}\hspace{-0.9mm}, \mathcal{V} \in \mathbb{H}^{\hspace{-0.4mm}N\hspace{-0.4mm}+\hspace{-0.4mm}1}\hspace{-0.5mm}, \\  \mathcal{Y}_k,  \mathcal{W}_k, \mathcal{Z} \in \mathbb{H}^{\hspace{-0.4mm}M\hspace{-0.4mm}N\hspace{-0.4mm}+\hspace{-0.4mm}1}\hspace{-0.4mm}, \;\textbf{w}_k, \\ \delta, \; \iota, \;
\textbf{U}_k, \;
\bar{\textbf{U}},\; s_k,\; \bar{s},\; \textbf{v}, \;\textbf{Y}_k, \; \textbf{y}_k, \; \textbf{Z}, \; \textbf{z}}} \hspace{-6mm} \underset{k \in \mathcal{K}}{\sum} \hspace{-0.7mm} \log_2 (\iota_k^{\rm N} \hspace{-0.7mm}+\hspace{-0.7mm} \iota_k^{\rm D} \hspace{-0.7mm}+\hspace{-0.7mm} \sigma_k^2) \hspace{-0.7mm}+\hspace{-0.7mm}\bar{R}(\hspace{-0.5mm}\iota_k^{\hspace{-0.3mm}\mathrm{D}(\hspace{-0.3mm}i_1\hspace{-0.3mm})}\hspace{-0.3mm}) \nonumber \\
\rm{s.t.} &({\rm C3})\hspace{-0.3mm}-\hspace{-0.3mm}({\rm C5}), (\overline{\overline{{\rm C6a}}}), (\overline{\overline{{\rm C6b}}}), (\overline{{\rm C7a}}), (\overline{{\rm C7b}}),(\overline{{\rm C8a}}), \nonumber \\ 
&(\overline{{\rm C8b}}), ({\rm C10a}),  (\overline{\overline{{\rm C10b}}}), ({\rm C11a}), (\overline{{\rm C11b}}),({\rm C11c}),\nonumber \\ 
& ({\rm C11d}), (\overline{\overline{{\rm C11e}}}), ({\rm C12a}), (\overline{{\rm C12b}}), ({\rm C12c}), ({\rm C12d}), \nonumber \\
&(\overline{\overline{{\rm C12e}}}), ({\rm C13a}), \text{and} \,(\overline{\overline{{\rm C13b}}}).
\end{align}
Note that the problem in~\eqref{eq: formu_subFin} is convex and can be solved by adopting a standard numerical solver for convex programming.

\section{PA Location Solution}
With fixed $\textbf{w}_k$, $\textbf{V}$, and $\textbf{P}$, the PA-positioning subproblem is:
\begin{align}
&\underset{\textbf{x}_{n}^{\rm Pin}} {\rm{maximize}} \underset{k \in \mathcal{K}}{\sum} \underset{||\bm{\triangle}\textbf{h}_{k} ||\leq \wp_{k}}{\rm{min}} \hspace{0.3mm} R_k \; \text{s. t.} \ ({\rm C1}), ({\rm C2}), \text{and} \ ({\rm C6}).
\label{eq:formulation222}
\end{align}
In fact, the PA locations simultaneously affect phase shifts, path loss, and blockage-induced channel coefficients, so these position-dependent terms are tightly coupled, making \eqref{eq:formulation222} challenging to handle directly.
To facilitate the presentation in the sequel, we decompose the PA-dependent channel into three separate components. Specifically, we define the path loss and phase shift vectors as $\textbf{f}_{n,u_k}=\big[\,\tfrac{1}{\|\bm{\phi}_k-\bm{\psi}^{\rm Pin}_{n,1}\|},\ldots,\tfrac{1}{\|\bm{\phi}_k-\bm{\psi}^{\rm Pin}_{n,M}\|}\big]^{\top}\in\mathbb{R}^{M\times 1}$
and
$\textbf{a}^{H}_{n,u_k}=\big[\,e^{-j\theta_{n,1,k}},\ldots,e^{-j\theta_{n,M,k}}\,\big]\in\mathbb{C}^{1\times M}$, $\forall n,k$, respectively,
where
$\theta_{n,m,k}=\tfrac{2\pi}{\lambda}\|\bm{\phi}_k-\bm{\psi}^{\rm Pin}_{n,m}\|+\tfrac{2\pi}{\hat{\lambda}}\|\bm{\psi}^{\rm FP}_n-\bm{\psi}^{\rm Pin}_{n,m}\|$, $\forall n,m,k$.
We further define $\textbf{a}^{H}_{k}=[\,\textbf{a}^{H}_{1,u_k},\ldots,\textbf{a}^{H}_{N,u_k}\,]\in\mathbb{C}^{1\times MN}$ and $\textbf{f}_k=\mathrm{vec}([\,\textbf{f}_{1,u_k},\ldots,\textbf{f}_{N,u_k}\,])\in\mathbb{R}^{MN\times 1}$. The blockage-impacted channel-coefficient vector for user $u_k$ is denoted by $\textbf{c}_k \in \mathbb{R}^{MN \times 1}$, where the $l(n,m)$-th element of $\textbf{c}_k$ is $\sqrt{\eta_k^{n,m}}$ and $l(n,m) = (n-1)M + m$. Similarly, we apply the same decomposition for each EA and characterize its path loss, phase shift, and blockage-impacted coefficients by $\textbf{f}^{\rm E}_g \in \mathbb{R}^{MNT \times 1}$, $\textbf{a}^{\rm E}_g \in \mathbb{C}^{MNT \times 1}$, and $\textbf{c}^{\rm E}_g \in \mathbb{C}^{MNT \times 1}$, respectively. Hence, the nominal channels can be written as $\hat{\textbf{h}}_k \hspace{-1mm}=\hspace{-1mm} \textbf{c}_k \hspace{-1mm} \odot \textbf{a}_k \hspace{-1mm} \odot \textbf{f}_k, \forall k$, and ${\rm{vec}}(\hat{\textbf{H}}^{\rm E}_g)= \textbf{c}^{\rm E}_g \odot \textbf{a}^{\rm E}_g \odot \textbf{f}^{\rm E}_g, \forall g$.

Note that all three terms are still coupled but operate on different physical scales: meter-level displacements affect path loss and channel coefficients affected by blockage, and wavelength-level displacements tune the phases. Motivated by this scale separation, we adopt a two-stage PA positioning strategy as a model-based decomposition for tractability. In the first stage, we determine a coarse large-scale PA placement dominated by path loss and blockage effects while keeping the phase terms fixed. In the second stage, starting from this coarse solution, we refine $\textbf{x}_{n}^{\rm Pin}$ within a few wavelengths to improve phase alignment and array gain. This separation significantly reduces the problem complexity while remaining physically meaningful.
In both stages, we employ the transformations in \eqref{eq: lower} and \eqref{eq: bounds} together with
$\textbf{Y}_k=\textbf{P}\textbf{W}_k\textbf{P}^{H}$ and $\textbf{Z}=\textbf{P}\textbf{V}\textbf{P}^H$
to obtain tractable performance lower bounds under constraints $({\rm C7})$ and $({\rm C8})$ in~\eqref{eq: bounds}. Hence, a lower bound of the problem in~\eqref{eq:formulation222} can be represented as:
\begin{align}\label{eq: boundsPA}
&\operatorname*{maximize}_{\substack{\iota_k^{\rm N }, \iota_k^{\rm D}, \hat{\textbf{H}}_k, \hat{\textbf{h}}_k \\ \textbf{a}_k \textbf{f}_k, \textbf{c}_k, \textbf{a}_g^{\rm E}, \textbf{f}_g^{\rm E}, \textbf{c}_g^{\rm E}}} {\sum} \log_2 (\iota_k^{\rm N} + \iota_k^{\rm D} + \sigma_k^2) - \underset{k \in \mathcal{K}}{\sum} \log_2 (\iota_k^{\rm D} + \sigma_k^2) \nonumber \\
& \text{s.t.}  ({\rm C1}), ({\rm C2})  \nonumber \\
&(\overline{{\rm C6}})\hspace{-1mm}: \hspace{-1.7mm}\underset{||\bm{\triangle}\textbf{H}_{{\rm E}_g}\hspace{-0.5mm}||_{\hspace{-0.2mm}F} \leq \wp_g^{\rm tot}}{\max} \hspace{-3mm} \textbf{H}_{{\rm E}_g}^H \hspace{-1mm} \textbf{Y}_k \hspace{-0.3mm} \textbf{H}_{{\rm E}_g} \hspace{-1.7mm} - \hspace{-1mm} \hat{R}^{\rm th}_{k,g} \hspace{-1.2mm} \left(\hspace{-0.5mm}\textbf{H}_{{\rm E}_g}^H \hspace{-1mm} \textbf{Z}\textbf{H}_{{\rm E}_g} \hspace{-1.5mm} + \hspace{-0.5mm} \sigma_{g}^2 \textbf{I}_{T}\hspace{-1mm}\right) \hspace{-1mm}\bm{\preceq} \hspace{-0.5mm} \bm{0}, \forall k,g, \nonumber \\ 
&(\widetilde{{\rm C7}})\hspace{-1mm}: \hspace{-0.5mm}\iota_k^{\rm 
N} \hspace{-1mm} + \hspace{-1.8mm} \underset{||\bm{\triangle}\textbf{h}_{k} ||\leq \wp_{k}}{\rm{max}} \hspace{-1.7mm} \left\{\hspace{-0.8mm} - \hspace{-0.7mm} (\hspace{-0.25mm}\hat{\textbf{h}}_{k} \hspace{-0.8mm}+ \hspace{-0.8mm} \bm{\triangle}\textbf{h}_{k}\hspace{-0.6mm})\hspace{-0.4mm}^H \hspace{-0.6mm} \textbf{Y}_k \hspace{-0.9mm}(\hat{\textbf{h}}_{k} \hspace{-1.0mm} + \hspace{-1.0mm} \bm{\triangle}\textbf{h}_{k} \hspace{-0.5mm})\hspace{-0.5mm} \hspace{-0.5mm} \right\} \hspace{-1mm} \leq \hspace{-1mm} 0,\hspace{-0.5mm} \forall k, \nonumber \\
&(\widetilde{{\rm C8}})\hspace{-1mm}: \hspace{-1mm} \underset{||\bm{\triangle}\textbf{h}_{k} ||\leq \wp_{k}}{\rm{max}} \hspace{-1mm} \left\{\hspace{-0.5mm} \hspace{-0.5mm}(\hat{\textbf{h}}_{k} \hspace{-1mm}+\hspace{-1mm}\bm{\triangle}\textbf{h}_{k})^H \hspace{-0.5mm} \bar{\textbf{Z}}_k (\hat{\textbf{h}}_{k} \hspace{-1mm}+\hspace{-1mm}\bm{\triangle}\textbf{h}_{k}\hspace{-0.4mm})\hspace{-0.3mm} \hspace{-0.3mm}\right\} \hspace{-0.7mm} \leq \hspace{-0.7mm} \iota_k^{\rm 
D}, \forall k,  \nonumber \\
&(\widetilde{{\rm C9}})\hspace{-1mm}: \hat{\textbf{h}}_k = \textbf{c}_k \odot \textbf{f}_k \odot \textbf{a}_k, \forall k, \nonumber \\ 
&(\widetilde{{\rm C10}})\hspace{-1mm}: \text{vec}(\hat{\textbf{H}}_g^{\rm E}) = \textbf{c}_g^{\rm E} \odot \textbf{f}_g^{\rm E} \odot \textbf{a}_g^{\rm E}, \forall g,
\end{align} 
where $\bar{\textbf{Z}}_k=\sum_{j\neq k}\textbf{Y}_{j}+\textbf{Z}, \forall k$. We will iteratively optimize $\{\textbf{f}_k, \textbf{c}_k, \textbf{f}_g^{\rm E}, \textbf{c}_g^{\rm E}, \textbf{x}^{\rm Pin}, \iota^{\rm N}_k, \iota^{\rm D}_k\}$ for meter-level PA positioning and $\{\textbf{a}_k, \textbf{a}_g^{\rm E}, \textbf{x}^{\rm Pin}, \iota^{\rm N}_k, \iota^{\rm D}_k\}$ for phase-shift-based wavelength-level adjustment.

\subsection{Large-Scale Movement with Blockage Effects}
In the first stage, we focus on the large-scale movement of the PAs. Since $\hat{\textbf{h}}_k$, $\textbf{f}_k$, and $\textbf{a}_k$ are highly coupled and enter quadratically in~\eqref{eq: boundsPA}, we first apply a lifting step for $\textbf{H}_k=\textbf{h}_k\textbf{h}_k^H$, $\textbf{C}_k=\textbf{c}_k\textbf{c}_k^{H}$,  $\textbf{F}_k=\textbf{f}_k\textbf{f}_k^{H}$, $\forall k$, and fixing the phase-shift matrix as $\textbf{A}_k^{\rm D}=\mathrm{diag}(\textbf{a}_k)$ and $\hat{\textbf{A}}_k=\textbf{A}_k^{\rm D}\odot \mathrm{diag}(\textbf{P}\textbf{w}_k^{*})$. Then, constraint $(\widetilde{\mathrm{C9}})$ is equivalent to:
\begin{align}\label{eq: C9seris}
&(\widetilde{{\rm C9a}})\hspace{-1mm}: \text{Tr}(\textbf{Y}_k\hat{\textbf{H}}_k) \hspace{-0.7mm}=\hspace{-0.7mm} \text{Tr}(\textbf{C}_k \hat{\textbf{A}}_k\textbf{F}_k \hat{\textbf{A}}_k^H), \forall k, \nonumber \\
&(\widetilde{{\rm C9b}})\hspace{-1mm}: \text{Rank}(\textbf{C}_k) \hspace{-0.7mm}=\hspace{-0.7mm} 1, \forall k, \; (\widetilde{{\rm C9c}})\hspace{-1mm}: \text{Rank}(\textbf{F}_k) \hspace{-0.7mm}=\hspace{-0.7mm} 1, \forall k, \nonumber \\
&(\widetilde{{\rm C9d}})\hspace{-1mm}: \text{Rank}(\mathcal{H}_k) \hspace{-0.7mm}=\hspace{-0.7mm} 1, \forall k, \nonumber \\
&(\widetilde{{\rm C9e}})\hspace{-1mm}: \mathcal{H}_k \hspace{-1mm}=\hspace{-1mm} \left[\hspace{-1mm}
 \begin{array}{cc}
    \hat{\textbf{H}}_k      & \hat{\textbf{h}}_k         \\
    \hat{\textbf{h}}_k^{H} & 1 
  \end{array}
     \hspace{-1mm} \right] \hspace{-1.2mm}\bm{\succeq 0}\hspace{-1.2mm}
\in \hspace{-1.2mm} \mathbb{C}^{(MN+1) \times (MN+1)},  \forall k.
\end{align}
Moreover, using the transformation in~\eqref{eq: C9seris} and the known selection matrices $\textbf{S}=[\,\textbf{I}_{MN}\ \bm{0}\,]\in\mathbb{R}^{MN\times(MN+1)}$ and $\textbf{s}=[\,\bm{0}_{MN}\ 1\,]^{\top}\in\mathbb{R}^{(MN+1)\times 1}$, constraints $(\mathrm{C7})$ and $(\mathrm{C8})$ in~\eqref{eq: boundsPA} can be equivalently converted by Lemma~1, as the following LMIs:
\begin{align}\label{eq: PA1_imperfectC7_C9}
&(\widetilde{{\rm C7a}}) \hspace{-1mm}:  \left[\hspace{-1mm}
\begin{array}{cc}\hspace{-1mm}
\textbf{Y}_k \hspace{-1mm}-\hspace{-0.7mm} \delta_k^{N} \textbf{I}_{MN} \hspace{-1mm}&
\textbf{Y}_k \textbf{S} \mathcal{H}_k \textbf{s} \\[0.4em]
\textbf{s}^H \mathcal{H}_k^H \textbf{S}^H \textbf{Y}_k \hspace{-1mm}&
\mathrm{Tr}(\textbf{Y}_k \textbf{S} \mathcal{H}_k \textbf{S}^H\hspace{-0.7mm})
 \hspace{-1mm}-\hspace{-1mm} \iota_k^{N} \hspace{-1mm}+\hspace{-1mm} \delta_k^{N}\hspace{-0.5mm} \wp_k^2 \hspace{-1mm}
\end{array}\hspace{-1mm}
\right]\hspace{-1mm}
\bm{\succeq} \hspace{-1mm}\bm{0}, \forall k, \nonumber \\
&(\widetilde{{\rm C8a}})\hspace{-1mm}:  \left[\hspace{-2mm}
\begin{array}{cc}\hspace{-1mm}
\delta_k^{D} \textbf{I}_{MN} \hspace{-0.7mm}-\hspace{-0.7mm} \bar{\textbf{Z}}_k \hspace{-1mm}&
-\bar{\textbf{Z}}_k\textbf{S}  \mathcal{H}_k \textbf{s} \\[0.4em]
-\textbf{s}^H \hspace{-0.7mm}\mathcal{H}_k^H \hspace{-0.7mm} \textbf{S}^H \hspace{-0.7mm} \bar{\textbf{Z}}_k \hspace{-1mm}&
\iota_k^{D} \hspace{-1.2mm}-\hspace{-0.7mm} \mathrm{Tr}(\bar{\textbf{Z}}_k \textbf{S} \mathcal{H}_k \textbf{S}^H\hspace{-0.5mm}) \hspace{-0.7mm}- \hspace{-0.7mm}\delta_k^{D} \wp_k^2 \hspace{-1mm}
\end{array}\hspace{-1mm}
\right]
\bm{\succeq}\hspace{-0.7mm} \bm{0}, \forall k, \nonumber \\
&(\widetilde{{\rm C7b}})\hspace{-1mm}: \delta_k^N \ge 0, \forall k, \; \; (\widetilde{{\rm C8b}})\hspace{-1mm}: \delta_k^D \hspace{-0.7mm}\ge\hspace{-0.7mm} 0, \forall k.
\end{align}
Then, the only non-convex constraints in~\eqref{eq: C9seris} and~\eqref{eq: PA1_imperfectC7_C9} are the rank-one constraints and the equality constraint $(\widetilde{{\rm C9a}})$.

We next shift our focus to constraint $(\overline{\rm C6})$ in~\eqref{eq: boundsPA} with defining $\textbf{M}^{\rm E}_{k,g}=\hat{R}^{\rm th}_{k,g}\textbf{Z}-\textbf{Y}_k$,  $\textbf{j}_g=\mathrm{vec}(\hat{\textbf{H}}_{\text{E}_g})\in\mathbb{C}^{MNT\times 1}$, and $\textbf{J}_g=\textbf{j}_g\textbf{j}_g^{H}, \forall g$.
Following the same lifting procedure as for the legitimate users, we set $\textbf{F}_g^{\rm E} = \textbf{f}_g^{\rm E}(\textbf{f}_g^{\rm E})^H$ and $\textbf{C}_g^{\rm E} = \textbf{c}_g^{\rm E}(\textbf{c}_g^{\rm E})^H$.
We further introduce a known nonzero vector
$\bar{\textbf{m}}\in \mathbb{C}^{MNT \times 1}$ and its rank-one matrix
$\bar{\textbf{M}} = \bar{\textbf{m}}\bar{\textbf{m}}^H$ to establish the relationship between $\textbf{j}_g$ and ${\textbf{c}_g^{\rm E}, \textbf{f}_g^{\rm E}, \textbf{a}_g^{\rm E}}$ in constraint $(\widetilde{{\rm C10}})$.
By fixing the phase-shift term as $\textbf{A}^{\rm D}_{{\rm E}g} = \text{diag}(\textbf{a}_g^{\rm E})$ and defining $\hat{\textbf{A}}_{g}^{\rm E}=\textbf{A}^{\rm D}_{{\rm E}_g}\odot\mathrm{diag}(\bar{\textbf{m}}^{*})$, constraint $(\widetilde{{\rm C10}})$ can be equivalently rewritten as:
\begin{align}\label{eq: transMM}
&(\widetilde{{\rm C10a}})\hspace{-1mm}: \text{Tr}(\overline{\textbf{M}}\textbf{J}_g) = \text{Tr}\left(\textbf{C}^{\rm E}_g \hat{\textbf{A}}^{\rm E}_g  \textbf{F}^{\rm E}_g  (\hat{\textbf{A}}^{\rm E}_g)^H\right), \forall g, \nonumber \\
&(\widetilde{{\rm C10b}})\hspace{-1mm}: \text{Rank} (\textbf{C}^{\rm E}_g) \hspace{-0.5mm}=\hspace{-0.5mm} 1, \forall g, \; (\widetilde{{\rm C10c}})\hspace{-1mm}: \text{Rank} (\textbf{F}^{\rm E}_g) \hspace{-0.5mm}=\hspace{-0.5mm} 1, \forall g, \nonumber \\
&(\widetilde{{\rm C10d}})\hspace{-1mm}: \text{Rank}(\mathcal{J}_g) = 1, \forall g, \nonumber \\
&(\widetilde{{\rm C10e}})\hspace{-1mm}: \mathcal{J}_g =\hspace{-1mm} \left[\hspace{-1mm}
 \begin{array}{cc}
    \textbf{J}_g      & \textbf{j}_g         \\
    \textbf{j}_g^{H} & 1 
  \end{array}
     \hspace{-1mm} \right] \bm{\succeq 0}
\in \mathbb{C}^{(MNT+1) \times (MNT+1)}, \forall g. 
\end{align}
Based on the transformation in~\eqref{eq: transMM}, we define the column selector $\bar{\textbf{s}}_t\in\mathbb{R}^{T\times 1}$ with $[\bar{\textbf{s}}_t]_t=1$ and zeros elsewhere, the selection matrix  as $\bar{\textbf{S}}_t=\bar{\textbf{s}}_t\otimes\textbf{I}_{MN}\in\mathbb{R}^{MNT\times MN}$, $\mathcal{S}_{k,g}=\big[\mathrm{Tr}(\bar{\textbf{S}}_t\textbf{M}^{\rm E}_{k,g}\bar{\textbf{S}}_{t'}^{H}\textbf{J}_g)\big]_{t,t'=1}^{T}\in\mathbb{H}^{T}$ and $\bar{\mathcal{S}}_{k,g}=[\,\textbf{M}^{\rm E}_{k,g}\bar{\textbf{S}}_{1}^{H}\textbf{j}_g,\ldots,\textbf{M}^{\rm E}_{k,g}\bar{\textbf{S}}_{T}^{H}\textbf{j}_g\,]\in\mathbb{C}^{MN\times T}$.
Therefore, by applying Lemma~2, $(\overline{\mathrm{C6}})$ can be equivalently written as:
\begin{align}
&(\widetilde{{\rm C6a}})\hspace{-1mm}:\hspace{-1mm}  \left[\hspace{-1mm}
\begin{array}{cc}\hspace{-1mm}
\hat{R}^{\rm th}_{k,g}\sigma_g^2 \textbf{I}_{T} \hspace{-0.7mm}+\hspace{-0.7mm} \mathcal{S}_{k,g} \hspace{-0.7mm}-\hspace{-0.7mm} \delta_{k,g}^{E_2} \textbf{I}_{T} \hspace{-7mm}&
 \bar{\mathcal{S}}_{k,g}^H \\[0.4em]
\bar{\mathcal{S}}_{k,g} \hspace{-7mm}&
\textbf{M}_{k,g}^{\rm E} \hspace{-1mm}-\hspace{-0.5mm} \frac{\delta_{k,g}^{E_2}}{(\wp_g^{\rm tot})^2} \textbf{I}_{MN}\hspace{-1mm}
\end{array}\hspace{-1mm}
\right]\hspace{-1mm}
\bm{\succeq}\hspace{-1mm}\bm{0}, \forall g,k, \nonumber \\
&(\widetilde{{\rm C6b}})\hspace{-1mm}: \delta_{k,g}^{E_2} \ge 0, \forall g,k.
\end{align}

Thus, stage 1 of problem~\eqref{eq:formulation222} can be equivalently written exploiting the auxiliary variables $\textbf{F}_k$, $\textbf{F}_g^{\rm E}$, $\textbf{C}_k$, and $\textbf{C}_g^{\rm E}$, as:
\begin{align}
\label{eq: formu2_firTran}
&\operatorname*{maximize}_{\substack{
   \textbf{x}_n^{\rm Pin},\; \textbf{F}_k,\; \textbf{C}_k,\\
   \textbf{F}_g^{\rm E},\; \textbf{C}_g^{\rm E}, \; \hat{\textbf{H}}_k,\;\hat{\textbf{h}}_k,  \\
 \mathcal{H}_g,\; \textbf{J}_g, \;\textbf{j}_g,\; \mathcal{J}_g, \; \iota, \; \delta 
}} \hspace{-2.5mm}\underset{k \in \mathcal{K}}{\sum} \hspace{-0.5mm} \log_2 \hspace{-0.3mm}(\hspace{-0.3mm}\iota_k^{\rm N} \hspace{-0.9mm}+\hspace{-0.7mm} \iota_k^{\rm D} \hspace{-0.9mm}+\hspace{-0.5mm} \sigma_k^2) \hspace{-0.7mm}-\hspace{-0.7mm} \underset{k \in \mathcal{K}}{\sum}\hspace{-0.5mm} \log_2 (\iota_k^{\rm D} \hspace{-0.7mm}+\hspace{-0.7mm} \sigma_k^2) \nonumber \\
\rm{s.t.}\ &({\rm C1})\hspace{-0.2mm}, ({\rm C2})\hspace{-0.2mm},  (\widetilde{{\rm C6a}}), (\widetilde{{\rm C6b}})\hspace{-0.2mm}, (\widetilde{{\rm C7a}})\hspace{-0.2mm},(\widetilde{{\rm C7b}})\hspace{-0.2mm}, (\widetilde{{\rm C8a}})\hspace{-0.2mm}, (\widetilde{{\rm C8b}})\hspace{-0.2mm}, \nonumber\\
&(\widetilde{{\rm C9a}})-(\widetilde{{\rm C9e}}), (\widetilde{{\rm C10a}})-(\widetilde{{\rm C10e}}), \nonumber \\
&(\widetilde{{\rm C11}})\hspace{-1mm}: \hspace{-0.7mm}\left(x_{n,m}^{\rm Pin} \hspace{-0.7mm}-\hspace{--.7mm} x_k\right)^2 \hspace{-2mm}= \tfrac{1}{\text{Diag} (\textbf{F}_k)_{l(n,m)}} \hspace{-1mm}- \hspace{-1mm}\hat{\beta}_{n,k}, \forall n, m, k, \nonumber \\
&(\widetilde{{\rm C12}})\hspace{-1mm}: \hspace{-0.7mm}\left(x_{n,m}^{\rm Pin} \hspace{-0.7mm}-\hspace{-0.7mm}x_{g,t}^{\rm E}\right)^{\hspace{-0.7mm}2} \hspace{-2mm}= \hspace{-0.7mm} \tfrac{1}{\text{Diag} (\textbf{F}_g^{\rm E})_{l(n,m,t)}} \hspace{-0.7mm}-\hspace{-0.7mm} \hat{\beta}_{n,g, t}^{\rm E}, \forall n, m, t, g, \nonumber \\
&(\widetilde{{\rm C13}})\hspace{-1mm}: \text{Diag}(\textbf{C}_k)_{l(n,m)} \hspace{-1mm}= \hat{\eta} \zeta_{k}^{n,m}(x_{n,m}^{\rm Pin}), \forall n,m,k, \nonumber \\
&(\widetilde{{\rm C14}})\hspace{-1mm}: \text{Diag}(\textbf{C}_g^{\rm E})_{\hspace{-0.3mm}l(\hspace{-0.5mm}n\hspace{-0.3mm},m\hspace{-0.3mm},t\hspace{-0.3mm})} \hspace{-0.7mm}= \hspace{-0.5mm}\hat{\eta} \zeta_{g,t}^{n,m}(x_{n,m}^{\rm Pin}), \forall n,m,t,g,
\end{align}
where $\iota = \{\iota_k^{\rm N},\iota_k^{\rm D}\}$, $\delta = \{\delta_k^{\rm D}, \delta_k^N, \delta_{k,g}^{E_2}\}$, $\hat{\beta}_{n,k}\hspace{-1mm}=\hspace{-1mm}(y_k \hspace{-0.7mm}-\hspace{-0.7mm} y_n^{ \rm FP})^2 \hspace{-0.7mm}+\hspace{-1mm} d^2$, $\hat{\beta}_{n,g,t}^{\rm E}\hspace{-1mm}=\hspace{-1mm}(y_{g,t} \hspace{-0.5mm}-\hspace{-0.5mm} y_n^{ \rm FP})^2 \hspace{-0.5mm}+\hspace{-0.5mm} d^2$, $\text{Diag}(\textbf{F}_g^{\rm E})_{l(n,m,t)}\hspace{-1mm}=\hspace{-1mm}[\textbf{F}_g^{\rm E}]_{l(n,m,t),l(n,m,t)}$, $\text{Diag}(\textbf{C}_g^{\rm E})_{l(n,m,t)}\hspace{-1mm}=\hspace{-1mm}[\textbf{C}_g^{\rm E}]_{l(n,m,t),l(n,m,t)}$, $l(n,m,t) = (t-1)MN+(n-1)M+m$, and $\text{Diag}(\textbf{F}_k)_{l(n,m)}\hspace{-1mm}=\hspace{-1mm}[\textbf{F}_k]_{l(n,m),l(n,m)}$, $\text{Diag}(\textbf{C}_k)_{l(n,m)}\hspace{-1mm}=\hspace{-1mm}[\textbf{C}_k]_{l(n,m),l(n,m)}$.
The problem in~\eqref{eq: formu2_firTran} is nonconvex due to the equality constraints, the rank-one constraints, the bilinear couplings, and the DC form objective. We first handle the equalities and bilinear terms by replacing each equality with two inequalities and adopting the Frobenius norm to decouple matrix couplings~\cite{conference,yu2020robust}. Taking $(\widetilde{\mathrm{C9a}})$ as an example, it can be equivalently represented as:
\begin{equation}
  \hspace{-4.4mm}\scalebox{0.85}{$\displaystyle
    (\widetilde{{\rm C9aa}}) \hspace{-1mm}: \hspace{-1mm}  \left\| \textbf{F}_{\hspace{-0.5mm}k} \hspace{-0.99mm}+\hspace{-0.8mm} \hat{\textbf{A}}_k^{\hspace{-0.2mm}H} \hspace{-0.99mm}\textbf{C}_k \hspace{-0.4mm} \hat{\textbf{A}}_k  \right\|_F^2 
\hspace{-2.2mm}-\hspace{-0.7mm} \left\| \textbf{F}_{\hspace{-0.5mm}k} \right\|_F^2 
\hspace{-1mm}-\hspace{-0.7mm} \left\|\hspace{-0.5mm} \hat{\textbf{A}}_k^{\hspace{-0.2mm}H} \hspace{-0.99mm}\textbf{C}_k \hspace{-0.2mm} \hat{\textbf{A}}_k\hspace{-0.5mm}\right\|_F^2 \hspace{-2mm}- \hspace{-1.3mm}2\text{Tr}(\textbf{Y}_k \hat{\textbf{H}}_k)
\hspace{-0.99mm}\leq\hspace{-0.99mm} 0, \forall k, \nonumber
  $}
\end{equation}
\begin{equation}\label{eq: Frobenius}
  \scalebox{0.85}{$\displaystyle
    (\widetilde{{\rm C9ab}})\hspace{-1mm} :\hspace{-1mm}  2\text{Tr}(\textbf{Y}_k \hat{\textbf{H}}_k\hspace{-0.7mm}) \hspace{-0.89mm}-\hspace{-0.69mm} \left\|\hspace{-0.5mm} \textbf{F}_{\hspace{-0.5mm}k} \hspace{-0.99mm}+\hspace{-0.8mm} \hat{\textbf{A}}_k^{\hspace{-0.2mm}H} \hspace{-0.79mm}\textbf{C}_k \hspace{-0.5mm} \hat{\textbf{A}}_k \hspace{-0.5mm} \right\|_F^2 
\hspace{-2.4mm}+\hspace{-0.7mm} \left\|\hspace{-0.4mm} \textbf{F}_{\hspace{-0.4mm}k} \hspace{-0.4mm}\right\|_{\hspace{-0.5mm}F}^2 
\hspace{-1mm}+\hspace{-0.8mm} \left\|\hspace{-0.5mm} \hat{\textbf{A}}_k^{\hspace{-0.2mm}H} \hspace{-0.99mm}\textbf{C}_k \hspace{-0.3mm} \hat{\textbf{A}}_k\hspace{-0.5mm} \right\|_{\hspace{-0.5mm}F}^2
\hspace{-0.99mm} \leq \hspace{-0.99mm} 0, \forall k, 
  $}
\end{equation}
both of which remain non-convex but admit a DC structure. 
Similarly, the equality constraints $(\widetilde{\mathrm{C11}})$, $(\widetilde{\mathrm{C12}})$, $(\widetilde{\mathrm{C13}})$, and $(\widetilde{\mathrm{C14}})$ are represented as two inequality constraints labeled “a” and “b”. Specifically, $(\widetilde{\mathrm{C10a}})$ analogously follows the same structure, adopting the Frobenius-norm identity in~\eqref{eq: Frobenius}. These transformations are omitted due to page limitations. Moreover, by directly applying Lemma~3, the rank-one constraints, $(\widetilde{\rm C9b}) - (\widetilde{\rm C9d})$ and $(\widetilde{\rm C10b}) - (\widetilde{\rm C10d})$, are also converted into their DC forms, identical in structure to~\eqref{eq: RankOne}.

However, the problem remains challenging due to the min-max nature of the constraints $(\widetilde{\mathrm{C13a}})$--$(\widetilde{\mathrm{C14b}})$. We take $(\widetilde{\mathrm{C13a}})$ and $(\widetilde{\mathrm{C13b}})$ as examples, which can be written as:
\begin{align}
(\widetilde{{\rm C13a}})\hspace{-1mm}:& \bar{C}_k^{l(n,m)} \hspace{-1.5mm} +\hspace{-0.3mm} \text{log}_2 \hspace{-1.2mm}\left(\hspace{-1mm}1\hspace{-0.7mm}+\hspace{-0.7mm} e^{-\vartheta \bar{\alpha}^{n,m}_{k}}\hspace{-0.8mm}\right) - \text{log}_2 (\hat{\eta}) \le 0, \forall n,m,k, \nonumber \\
(\widetilde{{\rm C13b}})\hspace{-1mm}:& \hspace{-0.1mm}\text{log}_2 \hspace{-0.5mm} (\hspace{-0.4mm}\hat{\eta}\hspace{-0.3mm})\hspace{-0.3mm}-\hspace{-0.6mm}\bar{C}_k^{l(\hspace{-0.4mm}n\hspace{-0.3mm},m\hspace{-0.4mm})}\hspace{-1.8mm} -\hspace{-0.5mm} \text{log}_2 \hspace{-1mm} \left(\hspace{-0.9mm}1\hspace{-1mm}+ \hspace{-0.6mm}e^{\hspace{-0.7mm}-\vartheta \hspace{-0.25mm}\bar{\alpha}^{\hspace{-0.25mm}n\hspace{-0.25mm},m}_{k}}\hspace{-0.8mm}\right) \hspace{-0.7mm}\le \hspace{-0.7mm}0\hspace{-0.25mm}, \forall n\hspace{-0.25mm},m\hspace{-0.25mm},k\hspace{-0.25mm},
\end{align}
where $\bar{C}_k^{l(n,m)} =\text{log}_2 \left(\text{Diag}(\textbf{C}_k)_{l(n,m)}\right)$ and $\bar{\alpha}^{n,m}_{k} = \tfrac{{\rm min}_{q \in \mathcal{Q}} \{d_{k,q}^{n,m}\}}{[\text{Diag}(\textbf{F}_k)_{l(n,m)}]^{-1/2}}$. 
Noting that in $(\widetilde{\rm C13a})$ and $(\widetilde{\rm C13b})$ the term $+\log\!\big(1+e^{\vartheta(\cdot)}\big)$ is convex and strictly increasing, whereas $-\log\!\big(1+e^{\vartheta(\cdot)}\big)$ is concave and strictly decreasing, we introduce the slack variables $\bar{\alpha}_{n,m,k}$ and $\bar{r}_{n,m,k}$ as:
\begin{align}
(\widetilde{\rm C13c})\hspace{-1mm}:& \bar{\alpha}_{n,m,k} \leq \tfrac{\mu_k^{n,m} (x_{n,m}^{\rm Pin})} {[\text{Diag}(\textbf{F}_k)_{l(n,m)}]^{-1/2}}, \forall n,m,k, 
\nonumber \\
(\widetilde{\rm C13d})\hspace{-1mm}:& \bar{\alpha}_{n,m,k} \geq \tfrac{\mu_k^{n,m} (x_{n,m}^{\rm Pin})} {[\text{Diag}(\textbf{F}_k)_{l(n,m)}]^{-1/2}}, \forall n,m,k,
\nonumber \\
(\widetilde{\rm C15a})\hspace{-1mm}:& [\text{Diag}(\textbf{F}_k)_{l(n,m)}]^{-1/2} \leq  \bar{r}_{n,m,k}, \forall n,m,k, \nonumber \\
(\widetilde{\rm C15b})\hspace{-1mm}:& [\text{Diag}(\textbf{F}_k)_{l(n,m)}]^{-1/2} \geq  \bar{r}_{n,m,k}, \forall n,m,k.
\end{align}
We then apply the quadratic identity, $\bar{r}_{n,m,k}\,\bar{\alpha}_{n,m,k}
=\tfrac{1}{4}\big[(\bar{r}_{n,m,k}+\bar{\alpha}_{n,m,k})^{2}-(\bar{r}_{n,m,k}-\bar{\alpha}_{n,m,k})^{2}\big]$, to decouple the two slack variables in constraints $(\widetilde{\rm C13a})$ and $(\widetilde{\rm C13b})$, as:
\begin{equation}
  \scalebox{0.88}{$\displaystyle
    (\widetilde{\rm C13a})\hspace{-1mm}: \text{log}_2\hspace{-1mm}\left(\hspace{-0.3mm}\text{Diag}\hspace{-0.3mm}(\hspace{-0.3mm}\textbf{C}_k\hspace{-0.3mm})_{l(\hspace{-0.3mm}n\hspace{-0.2mm},m\hspace{-0.2mm})}\hspace{-0.3mm}\right) \hspace{-0.5mm}+ \hspace{-1mm}\text{log}_2\hspace{-1mm} \left( \hspace{-0.5mm}1 \hspace{-0.5mm}+ \hspace{-0.5mm} e^{\hspace{-0.5mm}-\vartheta \hspace{-0.3mm}\bar{\alpha}_{\hspace{-0.2mm}n\hspace{-0.2mm},m\hspace{-0.2mm},k}}\hspace{-0.5mm}\right) \hspace{-0.5mm}- \hspace{-0.7mm}\text{log}_2 \hspace{-0.2mm} ( \hspace{-0.4mm}\hat{\eta} \hspace{-0.4mm})  \hspace{-0.9mm}\le  \hspace{-0.6mm}0\hspace{-0.4mm},\hspace{-0.5mm}\forall \hspace{-0.3mm} n\hspace{-0.3mm},\hspace{-0.5mm}m\hspace{-0.3mm},\hspace{-0.5mm}k\hspace{-0.3mm},\hspace{-0.3mm} \nonumber
  $}
\end{equation}
\begin{equation}
  \scalebox{0.88}{$\displaystyle
    (\widetilde{\rm C13b})\hspace{-1mm}:  -\text{log}_2\hspace{-1mm}\left(\hspace{-0.4mm}\text{Diag}\hspace{-0.3mm}(\hspace{-0.4mm}\textbf{C}_k\hspace{-0.4mm})_{l(\hspace{-0.3mm}n\hspace{-0.2mm},m\hspace{-0.3mm})}\hspace{-0.5mm}\right) \hspace{-0.7mm}- \hspace{-1mm}\text{log}_2\hspace{-1mm} \left( \hspace{-0.7mm}1 \hspace{-0.7mm}+ \hspace{-0.7mm} e^{\hspace{-0.7mm}-\vartheta \hspace{-0.3mm}\bar{\alpha}_{\hspace{-0.2mm}n\hspace{-0.2mm},m\hspace{-0.2mm},k}}\hspace{-0.7mm}\right) \hspace{-0.7mm} + \hspace{-0.7mm}\text{log}_2 \hspace{-0.5mm} ( \hspace{-0.4mm}\hat{\eta} \hspace{-0.4mm})  \hspace{-1mm}\le  \hspace{-0.9mm}0\hspace{-0.3mm},\hspace{-0.5mm}\forall \hspace{-0.4mm} n\hspace{-0.3mm},\hspace{-0.53mm}m\hspace{-0.3mm},\hspace{-0.5mm}k\hspace{-0.3mm},\hspace{-0.3mm} 
\nonumber
  $}
\end{equation}
\begin{equation}
  \scalebox{0.90}{$\displaystyle
    (\widetilde{\rm C13c})\hspace{-1mm}: \hspace{-0.3mm}\tfrac{(\bar{r}_{n,m,k}+ \bar{\alpha}_{n,m,k})^2 - (\bar{r}_{n,m,k} -\bar{\alpha}_{n,m,k})^2}{4} \hspace{-0.7mm}- \hspace{-1mm} \mu_k^{n,m} \hspace{-0.5mm} (\hspace{-0.3mm}x_{n,m}^{\rm Pin}\hspace{-0.3mm}) \hspace{-1mm}\leq \hspace{-1mm} 0, \hspace{-0.5mm}\forall \hspace{-0.3mm} n\hspace{-0.3mm},\hspace{-0.3mm}m\hspace{-0.3mm},k,
\nonumber
  $}
\end{equation}
\begin{equation}\label{eq: C13_seris}
  \scalebox{0.90}{$\displaystyle
    (\widetilde{\rm C13d})\hspace{-1mm}:\hspace{-0.7mm} \mu_k^{\hspace{-0.3mm}n,m}\hspace{-0.5mm} (\hspace{-0.3mm}x_{n,m}^{\rm Pin}\hspace{-0.3mm}) \hspace{-1mm}- \hspace{-0.7mm} \tfrac{(\bar{r}_{n,m,k}+ \bar{\alpha}_{n,m,k})^{\hspace{-0.2mm}2} \hspace{-0.7mm}-\hspace{-0.4mm} (\bar{r}_{n,m,k} - \bar{\alpha}_{n,m,k})^{\hspace{-0.2mm}2}}{4} \hspace{-0.7mm}\leq \hspace{-0.7mm}0, \hspace{-0.5mm}\forall \hspace{-0.3mm} n\hspace{-0.3mm},\hspace{-0.3mm}m\hspace{-0.3mm},k.
  $}
\end{equation}
Here, $(\widetilde{\rm C13a})$ and $(\widetilde{\rm C13b})$ are in DC form, whereas the remaining difficulty lies in $\mu_k^{n,m}(x_{n,m}^{\rm Pin})$ within $(\widetilde{\rm C13c})$--$(\widetilde{\rm C13d})$.
Following the same idea in~\eqref{eq: C13_seris}, $(\widetilde{\rm C14a})$--$(\widetilde{\rm C14b})$ can be equivalently transformed into $(\widetilde{\rm C14a})$--$(\widetilde{\rm C14d})$ with the aid of $(\widetilde{\rm C16a})$--$(\widetilde{\rm C16b})$ by introducing the slack variables $\bar{\alpha}_{n,m,g,t}^{\rm E}$ and $\bar{r}_{n,m,g,t}^{\rm E}$; details are omitted for brevity.


Nonetheless, the objective and constraints
$(\widetilde{\rm C9aa})$, $(\widetilde{\rm C9ab})$, $(\widetilde{\rm C9b})$-$(\widetilde{\rm C9d})$, $(\widetilde{\rm C10b})$-$(\widetilde{\rm C10d})$, $(\widetilde{\rm C10aa})$, $(\widetilde{\rm C10ab})$, $(\widetilde{\rm C10b})$, $(\widetilde{\rm C10c})$, $(\widetilde{\rm C11a})$, $(\widetilde{\rm C11b})$, $(\widetilde{\rm C12a})$, $(\widetilde{\rm C12b})$, $(\widetilde{\rm C13a})$--$(\widetilde{\rm C13d})$, and $(\widetilde{\rm C14a})$--$(\widetilde{\rm C14d})$ exhibit a DC structure, and constraints $(\widetilde{\rm C13c})$, $(\widetilde{\rm C13d})$, $(\widetilde{\rm C14c})$, and $(\widetilde{\rm C14d})$ involve min-max nonconvexity. Typically, we adopt Proposition~3 to obtain a performance lower bound for the objective at the $(i_2 + 1)$-th iteration. With Proposition~5, the DC format rank-one constraints take the same structure as in~\eqref{eq: RankOneSCA}. Specifically,
$ (\widetilde{\overline{\rm C9b}}) - (\widetilde{\overline{\rm C9d}}), (\widetilde{\overline{\rm C10b}}) - (\widetilde{\overline{\rm C10d}}) \Rightarrow (\widetilde{\rm C9b}) - (\widetilde{\rm C9d}), (\widetilde{\rm C10b}) - (\widetilde{\rm C10d})$, respectively.
The global underestimation for the remaining nonconvex terms via first-order Taylor expansions is:
\begin{align}
(\widetilde{\overline{\rm C9aa}})\hspace{-1mm}:& \hspace{-0.7mm}
\left\|\hspace{-0.2mm}\textbf{F}_{\hspace{-0.5mm}k} \hspace{-1mm}+\hspace{-1mm} \hat{\textbf{A}}_{k}^{\hspace{-0.8mm}H} \hspace{-0.7mm}\textbf{C}_{k}\hat{\textbf{A}}_{k}\hspace{-0.5mm}\right\|_{F}^{2}
\hspace{-3mm} -\hspace{-0.5mm} 2\text{Tr}(\hspace{-0.2mm}\textbf{Y}_{\hspace{-0.5mm}k}\hat{\textbf{H}}_{k}\hspace{-0.3mm}) \hspace{-0.9mm}+ \hspace{-1mm}\left\|\hspace{-0.3mm}\hat{\textbf{A}}_{k}^{\hspace{-0.5mm}H}\hspace{-0.7mm}\textbf{C}_{k}^{(\hspace{-0.3mm}i_{2}\hspace{-0.3mm})}\hspace{-0.7mm}\hat{\textbf{A}}_{k}\hspace{-0.3mm}\right\|_{\hspace{-0.3mm}F}^{2} \hspace{-2.5mm}+ \hspace{-1mm} \left\|\hspace{-0.2mm}\textbf{F}_{k}^{(\hspace{-0.2mm}i_{2}\hspace{-0.2mm})}\hspace{-0.5mm}\right\|_{\hspace{-0.2mm}F}^{\hspace{-0.2mm}2}
\nonumber \\
-\hspace{-0.5mm} &2 \text{Tr}\hspace{-0.8mm}\left(\hspace{-0.8mm}(\hspace{-0.2mm}\hat{\textbf{A}}_{k}\hat{\textbf{A}}_{k}^{\hspace{-0.8mm}H} \hspace{-0.5mm}\textbf{C}_{k}^{(i_{2})}\hspace{-0.5mm}\hat{\textbf{A}}_{k}\hat{\textbf{A}}_{k}^{\hspace{-0.8mm}H}\hspace{-0.5mm})^{\hspace{-0.3mm}H}\hspace{-0.5mm}\textbf{C}_{k}\hspace{-0.9mm}\right)
\hspace{-0.9mm}-\hspace{-1.2mm} 2\text{Tr}\hspace{-0.5mm}\left(\hspace{-0.9mm}(\textbf{F}_{k}^{(\hspace{-0.4mm}i_{2}\hspace{-0.4mm})}\hspace{-0.5mm})^{\hspace{-0.34mm}H}\textbf{F}_{k}\hspace{-0.7mm}\right)
\hspace{-1mm}
\le \hspace{-0.8mm}0, \hspace{-0.2mm} \forall k, \nonumber \\[2pt]
(\widetilde{\overline{\rm C9ab}})\hspace{-1mm}:&
2\text{Tr}(\textbf{Y}_{k}\hat{\textbf{H}}_{k}) 
\hspace{-0.8mm}-\hspace{-0.8mm} 2\text{Tr}\hspace{-1mm}\left(\hspace{-0.5mm}(\textbf{F}_{k}^{(i_{2})} \hspace{-1.3mm}+ \hspace{-0.8mm}\hat{\textbf{A}}_{k}^{H}\hspace{-0.5mm}\textbf{C}_{k}^{(i_{2})}\hat{\textbf{A}}_{k})^{H}\hspace{-0.5mm}(\textbf{F}_{k}\hspace{-0.5mm}-\hspace{-0.5mm}\textbf{F}_{k}^{(i_{2})})\right)
\nonumber \\
&
\hspace{-3mm}+\left\|\hat{\textbf{A}}_{k}^{H}\textbf{C}_{k}\hat{\textbf{A}}_{k}\right\|_{F}^{2} \hspace{-1mm}-\left\|\textbf{F}_{k}^{(i_{2})} \hspace{-1mm}+\hspace{-0.5mm} \hat{\textbf{A}}_{k}^{H}\textbf{C}_{k}^{(i_{2})}\hat{\textbf{A}}_{k}\right\|_{F}^{2} 
\hspace{-1mm}+\hspace{-0.5mm} \left\|\textbf{F}_{k}\right\|_{F}^{2} \nonumber \\
&\hspace{-3mm}- \hspace{-0.5mm}2\text{Tr}\!\left((\textbf{F}_{k}^{(\hspace{-0.25mm}i_{2}\hspace{-0.25mm})} \hspace{-1.2mm}+ \hspace{-1mm}\hat{\textbf{A}}_{k}^{H}\hspace{-0.8mm}\textbf{C}_{k}^{(\hspace{-0.25mm}i_{2}\hspace{-0.25mm})}\hspace{-0.8mm}\hat{\textbf{A}}_{k})^{H}\hspace{-0.5mm}\hat{\textbf{A}}_{k}^{H}\hspace{-0.5mm}(\textbf{C}_{k}\hspace{-0.8mm}-\hspace{-0.8mm}\textbf{C}_{k}^{(\hspace{-0.25mm}i_{2}\hspace{-0.25mm})})\hat{\textbf{A}}_{k}\hspace{-0.5mm}\right)\hspace{-0.8mm} \le\hspace{-0.8mm} 0\hspace{-0.3mm}, \forall k, \nonumber \\
(\widetilde{\overline{{\rm C11a}}})\hspace{-1mm}: &\hspace{-0.5mm}\left(\hspace{-0.5mm}x_{\hspace{-0.3mm}n\hspace{-0.2mm},\hspace{-0.2mm}m}^{\rm Pin} \hspace{-1.4mm}-\hspace{-0.9mm} x_{\hspace{-0.3mm}k}\hspace{-0.5mm}\right)^{\hspace{-0.6mm}2} \hspace{-1.7mm}-\hspace{-0.5mm} 
\tfrac{1}{z_{\hspace{-0.2mm}n\hspace{-0.2mm},\hspace{-0.2mm}m\hspace{-0.2mm},\hspace{-0.2mm}k}^{(i_2)}} \hspace{-0.6mm} + \hspace{-0.6mm} \tfrac{\left(\hspace{-0.9mm}z_{\hspace{-0.2mm}n\hspace{-0.2mm},\hspace{-0.2mm}m\hspace{-0.2mm},\hspace{-0.2mm}k} \hspace{-0.3mm} - \hspace{-0.3mm} z_{\hspace{-0.2mm}\hspace{-0.2mm}n\hspace{-0.2mm},\hspace{-0.2mm}m\hspace{-0.2mm},\hspace{-0.2mm}k}^{(i_2)}\hspace{-0.7mm}\right)}{\left(z_{n,m,k}^{(i_2)}\right)^2} \hspace{-0.6mm} 
+ \hspace{-0.5mm}\hat{\beta}_{n,k} \hspace{-0.9mm} \leq \hspace{-0.7mm}  0, \hspace{-0.3mm}  \forall \hspace{-0.4mm}n\hspace{-0.15mm},\hspace{-0.4mm}m\hspace{-0.15mm},\hspace{-0.4mm}k\hspace{-0.15mm}, \nonumber \\
(\widetilde{\overline{{\rm C11b}}})\hspace{-1mm}: &\tfrac{1}{z_{n\hspace{-0.1mm},\hspace{-0.1mm}m\hspace{-0.1mm},\hspace{-0.1mm}k}} \hspace{-0.2mm}-\hspace{-0.3mm} \hat{\beta}_{\hspace{-0.1mm}n\hspace{-0.1mm},\hspace{-0.1mm}k} \hspace{-0.5mm}-\hspace{-0.4mm}\left( \hspace{-0.3mm}x_{n,m}^{\text{Pin}\hspace{-0.1mm}(\hspace{-0.3mm}i_2\hspace{-0.2mm})} \hspace{-0.6mm} - \hspace{-0.3mm} x_{\hspace{-0.2mm}k} \hspace{-0.5mm}\right)^{ \hspace{-0.4mm}2}  \nonumber \\
&- 2 \hspace{-0.4mm}\left( \hspace{-0.4mm}x_{n,m}^{\text{Pin}(\hspace{-0.2mm}i_2\hspace{-0.2mm})} \hspace{-0.7mm} - \hspace{-0.4mm} x_{\hspace{-0.1mm}k} \hspace{-0.5mm}\right)  \hspace{-0.6mm}\left( \hspace{-0.5mm}x_{\hspace{-0.1mm}n\hspace{-0.1mm},\hspace{-0.1mm}m}^{\text{Pin}} \hspace{-0.7mm} - \hspace{-0.3mm} x_{n,m}^{\text{Pin}(\hspace{-0.2mm}i_2\hspace{-0.2mm})} \hspace{-0.4mm}\right) \hspace{-0.5mm}\leq \hspace{-0.4mm} 0\hspace{-0.2mm},\hspace{-0.2mm} \forall n\hspace{-0.1mm},\hspace{-0.1mm}m\hspace{-0.1mm},\hspace{-0.1mm}k\hspace{-0.1mm}, \nonumber \\
(\widetilde{\overline{{\rm C13a}}}) \hspace{-1.2mm}: & \log_2  \hspace{-1.2mm} \left(\hspace{-0.8mm}\text{Diag}(\hspace{-0.4mm}\textbf{C}_k^{(\hspace{-0.25mm}i_2\hspace{-0.25mm})}\hspace{-0.4mm})_{\hspace{-0.3mm}l(\hspace{-0.5mm}n,m\hspace{-0.5mm})} \hspace{-1mm}\right) \hspace{-1mm}+\hspace{-0.5mm} \tfrac{\hspace{-0.6mm}\text{Diag}(\hspace{-0.4mm}\textbf{C}_k\hspace{-0.4mm})_{\hspace{-0.4mm}l\hspace{-0.3mm}(\hspace{-0.3mm}n\hspace{-0.2mm},\hspace{-0.2mm}m\hspace{-0.3mm})} - \text{Diag}(\hspace{-0.4mm}\textbf{C}_k^{(\hspace{-0.2mm}i_2\hspace{-0.2mm})}\hspace{-0.4mm})_{l\hspace{-0.2mm}(\hspace{-0.3mm}n\hspace{-0.3mm},m\hspace{-0.4mm})}\hspace{-0.8mm} }{\text{ln}(2) \text{Diag}(\textbf{C}_k^{(i_2)})_{l(n,m)}} \nonumber \\
&\hspace{-1.5mm}+ \text{log}_2 \hspace{-0.8mm}\left(1\hspace{-0.8mm}+\hspace{-0.8mm} \text{exp}\hspace{-0.8mm} \left[-\vartheta \bar{\alpha}_{n,m,k}   \right]\right) \hspace{-0.8mm}-\hspace{-0.8mm} \log_2 (\hat{\eta})  \hspace{-0.8mm}\leq \hspace{-0.8mm}0, \hspace{-0.8mm}\forall n, m, k, \nonumber \\
(\widetilde{\overline{{\rm C13b}}})\hspace{-1mm}: & -\hspace{-0.8mm}\log_2 \hspace{-0.8mm}\left(\hspace{-0.3mm}\text{Diag}(\textbf{C}_k)_{l(n,m)}\hspace{-0.8mm}\right) \hspace{-0.4mm}  - \hspace{-0.8mm}\text{log}_2 \hspace{-0.8mm}\left(\hspace{-0.8mm}1 \hspace{-0.8mm}+\hspace{-0.8mm} \text{exp} \hspace{-0.8mm}\left[-\vartheta \bar{\alpha}_{n,m,k}^{(i_2)} \hspace{-0.4mm} \right] \hspace{-0.3mm}\right) \nonumber\\
 +& \log_2 (\hat{\eta}) \hspace{-0.8mm}+ \tfrac{\vartheta e^{-\vartheta \bar{\alpha}_{n,m,k}^{(i_2)}   }\hspace{-0.8mm}\left(\hspace{-0.8mm}\bar{\alpha}_{n,m,k} -\bar{\alpha}_{n,m,k}^{(i_2)}\hspace{-0.8mm}\right)}{\text{ln}(2) \left(1+ \text{exp} \left[-\vartheta \bar{\alpha}_{n,m,k}^{(i_2)}\right]\right)} \leq 0, \forall n, m, k, \nonumber \\
(\widetilde{\overline{\rm C15a}})\hspace{-1mm}: &z_{\hspace{-0.3mm}n\hspace{-0.2mm},\hspace{-0.2mm}m\hspace{-0.2mm},\hspace{-0.2mm}k}^{-1} \hspace{-1.3mm}-\hspace{-1mm} \left(\hspace{-1mm}(\hspace{-0.6mm}\bar{r}_{\hspace{-1mm}n\hspace{-0.2mm},\hspace{-0.2mm}m\hspace{-0.2mm},\hspace{-0.2mm}k}^{(i_2)}\hspace{-0.3mm})^{\hspace{-0.2mm}2} \hspace{-1.3mm}+\hspace{-0.8mm} 2\bar{r}_{\hspace{-0.8mm}n\hspace{-0.2mm},\hspace{-0.2mm}m\hspace{-0.2mm},\hspace{-0.2mm}k}^{(i_2)}\hspace{-0.4mm}(\hspace{-0.4mm}\bar{r}_{\hspace{-0.8mm}n\hspace{-0.2mm},m\hspace{-0.2mm},k} \hspace{-0.8mm}-\hspace{-0.8mm} \bar{r}_{\hspace{-0.8mm}n\hspace{-0.2mm},m\hspace{-0.2mm},k}^{(i_2)}) \hspace{-0.7mm} \right) \hspace{-0.7mm} \leq \hspace{-0.7mm}  0, \hspace{-0.3mm}  \forall \hspace{-0.4mm}n\hspace{-0.15mm},\hspace{-0.4mm}m\hspace{-0.15mm},\hspace{-0.4mm}k\hspace{-0.15mm}, \nonumber \\
(\widetilde{\overline{\rm C15b}})\hspace{-1mm}: &\bar{r}_{\hspace{-0.5mm}n,m,k}^{2} \hspace{-1mm}-\hspace{-1mm} \left(\hspace{-1.2mm}\tfrac{1}{z_{\hspace{-0.3mm}n\hspace{-0.2mm},\hspace{-0.2mm}m\hspace{-0.2mm},\hspace{-0.2mm}k}^{(i_2)}} \hspace{-0.3mm}- \tfrac{\left(\hspace{-0.6mm}z_{\hspace{-0.2mm}n\hspace{-0.2mm},\hspace{-0.2mm}m\hspace{-0.2mm},\hspace{-0.2mm}k} - z_{\hspace{-0.3mm}n,m,k}^{(i_2)}\hspace{-0.6mm}\right)}{(z_{\hspace{-0.3mm}n\hspace{-0.2mm},\hspace{-0.2mm}m\hspace{-0.2mm},\hspace{-0.2mm}k}^{(i_2)})^2} \hspace{-0.8mm} \right) \hspace{-0.9mm}\leq \hspace{-0.7mm} 0, \forall n,m,k, 
\end{align} 
where $\text{Diag}(\textbf{F}_k)_{l(n,m)}=z_{n,m,k}$ and $\text{Diag}(\textbf{F}_g^{\rm E})_{l(n,m,t)}=z_{n,m,g,t}^{\rm E}$ for notational simplicity. The terms “$(i_2)$” are the solutions obtained at iteration $i_2$ for the corresponding variables. Also, constraints $(\widetilde{\overline{{\rm C10aa}}})$, $(\widetilde{\overline{{\rm C10ab}}})$, $(\widetilde{\overline{{\rm C12a}}})$, $(\widetilde{\overline{{\rm C12b}}})$, $(\widetilde{\overline{{\rm C14a}}})$, $(\widetilde{\overline{{\rm C14b}}})$, $(\widetilde{\overline{{\rm C16a}}})$, and $(\widetilde{\overline{{\rm C16b}}})$ are omitted to avoid repetition, as they are analogous to $(\widetilde{\overline{{\rm C9aa}}})$, $(\widetilde{\overline{{\rm C9ab}}})$, $(\widetilde{\overline{{\rm C11a}}})$, $(\widetilde{\overline{{\rm C11b}}})$, $(\widetilde{\overline{{\rm C13a}}})$, $(\widetilde{\overline{{\rm C13b}}})$, $(\widetilde{\overline{{\rm C15a}}})$, and $(\widetilde{\overline{{\rm C15b}}})$, respectively.

To handle the sigmoid-involved constraints, we take $(\widetilde{\rm C13c})$ and $(\widetilde{\rm C13d})$ as examples. We first map the LoS availability metric $d_{k,q}^{n,m}$ of PA $m$ on waveguide $n$ for user $u_k$ with respect to blockage $q$ into a one-dimensional affine form as $d_{k,q}^{n,m} = \underset{i \in \mathcal{I}_{k,q}}{\max} \{[\textbf{a}_{k,q}^{i}]_1 x_{n,m}^{\rm Pin} + \bar{b}_{k,q}^{i}\}$, where $\bar{b}_{k,q}^{i} = [\textbf{a}_{k,q}^{i}]_2 y_n^{\rm FP} + [\textbf{a}_{k,q}^{i}]_3 d - b_{k,q}^{i}$ is a constant. Inspired by~\cite{yi2023trajectory}, we then identify, for each blockage $q$, the most active plane contributing to $d_{k,q}^{n,m}$ at the current PA position $x_{n,m}^{{\rm Pin}(i_2)}$ in the $i_2$-th iteration, i.e., $i^*(q) \in \underset{i \in \mathcal{I}_{k,q}}{\arg\max}  \{[\textbf{a}_{k,q}^{i}]_1 x_{n,m}^{{\rm Pin}(i_2)} + \bar{b}_{k,q}^{i} \}, \forall n,m,k$. Given these active planes ${i^*(q)}, \forall q$, we substitute them into $\mu_{k}^{n,m}(x_{n,m}^{\rm Pin})$ and, at the $(i_2+1)$-th iteration of the MM algorithm, construct a concave lower bound of $\mu_{k}^{n,m}(x_{n,m}^{\rm Pin})$ via a first-order Taylor expansion around $x_{n,m}^{{\rm Pin}(i_2)}$, namely:
\begin{align}\label{eq: activePlane}
&\mu_{k}^{n,m} (x_{n,m}^{\rm Pin}) \ge \mu_{k,{\rm LB}}^{n,m(i_2)} (x_{n,m}^{\rm Pin}) \nonumber \\
 \triangleq &\underset{q \in \mathcal{Q}}{\min} \Bigr\{\hspace{-0.8mm} d_{k,q}^{\hspace{-0.05mm}n\hspace{-0.2mm},\hspace{-0.2mm}m} \hspace{-0.7mm}(\hspace{-0.4mm}x_{n,m}^{{\rm Pin}(i_2)}\hspace{-0.4mm}) \hspace{-0.7mm}+\hspace{-0.7mm} [\textbf{a}_{k,q}^{i^*}]_1 \hspace{-0.3mm} ( \hspace{-0.4mm} x_{n,m}^{{\rm Pin}} \hspace{-1.2mm}-\hspace{-0.9mm} x_{n,m}^{{\rm Pin}(i_2)}\hspace{-0.4mm})\hspace{-0.9mm}\Bigr\}\hspace{-0.5mm}, \forall n,m,k. 
\end{align}
Therefore, a convex subset of $(\widetilde{\rm C13c})$ of the $(i_2+1)$-th iteration MM can be written as:
\begin{align}\label{eq: 13c}
&\widetilde{\overline{({\rm C13c})}}\hspace{-1mm}: \tfrac{(\bar{r}_{n,m,k}+ \bar{\alpha}_{n,m,k})^2 - (\bar{r}_{n,m,k}^{(i_2)}- \bar{\alpha}_{n,m,k}^{(i_2)})^2}{4} -\mu_{k,{\rm LB}}^{n,m(i_2)} (x_{n,m}^{\rm Pin}) \nonumber \\
&+ \hspace{-1mm}\tfrac{(\hspace{-0.3mm}\bar{r}_{\hspace{-0.5mm}n\hspace{-0.2mm},\hspace{-0.2mm}m\hspace{-0.2mm},\hspace{-0.2mm}k}^{(i_2)} \hspace{-1.1mm}-\hspace{-0.05mm} \bar{\alpha}_{\hspace{-0.5mm}n\hspace{-0.2mm},\hspace{-0.2mm}m\hspace{-0.2mm},\hspace{-0.2mm}k}^{(i_2)}\hspace{-0.3mm}) \hspace{-0.5mm} \left(\hspace{-0.9mm}(\hspace{-0.5mm}\bar{\alpha}_{\hspace{-0.3mm}n\hspace{-0.2mm},\hspace{-0.2mm}m\hspace{-0.2mm},\hspace{-0.2mm}k} \hspace{-0.5mm}- \bar{\alpha}_{\hspace{-0.3mm}n\hspace{-0.2mm},\hspace{-0.2mm}m,\hspace{-0.2mm}k}^{(i_2)}\hspace{-0.4mm}) \hspace{-0.3mm}-\hspace{-0.3mm} (\hspace{-0.5mm}\bar{r}_{\hspace{-0.3mm}n\hspace{-0.2mm},\hspace{-0.2mm}m\hspace{-0.2mm},\hspace{-0.2mm}k} \hspace{-0.5mm}- \bar{r}_{\hspace{-0.3mm}n\hspace{-0.2mm},\hspace{-0.2mm}m,\hspace{-0.2mm}k}^{(i_2)}\hspace{-0.4mm})\hspace{-0.8mm}\right)}{2} \hspace{-1mm}   \le\hspace{-1mm} 0\hspace{-0.2mm}, \hspace{-0.5mm}\forall n\hspace{-0.1mm},\hspace{-0.1mm}m\hspace{-0.1mm},\hspace{-0.1mm}k,
\end{align}
where $\widetilde{\overline{({\rm C13c})}} \Rightarrow \widetilde{({\rm C13c})}$. Similarly, to construct a convex upper bound for $\mu_{k}^{n,m}(x_{n,m}^{\rm Pin})$, we identify, for each PA–user pair, the most critical blockage at the current iterate $x_{n,m}^{{\rm Pin}(i_2)}$ as $q^*(i) \in \underset{q \in \mathcal{Q}}{\arg\min}  \Bigr\{d_{k,q}^{n,m} (x_{n,m}^{{\rm Pin}(i_2)})\Bigr\}, \forall n,m,k$. Using this active blockage index $q^*(i)$, we then establish a convex upper bound of $\mu_{k}^{n,m}(x_{n,m}^{\rm Pin})$ and hence obtain a convex inner approximation of constraint $(\widetilde{\overline{\rm C13d}})$ at the $(i_2+1)$-th MM iteration via a first-order Taylor expansion as:
\begin{align}\label{eq: 13d}
&\mu_{k}^{\hspace{-0.35mm}n\hspace{-0.2mm},\hspace{-0.2mm}m} \hspace{-0.6mm} (\hspace{-0.5mm}x_{\hspace{-0.4mm}n\hspace{-0.2mm},\hspace{-0.2mm}m}^{\rm Pin}\hspace{-0.4mm}) \hspace{-0.8mm} \le\hspace{-0.8mm} \mu_{\hspace{-0.4mm}k\hspace{-0.3mm},\hspace{-0.4mm}{\rm UB}}^{{n,m}} \hspace{-0.6mm}(\hspace{-0.6mm}x_{n,m}^{\rm Pin}\hspace{-0.4mm})\hspace{-0.9mm} = \hspace{-2mm}\underset{i \in \mathcal{I}_{k,q^*}}{\max} \hspace{-1.4mm} \Bigr\{\hspace{-0.8mm} [\textbf{a}_{k,q^*}^{i}]_1 \hspace{-0.8mm} (\hspace{-0.5mm}x_{\hspace{-0.4mm}n\hspace{-0.2mm},\hspace{-0.2mm}m}^{\rm Pin}\hspace{-0.4mm}) \hspace{-0.7mm}+\hspace{-0.7mm} \bar{b}_{\hspace{-0.3mm}k\hspace{-0.2mm},\hspace{-0.2mm}q^{\hspace{-0.15mm}*}}^{i} \hspace{-1.3mm}\Bigr\}\hspace{-0.7mm}, \hspace{-0.35mm}\forall n\hspace{-0.15mm},\hspace{-0.15mm}m\hspace{-0.15mm},\hspace{-0.15mm}k,\nonumber \\
&(\widetilde{\overline{{\rm C13d}}})\hspace{-1mm}: \mu_{k,{\rm UB}}^{{n,m}} (x_{n,m}^{{\rm Pin}}) + \tfrac{(\bar{r}_{n,m,k} -  \bar{\alpha}_{n,m,k})^2 -(\bar{r}_{n,m,k}^{(i_2)} + \bar{\alpha}_{n,m,k}^{(i_2)})^2 }{4} \nonumber \\
&-\hspace{-1mm} \tfrac{(\hspace{-0.3mm}\bar{r}_{\hspace{-0.5mm}n\hspace{-0.2mm},\hspace{-0.2mm}m\hspace{-0.2mm},\hspace{-0.2mm}k}^{(i_2)} \hspace{-1.1mm}+\hspace{-0.1mm} \bar{\alpha}_{\hspace{-0.5mm}n\hspace{-0.2mm},\hspace{-0.2mm}m\hspace{-0.2mm},\hspace{-0.2mm}k}^{(i_2)}\hspace{-0.3mm})  \hspace{-0.5mm} \left(\hspace{-0.9mm}(\hspace{-0.5mm}\bar{\alpha}_{\hspace{-0.3mm}n\hspace{-0.2mm},\hspace{-0.2mm}m\hspace{-0.2mm},\hspace{-0.2mm}k} \hspace{-0.5mm}- \bar{\alpha}_{\hspace{-0.3mm}n\hspace{-0.2mm},\hspace{-0.2mm}m,\hspace{-0.2mm}k}^{(i_2)}\hspace{-0.4mm}) \hspace{-0.3mm}+\hspace{-0.3mm} (\hspace{-0.5mm}\bar{r}_{\hspace{-0.3mm}n\hspace{-0.2mm},\hspace{-0.2mm}m\hspace{-0.2mm},\hspace{-0.2mm}k} \hspace{-0.5mm}- \bar{r}_{\hspace{-0.3mm}n\hspace{-0.2mm},\hspace{-0.2mm}m,\hspace{-0.2mm}k}^{(i_2)}\hspace{-0.4mm})\hspace{-0.8mm}\right)}{2}  \le\hspace{-1mm} 0\hspace{-0.2mm}, \hspace{-0.5mm}\forall n\hspace{-0.1mm},\hspace{-0.1mm}m\hspace{-0.1mm},\hspace{-0.1mm}k,
\end{align}
respectively. Note that we omit the details for $(\widetilde{\overline{\rm C14c}})$ and $(\widetilde{\overline{\rm C14d}})$, as they are analogous to $(\widetilde{\overline{\rm C13c}})$ and $(\widetilde{\overline{\rm C13d}})$ in  \eqref{eq: 13c} and \eqref{eq: 13d}, respectively.

Accordingly, the $(i_2+1)$-th MM iteration for stage~1 of PA positioning is as follows:
\begin{align}
\label{eq: formu2_Fin}
&\operatorname*{maximize}_{\substack{
   \textbf{x}_n^{\rm Pin},\; \textbf{F}_k,\; \textbf{C}_k, \; \iota,\\
   \textbf{F}_g^{\rm E},\; \textbf{C}_g^{\rm E}, \; \hat{\textbf{H}}_k,\;\hat{\textbf{h}}_k,\\
 \textbf{J}_g, \;\textbf{j}_g,\; \mathcal{J}_g, \; \mathcal{H}_k,\; \delta
}} \underset{k \in \mathcal{K}}{\sum} \log_2 (\iota_k^{\rm N} + \iota_k^{\rm D} + \sigma_k^2) +\hspace{-0.2mm}\bar{R}(\hspace{-0.2mm}\iota_k^{\hspace{-0.1mm}\mathrm{D}(\hspace{-0.1mm}i_2\hspace{-0.1mm})}\hspace{-0.3mm}) \nonumber \\
&\rm{s.t.} (\hspace{-0.3mm}{\rm C1}\hspace{-0.3mm})\hspace{-0.3mm}, \hspace{-0.3mm}(\hspace{-0.3mm}{\rm C2}\hspace{-0.3mm})\hspace{-0.3mm},  \hspace{-0.3mm}(\hspace{-0.25mm}\widetilde{{\rm C6a}}\hspace{-0.25mm})\hspace{-0.3mm}, \hspace{-0.3mm}(\hspace{-0.25mm}\widetilde{{\rm C6b}}\hspace{-0.25mm})\hspace{-0.3mm}, \hspace{-0.3mm}(\hspace{-0.25mm}\widetilde{{\rm C7a}}\hspace{-0.25mm})\hspace{-0.3mm}, \hspace{-0.3mm}(\hspace{-0.25mm}\widetilde{{\rm C7b}}\hspace{-0.25mm})\hspace{-0.3mm}, \hspace{-0.3mm}(\hspace{-0.25mm}\widetilde{{\rm C8a}}\hspace{-0.25mm})\hspace{-0.3mm}, \hspace{-0.3mm}(\hspace{-0.2mm}\widetilde{{\rm C8b}}\hspace{-0.2mm})\hspace{-0.3mm}, \hspace{-0.3mm}(\hspace{-0.2mm}\widetilde{\overline{{\rm C9aa}}}\hspace{-0.2mm})\hspace{-0.3mm}, 
\nonumber \\
&(\hspace{-0.2mm}\widetilde{\overline{{\rm C9ab}}}\hspace{-0.2mm})\hspace{-0.3mm}, \hspace{-0.3mm}(\hspace{-0.2mm}\widetilde{\overline{{\rm C9b}}}\hspace{-0.2mm})\hspace{-0.8mm}-\hspace{-0.8mm} (\hspace{-0.2mm}\widetilde{\overline{{\rm C9d}}}\hspace{-0.2mm})\hspace{-0.3mm}, \hspace{-0.3mm}(\hspace{-0.2mm}\widetilde{\overline{{\rm C10aa}}}\hspace{-0.2mm})\hspace{-0.3mm}, \hspace{-0.3mm}(\hspace{-0.2mm}\widetilde{\overline{{\rm C10ab}}}\hspace{-0.2mm})\hspace{-0.3mm}, \hspace{-0.3mm}(\hspace{-0.2mm}\widetilde{\overline{{\rm C10b}}}\hspace{-0.2mm})\hspace{-0.8mm}-\hspace{-0.8mm} (\hspace{-0.1mm}\widetilde{\overline{{\rm C10d}}}\hspace{-0.1mm})\hspace{-0.15mm}, \hspace{-0.15mm}  \nonumber \\
&(\hspace{-0.1mm}\widetilde{\overline{{\rm C11a}}}\hspace{-0.1mm})\hspace{-0.15mm}, \hspace{-0.15mm}(\hspace{-0.1mm}\widetilde{\overline{{\rm C11b}}}\hspace{-0.1mm})\hspace{-0.15mm}, \hspace{-0.15mm}(\hspace{-0.1mm}\widetilde{\overline{{\rm C12a}}}\hspace{-0.1mm})\hspace{-0.15mm}, \hspace{-0.15mm}(\hspace{-0.1mm}\widetilde{\overline{{\rm C12b}}}\hspace{-0.1mm})\hspace{-0.15mm}, (\hspace{-0.1mm}\widetilde{\overline{{\rm C13a}}}\hspace{-0.1mm}) \hspace{-0.8mm}-\hspace{-0.8mm} (\hspace{-0.1mm}\widetilde{\overline{{\rm C13d}}}\hspace{-0.1mm})\hspace{-0.15mm}, \nonumber \\
&(\hspace{-0.1mm}\widetilde{\overline{{\rm C14a}}}\hspace{-0.1mm}) \hspace{-0.8mm}-\hspace{-0.8mm} (\hspace{-0.1mm}\widetilde{\overline{{\rm C14d}}}\hspace{-0.1mm})\hspace{-0.15mm}, \hspace{-0.15mm}(\hspace{-0.1mm}\widetilde{\overline{{\rm C15a}}}\hspace{-0.1mm})\hspace{-0.15mm}, \hspace{-0.15mm}(\hspace{-0.1mm}\widetilde{\overline{{\rm C15b}}}\hspace{-0.1mm})\hspace{-0.15mm},\hspace{-0.15mm} (\hspace{-0.1mm}\widetilde{\overline{{\rm C16a}}}\hspace{-0.1mm})\hspace{-0.15mm}, \hspace{-0.15mm}(\hspace{-0.1mm}\widetilde{\overline{{\rm C16b}}}\hspace{-0.1mm}),
\end{align}
where $\bar{\alpha} = \{\bar{\alpha}_{n,m,k}, \bar{\alpha}_{n,m,g,t}^{\rm E}\}$ and $\bar{r} = \{\bar{r}_{n,m,k}, \bar{r}_{n,m,g,t}^{\rm E}\}$ to simplify notation. The resulting problem in~\eqref{eq: formu2_Fin} is convex and can be directly solved using CVX. 

\subsection{Phase-Shift-Based Small-Scale Movement}
With the coarse PA positions fixed, we next refine the phase shifts induced by PA displacements, i.e., we optimize ${\textbf{a}_k, \textbf{a}^{\rm E}_g, \textbf{x}^{\rm Pin}}$ given ${\textbf{f}_k, \textbf{f}^{\rm E}_g, \textbf{c}_k, \textbf{c}^{\rm E}_g}$. Define
$\hat{\boldsymbol{\chi}}_k=\textbf{f}_k\odot\textbf{c}_k$, such that $\hat{\textbf{h}}_k=\textbf{A}_k^{\rm D}\hat{\boldsymbol{\chi}}_k$, where $\textbf{A}_k^{\rm D}=\mathrm{diag}(\textbf{a}_k)$ is diagonal with unit-modulus entries. Then:
\begin{align}\label{eq: transA}
\textbf{h}_k \hspace{-0.8mm}= \hspace{-0.8mm} \textbf{A}_k^{\rm D} \hspace{-0.5mm}(\hspace{-0.4mm}\hat{\bm{\chi}}_k \hspace{-0.8mm}+\hspace{-0.8mm} \bm{\triangle} \bm{\chi}_k\hspace{-0.4mm})\hspace{-0.4mm}, ||\triangle \textbf{h}_k||_2 \hspace{-0.6mm}\leq\hspace{-0.6mm} \wp_k \hspace{-0.7mm} \Leftrightarrow \hspace{-0.7mm} ||\triangle \bm{\chi}_k||_2 \hspace{-0.5mm}\leq\hspace{-0.5mm} \wp_k\hspace{-0.3mm}, \hspace{-0.4mm} \forall k.
\end{align}
Recalling $({\rm C7})$ and $({\rm C8})$ in~\eqref{eq: boundsPA} and exploiting the transformation in~\eqref{eq: transA}, we lift the phase-shift variables as $\textbf{A}_k=\textbf{a}_k\textbf{a}_k^{H}$, $\forall k$, and $\textbf{A}_g^{\rm E}=\textbf{a}_g^{\rm E}(\textbf{a}_g^{\rm E})^{H}$, $\forall g$. Noting that $\textbf{A}_k^{{\rm D}H}\textbf{Y}_k\textbf{A}_k^{\rm D}=\textbf{A}_k^{\top}\odot\textbf{Y}_k$, the resulting performance lower bound and the associated constraints $({\rm C7})$ and $({\rm C8})$, after applying Lemma~1, can be written in terms of ${\textbf{A}_k,\textbf{A}_g^{\rm E}}$ as:
\begin{align}\label{eq: C78_pahse}
&\underset{k \in \mathcal{K}}{\sum} \hspace{-0.5mm} \log_2 \hspace{-0.3mm}(\hspace{-0.3mm}\iota_k^{\rm N} \hspace{-0.9mm}+\hspace{-0.7mm} \iota_k^{\rm D} \hspace{-0.9mm}+\hspace{-0.5mm} \sigma_k^2) \hspace{-0.7mm}-\hspace{-0.7mm} \underset{k \in \mathcal{K}}{\sum}\hspace{-0.5mm} \log_2 (\iota_k^{\rm D} \hspace{-0.7mm}+\hspace{-0.7mm} \sigma_k^2) \;\; \; \text{s.t.} \nonumber \\
&(\widehat{{\rm C7a}})\hspace{-1mm}:  \left[\hspace{-2mm}
\begin{array}{cc}
\textbf{A}^{\text{Y}}_k \hspace{-0.7mm}-\hspace{-0.7mm} \delta_k^{N} \textbf{I}_{MN} \hspace{-4.5mm}&
\textbf{A}^{\text{Y}}_k\hat{\bm{\chi}}_k  \\[0.4em]
\textbf{A}^{\text{Y}}_k\hat{\bm{\chi}}_k^H \hspace{-4.5mm}&
\hat{\bm{\chi}}_k^H \textbf{A}^{\text{Y}}_k \hat{\bm{\chi}}_k
\hspace{-0.7mm} -\hspace{-0.7mm} \iota_k^{N} \hspace{-0.7mm}+\hspace{-0.7mm} \delta_k^{N} \wp_k^2
\end{array}
\hspace{-1mm}\right]
\hspace{-0.5mm}\bm{\succeq} \hspace{-0.5mm}\bm{0},  \forall k, \nonumber \\
&(\widehat{{\rm C7b}})\hspace{-1mm}: \delta_k^N \ge 0, \forall k, \;\; (\widehat{{\rm C8b}})\hspace{-1mm}: \delta_k^D \ge 0, \forall k, \nonumber \\
&(\widehat{{\rm C8a}})\hspace{-1mm}:  \left[\hspace{-2mm}
\begin{array}{cc}
\delta_k^{D} \textbf{I}_{MN} \hspace{-0.5mm}-\hspace{-0.5mm} \textbf{A}^{\bar{\text{Z}}}_k \hspace{-3mm}&
-\textbf{A}^{\bar{\text{Z}}}_k\hat{\bm{\chi}}_k \\[0.4em]
-(\textbf{A}^{\bar{\text{Z}}}_k\hat{\bm{\chi}}_k)^H \hspace{-3mm}&
\iota_k^{D} \hspace{-1mm}-\hspace{-0.6mm} \hat{\bm{\chi}}_k^H \textbf{A}^{\bar{\text{Z}}}_k\hat{\bm{\chi}}_k \hspace{-0.5mm}-\hspace{-0.5mm} \delta_k^{D} \wp_k^2
\end{array}
\hspace{-1.5mm}\right] \hspace{-0.5mm}
\bm{\succeq} \hspace{-0.5mm} \bm{0}, \forall k, \nonumber \\
&(\hspace{-0.3mm}\widehat{{\rm C9a}}\hspace{-0.3mm})\hspace{-1mm}: \hspace{-0.5mm}\text{Rank}(\hspace{-0.3mm}\textbf{A}_k\hspace{-0.3mm}) \hspace{-0.8mm}=\hspace{-1mm} 1\hspace{-0.3mm}, \hspace{-0.3mm}\forall k, \;(\hspace{-0.3mm}\widehat{{\rm C9b}}\hspace{-0.3mm})\hspace{-1mm}:\hspace{-0.5mm}\textbf{A}_k \hspace{-0.7mm}\succeq \hspace{-0.7mm}\bm{0}, \text{diag}(\hspace{-0.3mm}\textbf{A}_k\hspace{-0.3mm}) \hspace{-1mm}=\hspace{-1mm} \bm{1}\hspace{-0.3mm}, \hspace{-0.3mm}\forall k\hspace{-0.3mm},
\end{align}
where $\textbf{A}^{\text{Y}}_k = \textbf{A}_k^\top \odot \textbf{Y}_k$ and $\textbf{A}^{\bar{\text{Z}}}_k = \textbf{A}_k^\top \odot \bar{\textbf{Z}}_k$. 

We then recall $(\overline{\rm C6})$ in~\eqref{eq: boundsPA} and, following the same idea as in~\eqref{eq: C78_pahse}, define $\hat{\bm{\chi}}^{\rm E}_{g,t} = \textbf{f}^{{\rm E}}_{g,t} \odot \textbf{c}^{{\rm E}}_{g,t} $ and $\textbf{A}_{{\rm E}_{g,t}}^{\rm D} = \text{diag}(\textbf{a}_{g,t}^{\rm E}) \in \mathbb{C}^{MN \times MN}, \forall g,t$, so that the nominal EA channel can be written as $\hat{\textbf{H}}_{{\rm E}_g} = [\textbf{A}_{{\rm E}_{g,1}}^{\rm D} \hat{\bm{\chi}}^{\rm E}_{g,1}, ..., \textbf{A}_{{\rm E}_{g,T}}^{\rm D} \hat{\bm{\chi}}^{\rm E}_{g,T}] \in \mathbb{C}^{MN \times T}$. The deterministic and uncertain parts are collected as $\hat{\textbf{X}}_{{\rm E}_g} = [\hat{\bm{\chi}}^{\rm E}_{g,1}, ..., \hat{\bm{\chi}}^{\rm E}_{g,T}]$ and $\bm{\triangle}\textbf{X}_{{\rm E}_g} =  [\bm{\triangle}\bm{\chi}^{\rm E}_{g,1}, ..., \bm{\triangle}\bm{\chi}^{\rm E}_{g,T}] \in \mathbb{C}^{MN \times T}$, respectively.  We further  define the known matrix $\hat{\textbf{X}}_{{\rm E}_g}^{\rm D} = \text{blkdiag} (\hat{\bm{\chi}}^{\rm E}_{g,1}, ..., \hat{\bm{\chi}}^{\rm E}_{g,T})$ and $ \bm{\triangle}\textbf{X}_{{\rm E}_g}^{\rm D} = \text{blkdiag} (\bm{\triangle}\bm{\chi}^{\rm E}_{g,1}, ..., \bm{\triangle}\bm{\chi}^{\rm E}_{g,T}) \in \mathbb{C}^{MNT \times T}$, we have $||\bm{\triangle} \textbf{H}_{E_g}||_F \leq \wp_g^{\rm tot} \Leftrightarrow || \bm{\triangle}\textbf{X}_{{\rm E}_g}||_F \leq \wp_g^{\rm tot} $.
With $\textbf{A}_g^{\rm E} = \textbf{a}_g^E (\textbf{a}_g^E)^H$ and $\widetilde{\textbf{M}}_{k,g}^{\rm E} = (\bm{1}_T \bm{1}_T^\top) \otimes{\textbf{M}}_{k,g}^{\rm E} \in \mathbb{C}^{MNT \times MNT}$, the LMIs associated with the EA constraints, after applying Lemma~2, can be written as:
\begin{equation}
\scalebox{0.91}{$(\hspace{-0.3mm}\widehat{{\rm C6a}}\hspace{-0.3mm})\hspace{-1.1mm}:  \hspace{-0.7mm}\left[\hspace{-2.2mm}
\begin{array}{cc}
(\hspace{-0.25mm}\hat{\textbf{X}}_{{\rm E}_g}^{\rm D}\hspace{-0.35mm})^{\hspace{-0.25mm}H} \hspace{-1.1mm}\textbf{A}^{\text{M}}_{k,g}\hat{\textbf{X}}_{{\rm E}_g}^{\rm D} \hspace{-2.2mm}+\hspace{-0.7mm} (\hspace{-0.4mm}\hat{R}^{\rm th}_{\hspace{-0.3mm}k\hspace{-0.3mm},\hspace{-0.2mm}g}\hspace{-0.2mm}\sigma_{\hspace{-0.3mm}g}^2\hspace{-1mm} -\hspace{-0.8mm} \delta_{\hspace{-0.2mm}k\hspace{-0.2mm},\hspace{-0.1mm}g}^{E_3}\hspace{-0.5mm}) \textbf{I}_{T} \hspace{-5mm}&
 (\textbf{A}^{\text{M}}_{k,g} \hat{\textbf{X}}_{{\rm E}_g}^{\rm D})^H \\[0.4em]
\textbf{A}^{\text{M}}_{k,g} \hat{\textbf{X}}_{{\rm E}_g}^{\rm D} \hspace{-5mm}&
\textbf{A}^{\text{M}}_{\hspace{-0.15mm}k\hspace{-0.3mm},\hspace{-0.2mm}g} \hspace{-1.1mm}+\hspace{-0.7mm} \frac{\delta_{k,g}^{E_3}}{(\hspace{-0.35mm}\wp_g^{\rm tot}\hspace{-0.4mm})^2} \textbf{I}_{\hspace{-0.2mm}M\hspace{-0.5mm}NT}
\end{array}\hspace{-2.2mm}
\right] \hspace{-1.4mm}
\bm{\succeq} \hspace{-0.5mm} \bm{0}, \forall k,g, \nonumber
  $}
\end{equation}
\begin{equation}
\scalebox{0.95}{$\hspace{-21.7mm}(\hspace{-0.3mm}\widehat{{\rm C6b}}\hspace{-0.3mm})\hspace{-0.99mm}: \delta_{k,g}^{E_3} \ge 0, \forall k, g, \;\;
\widehat{({\rm C10a})}\hspace{-0.8mm}: \text{Rank}(\textbf{A}_g^{\rm E}) = 1, \forall g, \nonumber
  $}
\end{equation}
\begin{equation}
\scalebox{0.98}{$\hspace{-29.1mm}\widehat{({\rm C10b})}\hspace{-1mm}: \textbf{A}_g^{\rm E} \bm{\succeq 0}, \;\; \text{diag}(\textbf{A}_g^{\rm E}) = \bm{1}_{MNT}, \forall g, 
  $}
\end{equation}
where $\textbf{A}^{\text{M}}_{k,g} = (\textbf{A}_g^{\rm E})^\top \odot \widetilde{\textbf{M}}_{k,g}^{\rm E}$.
Thus, the second phase of the PA positioning problem, with variables $\textbf{A}_k$ and $\textbf{A}_g^{\rm E}$ to characterize the pure phase-alignment components for legitimate user $k$ and EA $g$, respectively, can be formulated as:
\begin{align}
\label{eq: before_Pen_formu2_firTran}
&\operatorname*{maximize}_{\substack{
   \textbf{x}_n^{\rm Pin},\; \textbf{A}_k, \;\iota, \; \delta,\\
    \textbf{A}_g^{\rm E}, \; \theta_{l(n,m),k}, \; \theta_{l(n,m,t),g}^{\rm E} 
}} \hspace{-7.5mm}\underset{k \in \mathcal{K}}{\sum} \hspace{-0.7mm} \log_2 (\iota_k^{\rm N} \hspace{-0.8mm}+\hspace{-0.5mm} \iota_k^{\rm D} \hspace{-0.8mm}+\hspace{-0.5mm} \sigma_k^2) \hspace{-0.6mm}-\hspace{-1.1mm} \underset{k \in \mathcal{K}}{\sum} \hspace{-0.7mm} \log_2 (\iota_k^{\rm D} \hspace{-0.8mm}+\hspace{-0.5mm} \sigma_k^2) \nonumber \\
&\rm{s.t.}\ ({\rm C1}), ({\rm C2}),  (\widehat{{\rm C6a}}), (\widehat{{\rm C6b}}), (\widehat{{\rm C7a}}), (\widehat{{\rm C7b}}), (\widehat{{\rm C8a}}),\nonumber \\
&(\widehat{{\rm C8b}}), (\widehat{{\rm C9a}}), (\widehat{{\rm C9b}}), (\widehat{{\rm C10a}}), (\widehat{{\rm C10b}}), \nonumber \\
&(\hspace{-0.4mm}\widehat{{\rm C11}}\hspace{-0.4mm})\hspace{-1mm}: \hspace{-0.7mm} \theta_{l(n,m),k} \hspace{-0.8mm}=\hspace{-0.8mm} \tfrac{2\pi \sqrt{\left(x_{n,m}^{\rm Pin} \hspace{-0.4mm}- x_k\right)^2 \hspace{-0.5mm}+ \hat{\beta}_{n,k}}}{\lambda} + \hspace{-0.8mm}\tfrac{2 \pi x_{n,m}^{\rm Pin}}{\hat{\lambda}}\hspace{-0.4mm}, \forall \hspace{-0.4mm}n, \hspace{-0.4mm}m, \hspace{-0.4mm}k, \nonumber \\
&(\hspace{-0.4mm}\widehat{{\rm C12}}\hspace{-0.4mm})\hspace{-1mm}:\hspace{-0.7mm}
[\hspace{-0.3mm}\textbf{A}_k\hspace{-0.3mm}]_{\hspace{-0.3mm}l\hspace{-0.4mm}(\hspace{-0.4mm}n,m\hspace{-0.4mm}), l(\hspace{-0.4mm}n'\hspace{-0.4mm},m'\hspace{-0.4mm})} \hspace{-1mm}=\hspace{-1mm} e^{\hspace{-1mm}-\hspace{-0.3mm}j\hspace{-0.3mm}\left(\hspace{-0.6mm} \theta_{\hspace{-0.3mm}l\hspace{-0.3mm}(\hspace{-0.3mm}n,m\hspace{-0.3mm})\hspace{-0.3mm},\hspace{-0.3mm}k} -\theta_{\hspace{-0.3mm}l\hspace{-0.3mm}(\hspace{-0.3mm}n'\hspace{-0.3mm},m'\hspace{-0.3mm}),\hspace{-0.1mm}k\hspace{-0.3mm}} \hspace{-0.3mm} \right)}\hspace{-0.8mm}, \forall \hspace{-0.15mm}n, \hspace{-0.15mm}m\hspace{-0.25mm}, \hspace{-0.15mm}n'\hspace{-0.95mm}, \hspace{-0.15mm}m'\hspace{-0.95mm}, \hspace{-0.15mm}k, \nonumber \\
&(\hspace{-0.4mm}\widehat{{\rm C13}}\hspace{-0.4mm})\hspace{-1mm}: \hspace{-0.7mm}
\theta_{l(n,m,t),g}^{\rm E} \hspace{-1.2mm}=\hspace{-0.8mm} \tfrac{2\pi \hspace{-0.4mm} \sqrt{\hspace{-0.4mm}\left(\hspace{-0.4mm}x_{n,m}^{\rm Pin} \hspace{-0.7mm}- x_{g,t}^{\rm E}\hspace{-0.4mm}\right)^2 \hspace{-0.8mm}+ \hat{\beta}_{n,g, t}^{\rm E}}}{\lambda}   + \hspace{-0.8mm}\tfrac{2 \pi x_{n,m}^{\rm Pin}}{\hat{\lambda}}\hspace{-0.4mm}, \hspace{-0.6mm} \forall \hspace{-0.1mm}n, \hspace{-0.4mm}m, \hspace{-0.4mm}t, \hspace{-0.4mm}g\hspace{-0.2mm}, \nonumber \\
&(\hspace{-0.4mm}\widehat{{\rm C14}}\hspace{-0.4mm})\hspace{-1mm}: \hspace{-0.7mm}
[\hspace{-0.3mm}\textbf{A}_g^{\rm E}\hspace{-0.3mm}]_{\hspace{-0.1mm}l\hspace{-0.15mm}(\hspace{-0.14mm}n,m\hspace{-0.12mm},t\hspace{-0.14mm}), l(\hspace{-0.14mm}n'\hspace{-0.14mm},m'\hspace{-0.18mm},t'\hspace{-0.4mm})} \hspace{-11mm} \nonumber \\
& \hspace{15mm}= e^{\hspace{-1mm}-\hspace{-0.3mm}j\hspace{-0.3mm}\left(\hspace{-0.6mm} \theta^{\rm E}_{\hspace{-0.3mm}l\hspace{-0.3mm}(\hspace{-0.3mm}n,m\hspace{-0.3mm},t\hspace{-0.3mm})\hspace{-0.3mm},\hspace{-0.3mm}g} -\theta^{\rm E}_{\hspace{-0.3mm}l\hspace{-0.3mm}(\hspace{-0.3mm}n'\hspace{-0.3mm},m'\hspace{-0.3mm}, t'\hspace{-0.3mm}),\hspace{-0.1mm}g\hspace{-0.3mm}} \hspace{-0.3mm} \right)}\hspace{-0.8mm}, \forall \hspace{-0.15mm}n, \hspace{-0.15mm}m, \hspace{-0.15mm}t\hspace{-0.25mm}, \hspace{-0.15mm}n'\hspace{-0.95mm}, \hspace{-0.15mm}m'\hspace{-0.95mm}, \hspace{-0.15mm}t'\hspace{-0.95mm}, \hspace{-0.15mm}g.
\end{align}

Noting the multiple nonconvex equality constraints in~\eqref{eq: before_Pen_formu2_firTran}, we handle them via the procedure in~\eqref{eq: Frobenius}: $(\widehat{\rm C11a})$ and $(\widehat{\rm C11b})$ replace the equality in $(\widehat{\rm C11})$ with “$\leq$” and “$\geq$,” respectively, and $(\widehat{\rm C13a})$ and $(\widehat{\rm C13b})$ do the same for $(\widehat{\rm C13})$.
Moreover, since $\operatorname{Rank}(\textbf{A}_k)=1$, $\textbf{A}_k\succeq\bm{0}$, $\forall k$, and $\operatorname{Rank}(\textbf{A}_g^{\rm E})=1$, $\textbf{A}_g^{\rm E}\succeq\bm{0}$, $\forall g$, matrices $\textbf{A}_k$ and $\textbf{A}_g^{\rm E}$ are fully characterized by any nonzero column. Without loss of generality, we adopt their first columns to simplify notation~\cite{conference}. By trigonometric identities, $(\widehat{\rm C12})$ can be equivalently transformed into:
\begin{align}\label{eq: C12}
(\widehat{{\rm C12a}})\hspace{-0.8mm}:& \Re\left([\textbf{A}_k]_{i,1}\right) = \cos\left(\hat{\theta}_{i,k}\right), \forall i \in \{2, ..., MN\}, \nonumber \\
(\widehat{{\rm C12b}})\hspace{-0.8mm}:& \Im\left([\textbf{A}_k]_{i,1}\right) = -\sin\left(\hat{\theta}_{i,k}\right), \forall i \in \{2, ..., MN\}, 
\end{align}
where $\hat{\theta}_{i,k} = \theta_{i,k} - \theta_{1,k}$ and index $i$ maps to the pair $(n,m)$ through $i = (n-1)M + m$, and $\hat{\theta}^{\rm E}_{\bar{i},g} = \theta_{\bar{i},g}^{\rm E} - \theta_{1,g}^{\rm E}$ with $\bar{i} = (t-1)MN + (n-1)M +m$. Note that $(\widehat{\rm C14a})$ and $(\widehat{\rm C14b})$ are analogous to~\eqref{eq: C12}, thus their details are omitted.

Currently, nonconvexity arises from $(\widehat{\rm C9a})$, $(\widehat{\rm C10a})$, $(\widehat{\rm C11a})$, $(\widehat{\rm C13a})$, and the objective due to their DC form. More challengingly, $(\widehat{\rm C12a})$, $(\widehat{\rm C12b})$, $(\widehat{\rm C14a})$, and $(\widehat{\rm C14b})$ suffer from periodicity. We adopt Proposition~3 to handle the objective’s nonconvex terms and Proposition~5 to handle the rank-one DC constraints, deriving convex subsets with $(\widehat{\overline{\rm C9a}})\Rightarrow(\widehat{\rm C9a})$ and $(\widehat{\overline{\rm C10a}})\Rightarrow(\widehat{\rm C10a})$. Moreover, the convex subset of $(\widehat{\rm C11a})$ obtained via first-order Taylor expansion is:
\begin{align}\label{eq: C11a}
(\widehat{\overline{{\rm C11a}}})\hspace{-1.3mm}&: \hspace{-0.5mm}  \theta_{\hspace{-0.3mm}l(\hspace{-0.5mm}n\hspace{-0.3mm},m\hspace{-0.3mm})\hspace{-0.3mm},k} \hspace{-1mm} - \hspace{-1.5mm} \left(\hspace{-1.2mm} \tfrac{2\pi(x_{n,m}^{{\rm Pin}(i_3)} \hspace{-0.8mm}- x_k)}{\lambda\sqrt{\hspace{-0.7mm}\left(\hspace{-0.7mm}x_{n,m}^{{\rm Pin}(i_3)} \hspace{-0.7mm}- x_k\hspace{-0.7mm}\right)^2 \hspace{-1mm}+ \hat{\beta}_{n,k}}} \hspace{-0.8mm} + \hspace{-0.8mm}\tfrac{2 \pi}{\hat{\lambda}} \hspace{-1.1mm}\right) \hspace{-1.5mm} \left(\hspace{-0.7mm}x_{n,m}^{\rm Pin} \hspace{-0.8mm} - \hspace{-0.8mm} x_{n,m}^{{\rm Pin}(i_3)}\hspace{-0.7mm}\right)\hspace{-0.4mm} \nonumber \\ 
& \hspace{-3mm}- \hspace{-0.3mm} \tfrac{2\pi \sqrt{(x_{n,m}^{{\rm Pin}(i_3)} \hspace{-0.8mm}- x_k)^2 + \hat{\beta}_{n,k}}}{\lambda}  \hspace{-0.8mm}- \hspace{-0.8mm} \tfrac{2 \pi x_{n,m}^{{\rm Pin}(i_3)}}{\hat{\lambda}} \hspace{-0.8mm}\leq 0, \forall n,m,k,
\end{align}
where details of $(\widehat{\overline{\rm C13a}})$ are omitted as it is analogous to $(\widehat{\overline{\rm C11a}})$, $(\widehat{\overline{{\rm C11a}}}) \Rightarrow (\widehat{{\rm C11a}})$ and $(\widehat{\overline{{\rm C13a}}}) \Rightarrow (\widehat{{\rm C13a}})$. The terms with ``$(i_3)$'' are the solutions of the $i_3$-th iteration for the respective variables.
To further address the intrinsic nonconvexity, we introduce a penalty-based MM algorithm~\cite{mao2023amplitude,yu2021irs}.
We let  {\small $\Gamma\hspace{-0.8mm}\left(\hspace{-0.6mm}[\textbf{A}_k]_{\hspace{-0.3mm}i,\hspace{-0.3mm}1}\hspace{-0.3mm}, \hat{\theta}_{\hspace{-0.3mm}i\hspace{-0.3mm},k}\hspace{-0.85mm}\right) \hspace{-0.85mm}=\hspace{-0.6mm} \left|\hspace{-0.3mm}\Re\hspace{-0.6mm}\left(\hspace{-0.2mm}[\textbf{A}_k]_{i,1}\hspace{-0.6mm}\right) \hspace{-0.6mm}-\hspace{-0.6mm} \cos\hspace{-0.6mm}\left(\hspace{-0.6mm}\hat{\theta}_{i,k}\hspace{-0.6mm}\right) \hspace{-0.6mm}\right|^2 \hspace{-1mm}+\hspace{-0.8mm} \left|\hspace{-0.3mm}\Im\hspace{-0.6mm}\left(\hspace{-0.2mm}[\hspace{-0.2mm}\textbf{A}_k]_{i,1}\hspace{-0.4mm}\right)\hspace{-0.6mm} + \hspace{-0.6mm}\sin\hspace{-0.6mm}\left(\hspace{-0.6mm}\hat{\theta}_{i,k}\hspace{-0.6mm}\right)\hspace{-0.5mm} \right|^2$} and
{\small$\Gamma\hspace{-0.8mm}\left(\hspace{-0.6mm}[\textbf{A}_g^{\rm E}]_{\hspace{-0.3mm}\bar{i},\hspace{-0.3mm}1}\hspace{-0.3mm}, \hat{\theta}^{\rm E}_{\hspace{-0.3mm}\bar{i}\hspace{-0.3mm},g}\hspace{-0.85mm}\right) \hspace{-0.85mm}=\hspace{-0.6mm} \left|\hspace{-0.3mm}\Re\hspace{-0.6mm}\left(\hspace{-0.2mm}[\textbf{A}_g^{\rm E}]_{\bar{i},1}\hspace{-0.6mm}\right) \hspace{-0.6mm}-\hspace{-0.6mm} \cos\hspace{-0.6mm}\left(\hspace{-0.6mm}\hat{\theta}_{\bar{i},g}^{\rm E}\hspace{-0.6mm}\right) \hspace{-0.6mm}\right|^2 \hspace{-1mm}+\hspace{-0.8mm} \left|\hspace{-0.3mm}\Im\hspace{-0.6mm}\left(\hspace{-0.2mm}[\hspace{-0.2mm}\textbf{A}_g^{\rm E}]_{\bar{i},1}\hspace{-0.4mm}\right)\hspace{-0.6mm} + \hspace{-0.6mm}\sin\hspace{-0.6mm}\left(\hspace{-0.6mm}\hat{\theta}_{\bar{i},g}^{\rm E}\hspace{-0.6mm}\right)\hspace{-0.5mm} \right|^2$}.
Variables $\rho_1, \rho_2 > 0$ are penalty coefficients imposing costs on violations of $(\widehat{{\rm C12a}})$, $(\widehat{{\rm C12b}})$, and $(\widehat{{\rm C14a}})$, $(\widehat{{\rm C14b}})$. Specifically, the penalty formulation is equivalent to the original problem, when $\rho_1$ and $\rho_2$ are sufficiently large, i.e., $\rho_1, \rho_2 \rightarrow \infty$~\cite{yu2021irs, mao2023amplitude}. 
We then apply MM with a Lipschitz-gradient surrogate to establish a global upper bound~\cite{mairal2015incremental} for {\small$\Gamma\left([\textbf{A}_k]_{i,1}, \hat{\theta}_{i,k}\right)$}: 
\begin{equation}\label{eq: Surrog}
  \scalebox{0.8}{$\displaystyle
    \begin{aligned}
      \overline{\Gamma}^{(\hspace{-0.3mm}i_3\hspace{-0.3mm})}_{i,k} \hspace{-1mm}
      &\hspace{-0.5mm}= \hspace{-0.9mm} 2\hspace{-0.9mm}\left(\hspace{-0.5mm}\hat{\Re}_{i,k}^{(\hspace{-0.3mm}i_3\hspace{-0.3mm})} \hspace{-0.5mm}\sin \hat{\theta}_{i,k}^{(\hspace{-0.3mm}i_3\hspace{-0.3mm})} \hspace{-0.8mm}+\hspace{-0.8mm} \hat{\Im}_{i,k}^{(\hspace{-0.3mm}i_3\hspace{-0.3mm})}\hspace{-0.7mm}\cos \hat{\theta}_{i,k}^{(\hspace{-0.3mm}i_3\hspace{-0.3mm})}\hspace{-0.8mm}\right)\hspace{-0.8mm}\triangle \hat{\theta}_{i,k} \hspace{-0.6mm} + \hspace{-0.6mm}2\hat{\Re}_{i,k}^{(\hspace{-0.3mm}i_3\hspace{-0.3mm})}\hspace{-0.5mm}\triangle \Re_{i,k}\hspace{-0.9mm}+\hspace{-1mm} 2\hat{\Im}_{i,k}^{(\hspace{-0.3mm}i_3\hspace{-0.3mm})}\hspace{-0.5mm}\triangle \hspace{-0.3mm} \Im_{i,k} \\
      &+ \frac{L_\text{AR}}{2}(\triangle \Re_{i,k})^2 \hspace{-0.3mm}+\hspace{-0.3mm} \frac{L_\text{AI}}{2}(\triangle \Im_{i,k})^2 \hspace{-0.3mm}+\hspace{-0.3mm} \frac{L_{\rm TH}}{2} (\triangle \hat{\theta}_{i,k})^2 \hspace{-0.3mm}+\hspace{-0.3mm} \Phi^{(i_3)} , \forall i, k,
    \end{aligned}
  $}
\end{equation}
where {\small$\hat{\Re}_{i,k}^{(\hspace{-0.3mm}i_3\hspace{-0.3mm})} \hspace{-0.5mm}= \hspace{-0.5mm}\Re([\textbf{A}_k]_{i,1}^{(\hspace{-0.3mm}i_3\hspace{-0.3mm})}) \hspace{-0.5mm}-\hspace{-0.5mm} \cos\hspace{-0.5mm} \hat{\theta}_{i,k}^{(\hspace{-0.3mm}i_3\hspace{-0.3mm})}$, $\hat{\Im}_{i,k}^{(\hspace{-0.3mm}i_3\hspace{-0.3mm})} \hspace{-0.5mm}=\hspace{-0.5mm} \Im([\textbf{A}_k]_{i,1}^{(\hspace{-0.3mm}i_3\hspace{-0.3mm})}) \hspace{-0.5mm}+\hspace{-0.5mm} \sin\hspace{-0.5mm} \hat{\theta}_{i,k}^{(\hspace{-0.3mm}i_3\hspace{-0.3mm})}$, $\triangle \hat{\theta}_{i,k} \hspace{-0.5mm}=\hspace{-0.5mm} \hat{\theta}_{i,k} \hspace{-0.5mm}- \hat{\theta}_{i,k}^{(\hspace{-0.3mm}i_3\hspace{-0.3mm})}$, $\triangle \Re_{i,k}=\Re([\textbf{A}_k]_{i,1}) - \Re([\textbf{A}_k]_{i,1}^{(i_3)}) $, $\triangle \Im_{i,k} = \Im([\textbf{A}_k]_{i,1}) - \Im([\textbf{A}_k]_{i,1}^{(i_3)})$}. 
Note that the global upper bound for {\small$\Gamma \left(\hspace{-0.6mm}[\textbf{A}_g^{\rm E}]_{\hspace{-0.3mm}\bar{i},\hspace{-0.3mm}1}\hspace{-0.3mm}, \hat{\theta}^{\rm E}_{\hspace{-0.3mm}\bar{i}\hspace{-0.3mm},g}\hspace{-0.85mm}\right)$} is denoted by $\overline{\Gamma}^{(\hspace{-0.3mm}i_3\hspace{-0.3mm})}_{\bar{i},g}$ and follows the same structure as~\eqref{eq: Surrog}; details are omitted due to page limits.
The Lipschitz constants are {\small$L_{\rm AR}=L_{\rm AI}=2, L_{\rm TH}=4$}, obtained from the maximum curvature~\cite{mairal2015incremental}.
Accordingly, the $(i_3+1)$-th iteration of the penalty-based MM method for stage 2 of the PA positioning problem can be expressed as:
\begin{align}
\label{eq: penalty_final}
&\operatorname*{maximize}_{\substack{
   \textbf{x}_n^{\rm Pin},\; \textbf{A}_k, \; \iota, \; \delta, \; \textbf{A}_g^{\rm E},\\
   \hat{\theta}_{i,k}, \; \hat{\theta}^{\rm E}_{\bar{i},g}, \; \theta_{l(n,m),k}, \; \theta_{l(n,m,t),g}^{\rm E}
}} \hspace{-7mm}\underset{k \in \mathcal{K}}{\sum} \log_2 (\iota_k^{\rm N} \hspace{-0.8mm}+\hspace{-0.6mm} \iota_k^{\rm D} \hspace{-0.8mm}+\hspace{-0.6mm} \sigma_k^2)  \hspace{-0.8mm}+\hspace{-0.8mm} \bar{R}({\iota_k^{{\rm D}(i_3)}}) \nonumber \\
&\hspace{25mm}-\rho_1 \sum_{k=1}^{K}\sum_{i=2}^{MN} \hat{\Gamma}^{(\hspace{-0.3mm}i_3\hspace{-0.3mm})}_{i,k} - \rho_2 \sum_{g=1}^{G}\sum_{\bar{i}=2}^{MNT} \overline{\Gamma}^{(\hspace{-0.3mm}i_3\hspace{-0.3mm})}_{\bar{i},g} \nonumber \\
\rm{s.t.}  &({\rm C1})\hspace{-0.3mm}, ({\rm C2})\hspace{-0.3mm},  (\widehat{{\rm C6a}}), (\widehat{{\rm C6b}}), (\widehat{{\rm C7a}}), (\widehat{{\rm C7b}}), (\widehat{{\rm C8a}}), \nonumber \\
& (\widehat{{\rm C8b}}), (\widehat{\overline{{\rm C9a}}}), (\widehat{{\rm C9b}}), (\widehat{\overline{{\rm C10a}}}), (\widehat{{\rm C10b}}), (\widehat{\overline{{\rm C11a}}}),  \nonumber \\ & (\widehat{{\rm C11b}}), (\widehat{\overline{{\rm C13a}}}), \text{and} \ (\widehat{{\rm C13b}}),
\end{align}
where the resulting problem in~\eqref{eq: penalty_final} is convex and can be directly solved using standard convex optimization solvers. This yields a suboptimal solution to the problem.

At each outer BCD iteration, one block of optimization variables is optimized until convergence to a predefined tolerance, while other blocks remain fixed. The resulting procedure is summarized in \textbf{Algorithm 1}, which is guaranteed to converge to a suboptimal solution of~\eqref{eq:formulation} and its overall computational complexity is given by~\cite{hu2021robust, yu2020robust} 
\begin{align}
&\mathcal{O} \Big( I_{\rm BCD} \Big[ I_1\left(K(MN+N+1)^3+KG(MN+T)^3\right) \nonumber \\
&\ + \hspace{-0.6mm} I_2\hspace{-0.6mm}\left(K(MN \hspace{-0.6mm}+\hspace{-0.6mm} 1)^3 \hspace{-0.6mm}+\hspace{-0.6mm} G(MNT\hspace{-0.6mm}+\hspace{-0.6mm}1)^3\hspace{-0.6mm}+\hspace{-0.6mm}KG(MN\hspace{-0.6mm}+\hspace{-0.6mm}T)^3 \right) \nonumber \\
&\ +\hspace{-0.6mm} I_3\hspace{-0.6mm}\left(K(MN\hspace{-0.5mm}+\hspace{-0.5mm}1\hspace{-0.5mm})^3 \hspace{-0.6mm}+\hspace{-0.6mm} G(MNT)^3 \hspace{-0.8mm}+\hspace{-0.6mm} KG(MNT\hspace{-0.6mm}+\hspace{-0.6mm}T)^3 \right)\hspace{-0.5mm} \Big] \hspace{-0.5mm}\Big)\hspace{-0.5mm}.
\end{align}
Note that $I_{\rm BCD}$, $I_1$, $I_2$, and $I_3$ denote the numbers of outer BCD and MM iterations required for solving subproblem 1 in (41), the coarse PA-positioning problem in (57), and the phase-refinement problem in (65), respectively. This expression reveals that the computational complexity scales polynomially with respect to the key system dimension, which is suitable for practical implementation~\cite{yu2021irs}.

\section{Simulation Results}
\begin{figure}[t]
\begin{algorithm}[H] 
\caption{Overall BCD Algorithm}
\begin{algorithmic}[1]
\State \textbf{Initialize:} Set the maximum number of iterations $\tau_{\max}$ and the convergence tolerance $\tau_{\rm tol} \rightarrow 0$. Initialize the iteration index $\tau = 0$ and the optimization variables $\textbf{W}_k^{(\tau)}, \forall k$, $\textbf{V}^{(\tau)}$, $\textbf{P}^{(\tau)}$, and $\textbf{x}^{{\rm Pin}(\tau)}$.
\State \textbf{repeat} \hfill \texttt{$\triangleright$ Main Loop: BCD}
    \State \hspace{1em} \textbf{Subproblem 1:} For fixed $\textbf{x}^{{\rm Pin}(\tau)}$, solve the problem in~\eqref{eq: formu_subFin} to obtain $\{\textbf{W}^{(\tau+1)}_k, \textbf{V}^{(\tau+1)}, \textbf{P}^{(\tau+1)}\}$.
    \State \hspace{1em} \textbf{Subproblem 2:} With $\{\textbf{W}_k^{(\tau+1)}$,$ \textbf{V}^{(\tau+1)}, \textbf{P}^{(\tau+1)}\}$, solve the following two stages:
    \State \hspace{2em} \textit{Stage 1:} Solve the problem in~\eqref{eq: formu2_Fin} to obtain a coarse solution $\textbf{x}_{{\rm c}}^{{\rm Pin}(\tau+1)}$.
    \State \hspace{2em} \textit{Stage 2:} Starting from $\textbf{x}_{{\rm c}}^{{\rm Pin}(\tau+1)}$, solve the phase-refinement problem in~\eqref{eq: penalty_final} to obtain $\textbf{x}^{{\rm Pin}(\tau+1)}$.
    \State \hspace{1em} Update $\tau \leftarrow \tau + 1$.
\State \textbf{until} $\tau = \tau_{\max}$ or performance gap $\leq \tau_{\rm tol}$.
\State \textbf{return} $\{\textbf{W}^*_k, \textbf{V}^*\hspace{-0.3mm}, \textbf{P}^*\hspace{-0.3mm}, \textbf{x}^{{\rm Pin}*}\} 
= \{\textbf{W}^{(\tau)}_k\hspace{-0.5mm}, \textbf{V}^{(\tau)}\hspace{-0.5mm}, \textbf{P}^{(\tau)}\hspace{-0.5mm}, \textbf{x}^{{\rm Pin}(\tau)}\}$.
\end{algorithmic}
\end{algorithm}
\end{figure}
This section evaluates the performance of the proposed multi-waveguide, multi-PA-assisted secure and robust system at the carrier frequency $f_c = 28$ GHz via numerical simulations. Unless otherwise stated, all results are averaged over multiple random realizations of user locations within a $15 \ \text{m} \times 15 \ \text{m}$ area, based on the configuration depicted in Fig.~1. Moreover, we set $\vartheta = 500$, $Q = 2$, $d = 5$ m, $N = 5$, $M = 2$, $G = 1$, $T =2$, $K = 2$, $P_{\rm max} = 20$ dBm,  $\wp_g^{\rm arc} = 1^\circ$, $\wp_g^{\rm E} = 1$ cm, and fix the threshold of achievable rate for EA to $R_{\rm th} = R^{\rm th}_{k,g} = 1$ bits/s/Hz, $\forall k,g$. The dielectric waveguides are fabricated from polytetrafluoroethylene (PTFE)~\cite{suzuki2022pinching} with parameters: $\eta_{\rm eff}=1.42$, $\epsilon_r=2.1$, and ${\rm tan}(\delta_e)=2\times10^{-4}$. The blockages are randomly generated in each realization: their length along the $y$-coordinate is fixed at $2.8$ m, while their $x$-coordinate and height are drawn randomly from $[3,5]$ m and $[5,8]$ m, respectively. To quantify the channel-estimation uncertainty, we define the maximum normalized estimation error of the legitimate channels as $\kappa_k = \wp_k / \left|\textbf{h}_k \right|_2$ and set $\kappa_k = \kappa, \forall k$, with $\kappa^2 = 0.1$. Simulation-specific parameters are detailed in the captions of the respective figures.

\begin{figure}[t] 
	\centering
	\subfigure[$\wp^{\rm E}_g = 0$ cm.]{
		\includegraphics[width=1.55in]{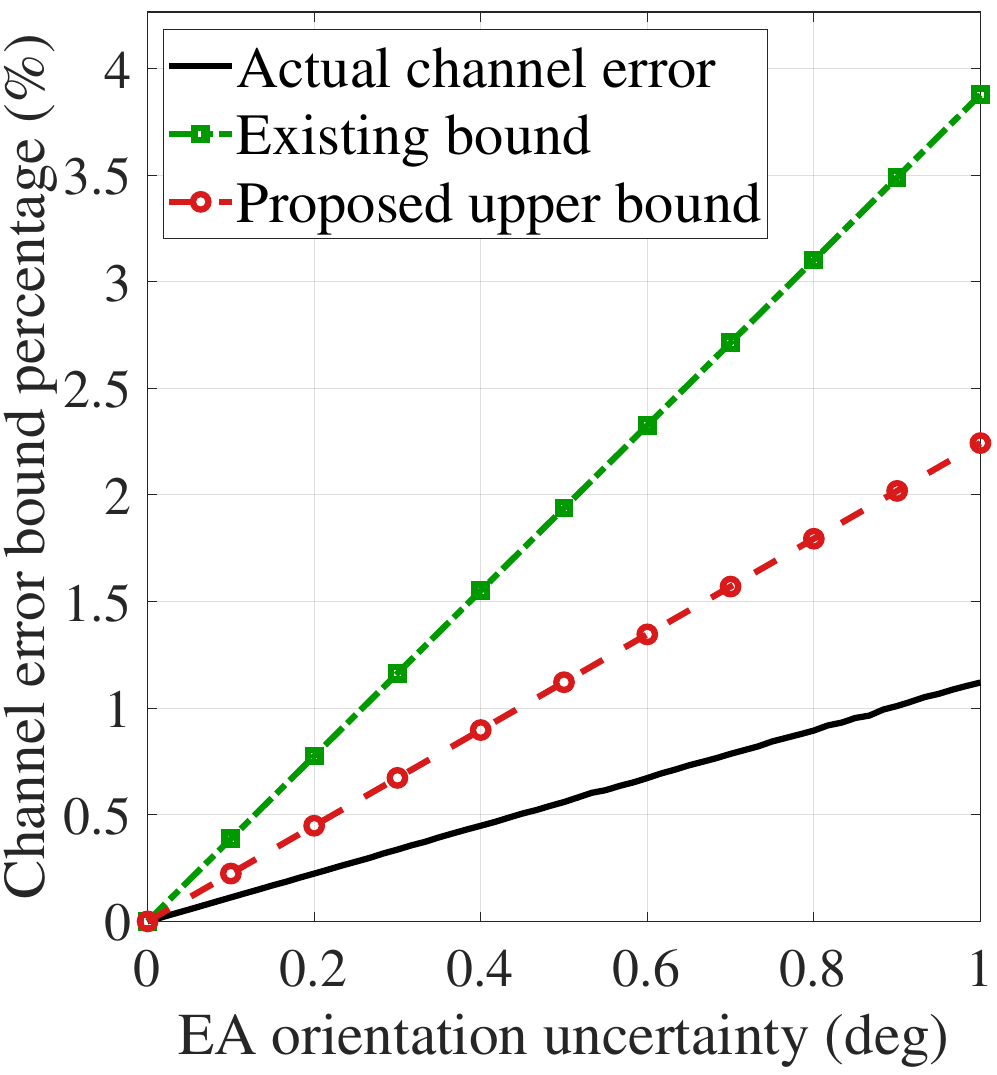}
	}\hspace{1mm}
	\subfigure[$\wp^{\rm arc}_g = 0^\circ$.]{
		\includegraphics[width=1.55in]{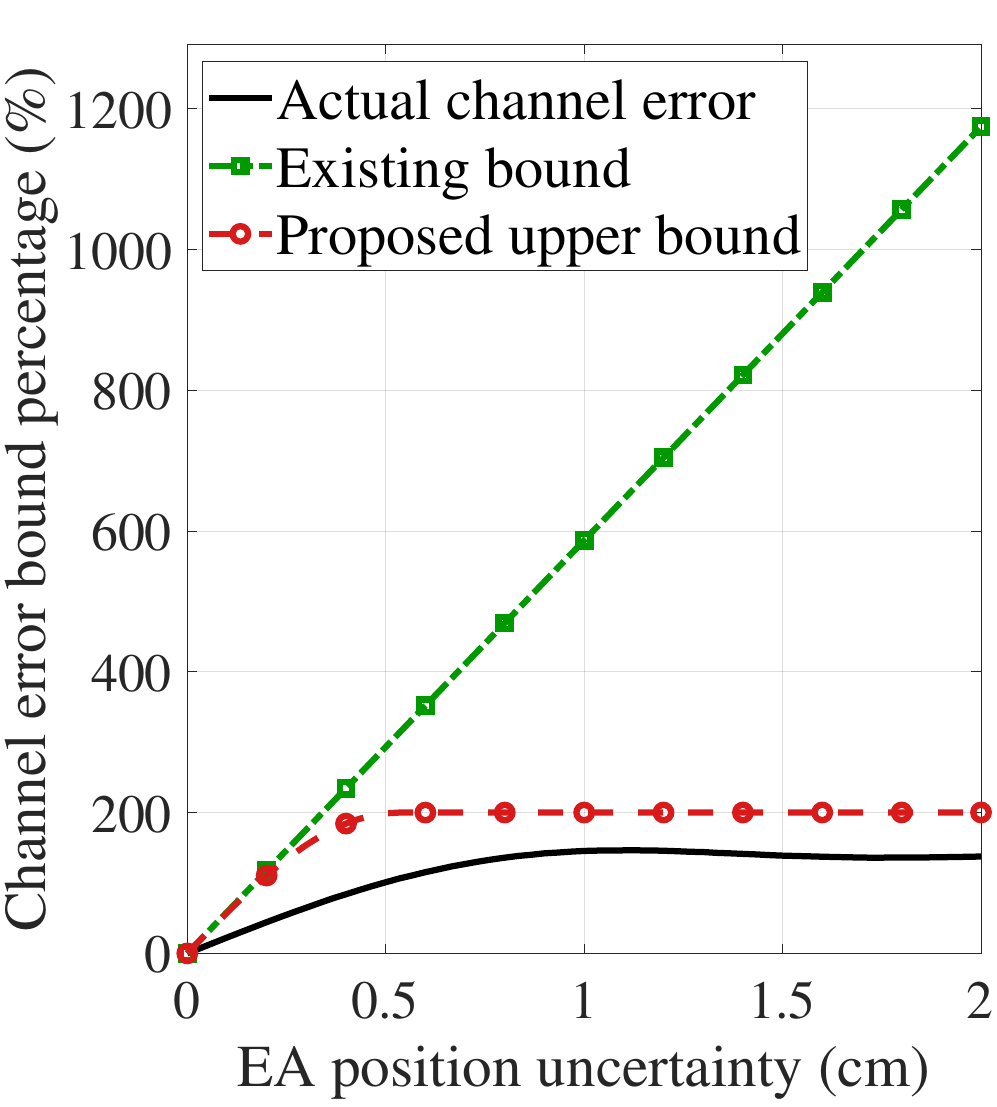}
	}
    \vspace{-0.5em}
	\caption{Proposed channel error bound when $N = 5$, $M = 2$, and $T = 2$.}
	\label{fig: CSI_bound}
    \vspace{-1em}
\end{figure}

To begin with, we evaluate the effectiveness of the proposed geometry-aware upper bound on channel errors derived in~\eqref{eq: CHannelBouns}, exploiting $10000$ random channel realizations drawn from the prescribed uncertainty regions. We compare the following three quantities: (i) the actual channel error computed via Monte Carlo using~\eqref{eq: profGamma}, (ii) the conventional near‑field linear error bound from~\cite{ren2025robust}, and (iii) the proposed upper bound obtained from~\eqref{eq: CHannelBouns}. The performance metric is the normalized channel‑error percentage, $\|\textbf{H}_{{\rm E}_g}-\hat{\textbf{H}}_{{\rm E}_g}\|/\|\hat{\textbf{H}}_{{\rm E}_g}\|$. Fig.~\ref{fig: CSI_bound} plots the normalized error versus orientation error and location error in subplots (a) and (b), respectively. Owing to the distributed PA locations, the existing linear bound~\cite{ren2025robust, xiu2024robust} is excessively loose. This can be attributed that phase contributions from different PAs may partially cancel under small positional or angular perturbations, producing a sublinear growth of CSI error. Consequently, as the EA moves farther away and path phases become largely uncorrelated with the nominal geometry, the error saturates, yielding a flattening of the curve at large position errors. These behaviors demonstrate that the EA CSI mismatch is highly nonlinear, validating the need for the proposed geometry-aware uncertainty bound.

For comparison, we evaluate the proposed scheme in the presence and absence of blockages, labeled as ``\textit{Pro. w blk}'' and ``\textit{Pro. w/o blk}'', respectively, against five benchmark schemes. Specifically, we consider: (i) ``\textit{Upper bound}'', which shares the same setting as the considered model except that both blockages and EAs are removed and the power attenuation along the waveguides is ignored, thus providing an optimistic performance upper bound; (ii) ``\textit{BM1 w blk}'', which adopts a fixed uniformly distributed beamforming strategy with half of the transmit power allocated to AN and the other half to the information signal, while the total power is equally distributed among all PAs, so that performance improvements only stem from optimizing the PA positions under realistic blockage conditions; (iii) ``\textit{BM1 w/o blk}'', which shares the same configuration as ``\textit{BM1 w blk}'' but assumes an ideal environment without blockages; (iv) ``\textit{BM2 w blk}'', which employs a conventional BS precoder design while all PAs are fixed at their feed points and blockage effects are taken into account; and (v) ``\textit{BM2 w/o blk}'', which is identical to ``\textit{BM2 w blk}'' except that it assumes no blockages. For completeness, we also implement a ``\textit{Naive}'' scheme that first optimizes the resource allocation while neglecting waveguide attenuation and blockage impacts, and then re-evaluates the resulting solution under the realistic attenuation and blockage conditions; if the obtained solution violates the power constraint (C3), we enforce $p_{n,m} = p_{n,m}^{\max}, \forall n,m$ and if it violates the information-leakage constraint (C6), the achievable sum rate is penalized to zero. Note that this naive scheme almost always yields zero performance due to violations of the power and information-leakage constraints; its results are omitted from the plots.

\begin{figure}[t]
\centerline{\includegraphics[width=3.4in]{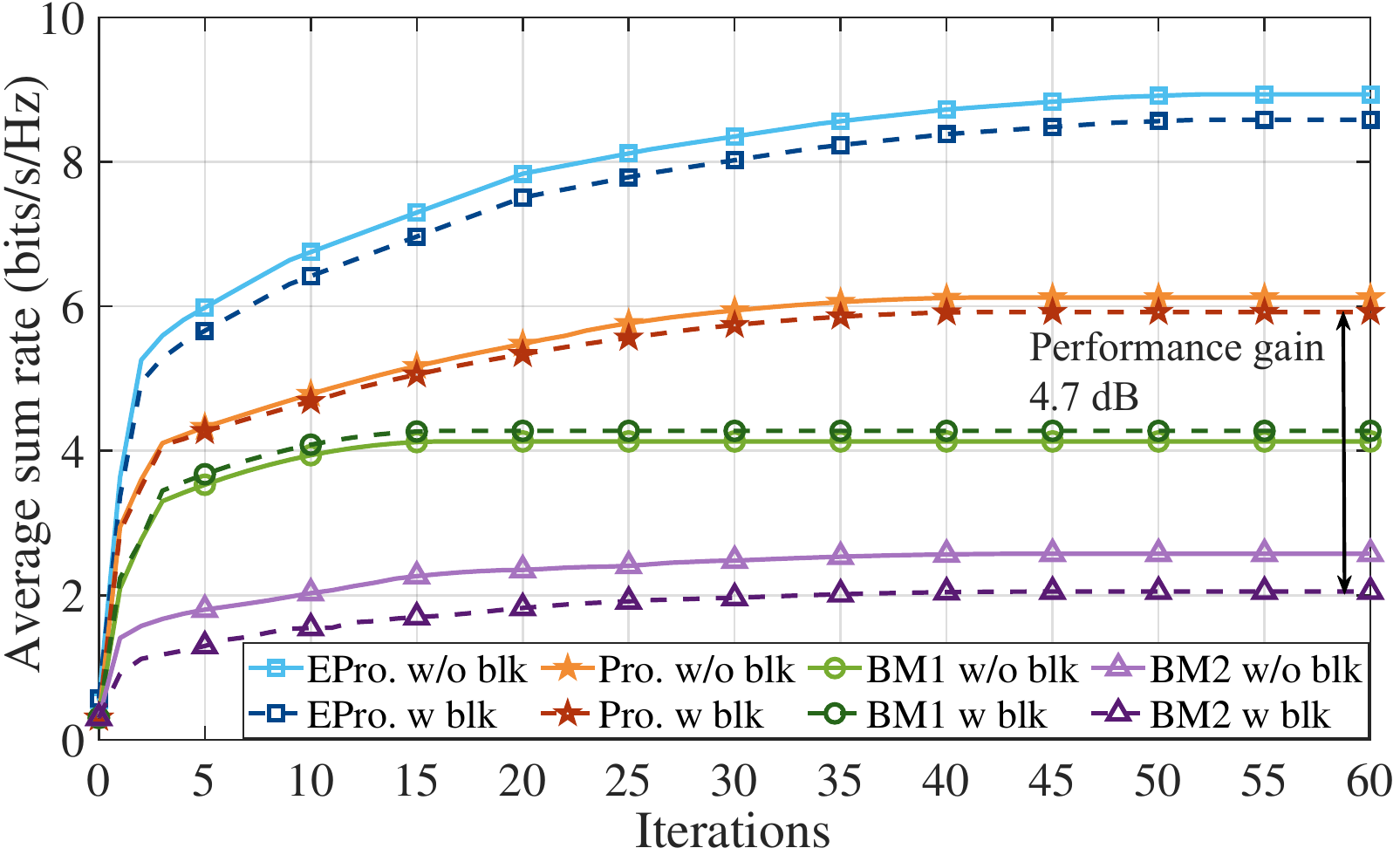}}
\vspace{-0.5em}
\caption{Convergence of the proposed algorithm against multiple benchmark schemes under different system configurations.}
\vspace{-1.5em}
\label{fig:convergence}
\end{figure}

Fig.~\ref{fig:convergence} shows the convergence behavior of the proposed schemes against benchmarks under both blockage-aware and blockage-free settings. Besides the default setup, we also consider an enlarged system with $N=10$, $M=3$, and $K=4$, denoted by ``\textit{EPro.}'' It is observed that although the proposed schemes require slightly more iterations, they converge stably within about $40$ iterations for the default setting and $45$ iterations for the enlarged system, which confirms the robustness and scalability of the proposed BCD-based design. In addition, the proposed schemes consistently outperform both benchmarks and achieve a notable gain of $4.7$ dB over the conventional fixed-position antenna scheme. Another important observation is that the performance gap between the blockage-aware and blockage-free cases is much smaller for the proposed schemes than for the fixed-antenna benchmark. More interestingly, for BM1, which only optimizes PA positioning without beamforming design, the blockage-aware case even slightly outperforms the blockage-free case. These results indicate that PA flexibility mitigates blockage-induced degradation through adaptive antenna repositioning, and further reveal that blockages can be exploited as natural shields to attenuate eavesdropping channels.

To evaluate the practical implementation cost, we further measure the wall-clock time on a platform with an Intel Core i7-14700 processor (2.10 GHz), 32 GB RAM, and CVX with the SDPT3 solver. Note that the enlarged setting $(N,M,K)=(10,3,4)$ substantially increases the size of the convex subproblems compared with the default setting $(5,2,2)$. In particular, the key antenna-related dimensions grow from $MN=10$ to $30$ and from $MNT=20$ to $60$, which enlarges the dominant lifted LMI blocks from $11$ and $21$ to $31$ and $61$, respectively. Since the complexity in (66) grows superlinearly with these effective matrix dimensions, a noticeable increase in runtime is expected. Indeed, the average CPU time per BCD iteration per realization increases from $745.5$ s to $5012.3$ s, corresponding to a normalized runtime increase of $6.72$. We emphasize that these runtime results are obtained from a general-purpose CVX/SDPT3 implementation and are intended to reflect computational scaling rather than optimized execution speed. In practice, the runtime could be further reduced by exploiting problem structure, developing customized solvers, or using dedicated hardware acceleration. Considering the substantial enlargement of the lifted variables and LMI constraints, this scaling remains acceptable for static or semi-static base-station deployments.

\begin{figure}[t]
    \centering
    \subfigure[Average achievable sum rate versus the total transmit power budget, $P_{\max}$, of different antenna configurations.]{
        \includegraphics[width=3.4in]{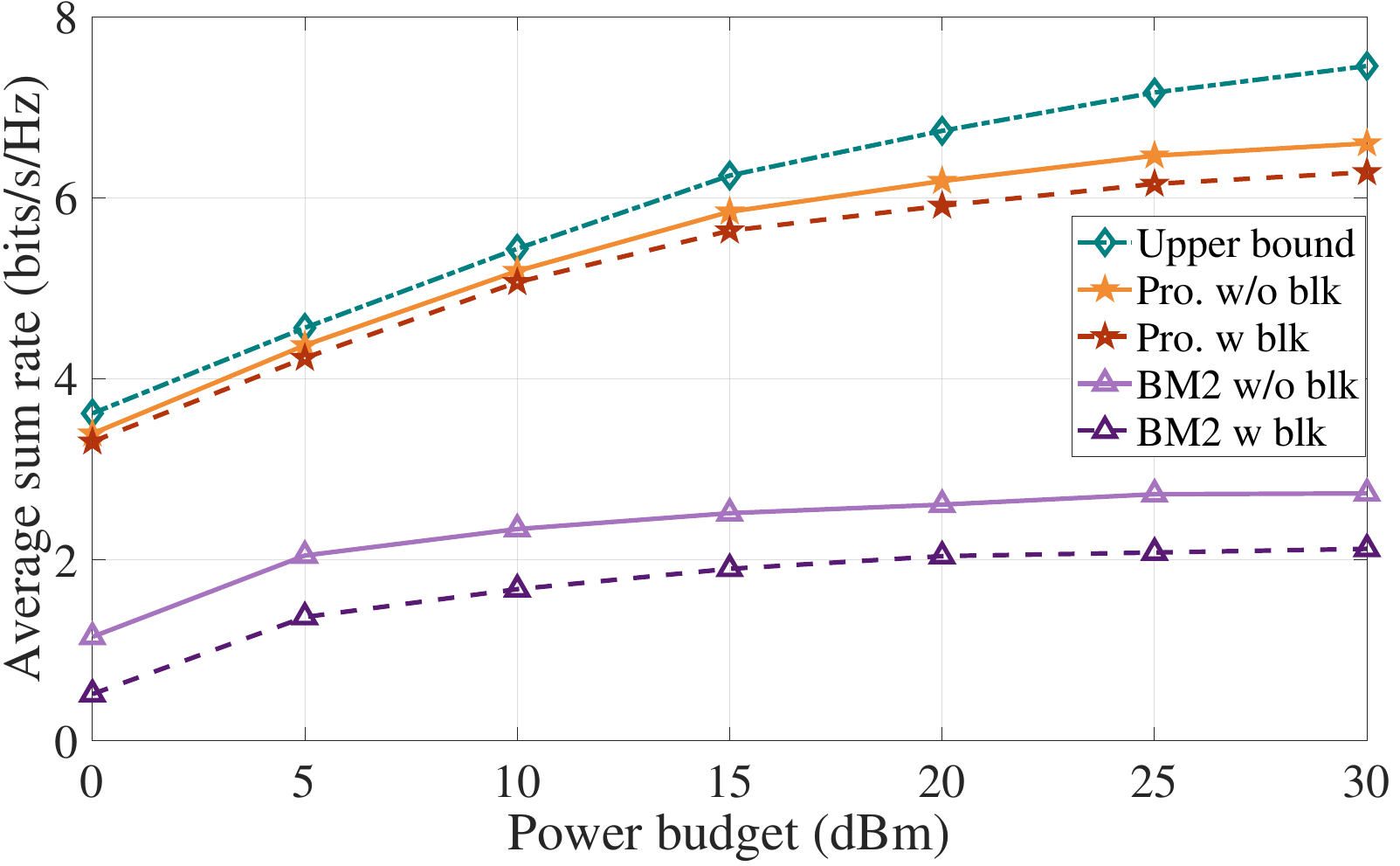}
    }
    \vspace{0.25em}
    \subfigure[Representative snapshot of the optimized PA locations and per-PA power allocations for $P_{\max} = 30$ dBm under blockage-aware and blockage-free designs.]{
        \includegraphics[width=3in]{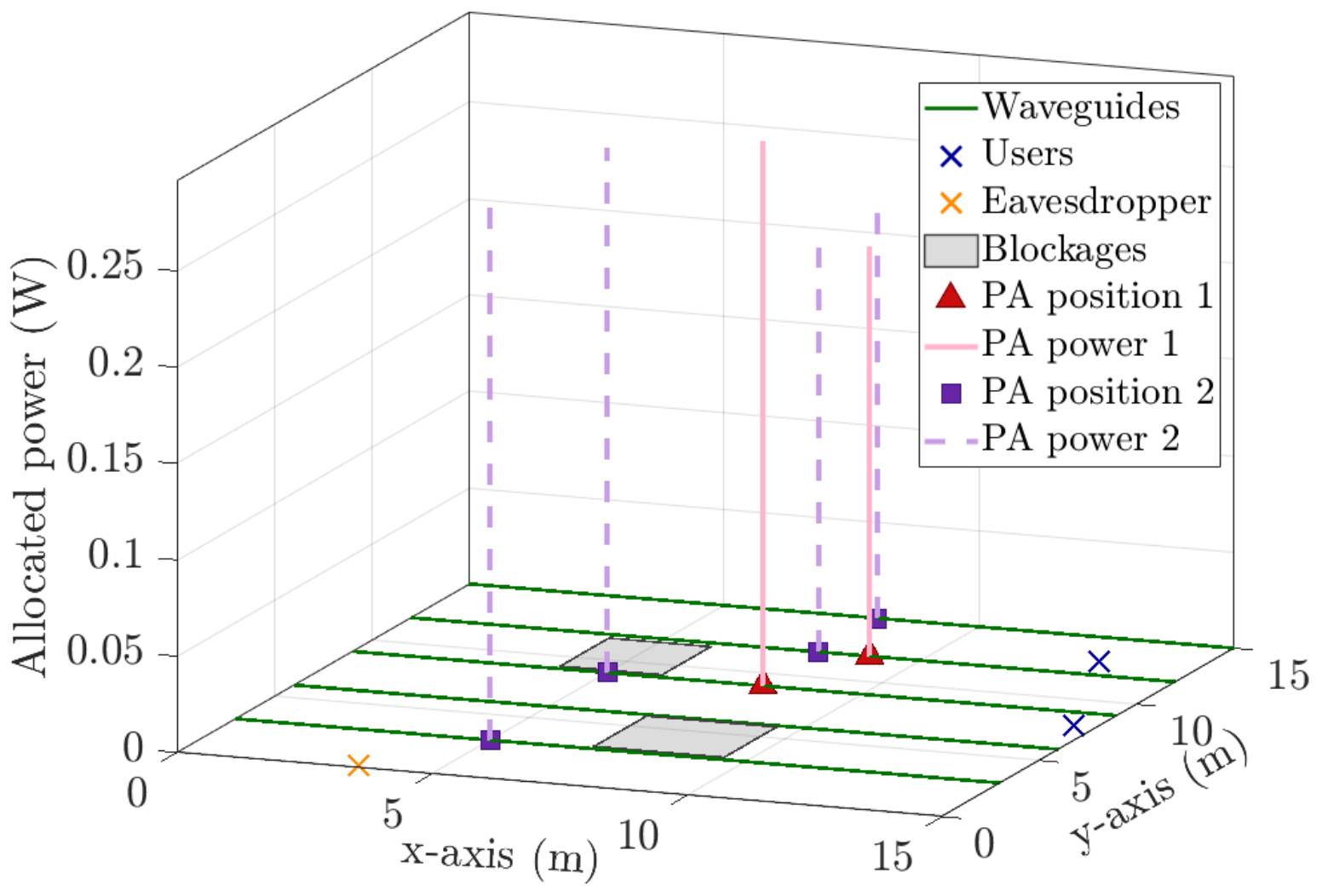}
    }
     \vspace{-0.5em}
    \caption{Average achievable sum rate versus the total transmit power budget of different antenna configurations with a snapshot example of optimized PA locations and power allocations.}
    \label{fig: powerInc}
    \vspace{-1em}
\end{figure}
We next examine the impact of the total power budget and the PA configuration, as illustrated in Fig.~\ref{fig: powerInc}(a). As $P_{\max}$ increases, all schemes exhibit a monotonic increase in the average sum rate, but with markedly different growth rates. The fixed-antenna baselines (BM2 with and without blockages) show only modest improvements, since their performance is fundamentally limited by the lack of spatial flexibility at the PA side. In contrast, both proposed designs, with and without blockages, can more effectively exploit the additional transmit power, which leads to a significantly larger rate increase as $P_{\max}$ grows. The upper-bound curve, obtained from an idealized system with no eavesdropper threat, waveguide attenuation, or blockage effects, achieves the highest rates. It can be observed that the sum rate gap between proposed schemes and the upper bound is slightly enlarged as $P_{\max}$ increases. This is because, although a larger total power budget provides more available resources, the presence of an eavesdropper requires allocating a portion of the power to artificial noise rather than entirely devoting it to information beamforming.
Notably, the gap between the two proposed curves (with and without blockages) remains relatively small over the entire considered power range. In contrast, the corresponding gap between the fixed-antenna baselines is much larger. This indicates that PA flexibility substantially mitigates the detrimental impact of blockages and enables the system to convert the available transmit power budget into sum-rate gains more efficiently. As the presence of blockages inevitably removes some PA-position candidates, the blockage-aware design still performs slightly below the ideal blockage-free case.

Fig.~\ref{fig: powerInc}(b) illustrates a representative snapshot of the optimized PA deployment and power allocation for $P_{\max}=30$ dBm, with two rectangular blockages whose lengths are $4.34$ m and $3.19$ m and whose heights are $7.75$ m and $5.36$ m, respectively. For clarity, only their ground projections are shown as grey rectangles. It can be noted that the active PAs are generally located near the centroid of the user cluster, balancing waveguide and wireless path losses. Specifically, among the $10$ available PAs, the blockage-aware design ``\textit{Pro. w blk}'', given by PA position~1 and PA power~1, activates two elements to serve the $K=2$ users, whereas the design optimized for an ideal blockage-free environment, denoted by ``\textit{Pro. w/o blk}'' with PA position~2 and PA power~2, activates four. Although the blockages constrain feasible PA locations for serving legitimate users, they can serve as physical shields that partially obstruct the LoS to EAs, allowing the blockage-aware design to concentrate more transmit power on the information-carrying beams and rely less on AN generation. 

\begin{figure}[t]
\centerline{\includegraphics[width=3.3in]{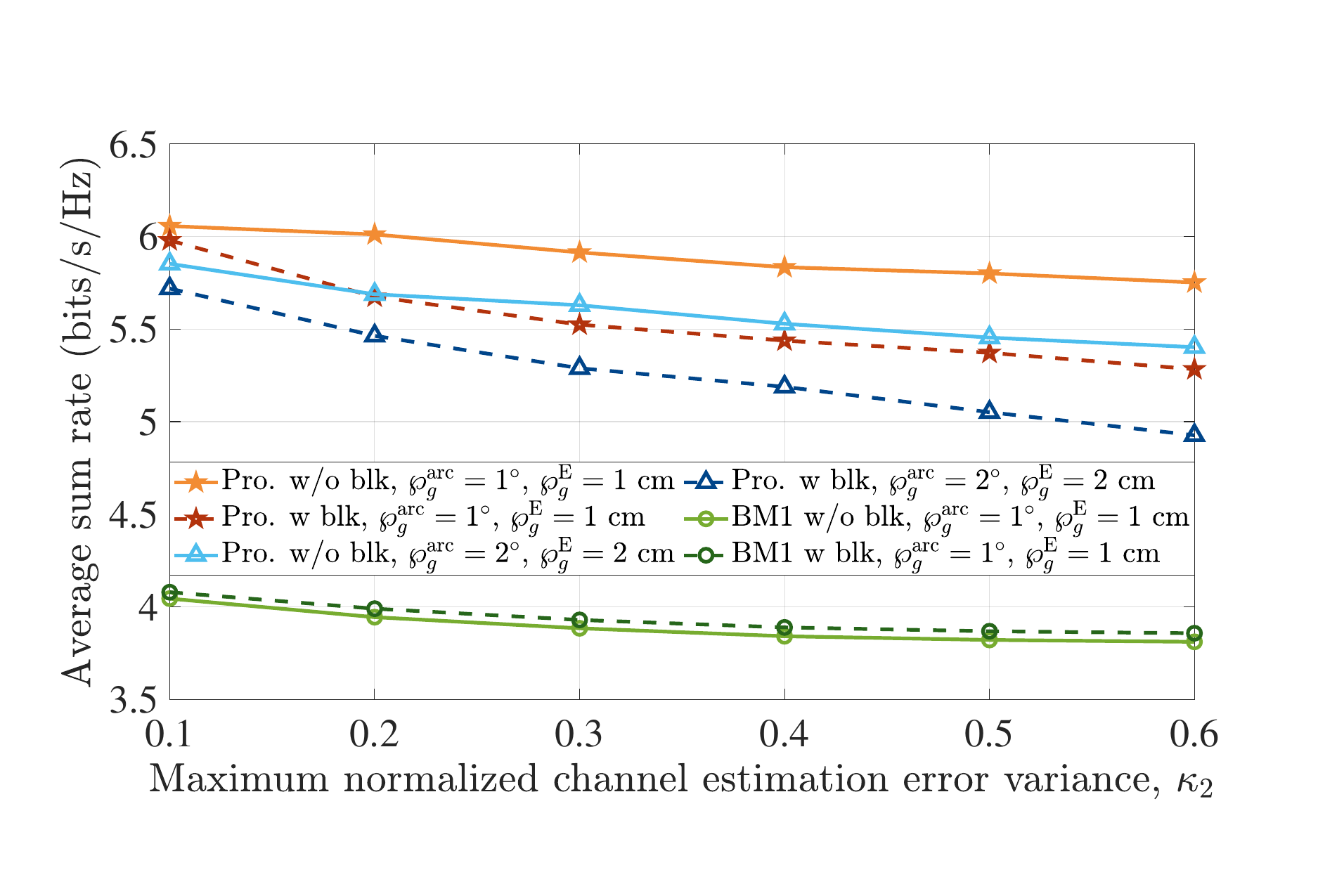}}
\caption{Average sum rate versus the maximum normalized
channel estimation error variance.}
\vspace{-1em}
\label{fig: bounds}
\end{figure}
Fig.~\ref{fig: bounds} shows the average system sum rate versus the maximum normalized users' channel estimation error variance~$\kappa^{2}$ for different configurations and EA uncertainty levels, comparing the proposed design with the two BM1 baselines. For all schemes, the average sum rate decreases as~$\kappa^{2}$ increases, since a larger estimation error variance yields less accurate CSI at the BS. As a result, the information beamformer and AN precoder become more conservative in forming energy-focusing beams, and the optimized PA positions cannot ensure precise phase alignment with the legitimate users, thereby jointly reducing the received signal power and increasing residual interference.
Despite this degradation, the proposed scheme consistently achieves a much higher sum rate than both BM1 baselines over the entire range of~$\kappa^{2}$ and for all blockage configurations. This confirms that the joint optimization of BS beamforming, PA power-ratio and PA positions exploits the available spatial degrees of freedom for secure transmission more efficiently, even under CSI and positioning uncertainties. Moreover, the rate gap between ``\textit{Pro. w blk}'' and ``\textit{Pro. w/o blk}'' widens as~$\kappa^{2}$ grows, because inaccurate CSI limits the ability of the BS to compensate for blockage‑induced power loss adaptively. In contrast, the curves for ``\textit{BM1 w blk}'' and ``\textit{BM1 w/o blk}'' almost overlap, since BM1 uses a fixed beamformer with uniform power allocation and thus depends only weakly on CSI; this makes BM1 less sensitive to CSI errors but also explains its significantly lower sum rate compared with the proposed joint design. On the other hand, as mentioned in \textit{Remark 1}, larger values of $\wp_g^E$ and $\wp_g^{\mathrm{arc}}$ can be interpreted as larger mobility budgets, or equivalently, longer CSI update intervals for passive EAs.

\section{Conclusion}
In this paper, we proposed a blockage-aware PA-assisted multi-waveguide architecture for secure and robust downlink transmission under imperfect CSI. By developing a 3D blockage-aware channel model that jointly captures waveguide attenuation, free-space propagation, and geometry-based LoS conditions, together with a geometry-aware CSI uncertainty bound for distributed multi-antenna eavesdroppers, we formulated a robust sum-rate maximization problem that jointly optimizes beamforming, AN covariance, PA power-ratio allocation, and PA positions under secrecy constraints. The resulting highly non-convex design was tackled via a BCD-based algorithm leveraging S-procedure and MM-based reformulations, including a physically motivated two-stage PA positioning strategy that separates large-scale movement for path loss and blockage-aware channel coefficients from small-scale phase-shift alignment. Simulation results verified the tightness of the proposed channel uncertainty model and demonstrated substantial performance gains compared with conventional fixed-antenna baselines. These findings highlight the value of incorporating spatial blockages into secure communication design, especially revealing that ignoring blockage can cause severe performance degradation and even infeasible secrecy constraints. In contrast, adaptive PA positioning can preserve LoS for legitimate users, disrupt eavesdroppers’ channels, and reduce the AN power required to meet secrecy targets.

Note that the present work focuses on theoretical design and analysis. Accordingly, experimental or ray-tracing-based validation of near-field blockage-aware PA channels constitutes an important direction for future work.


\begin{IEEEbiography}
[{\includegraphics[width=1.05in,height=1.3in]{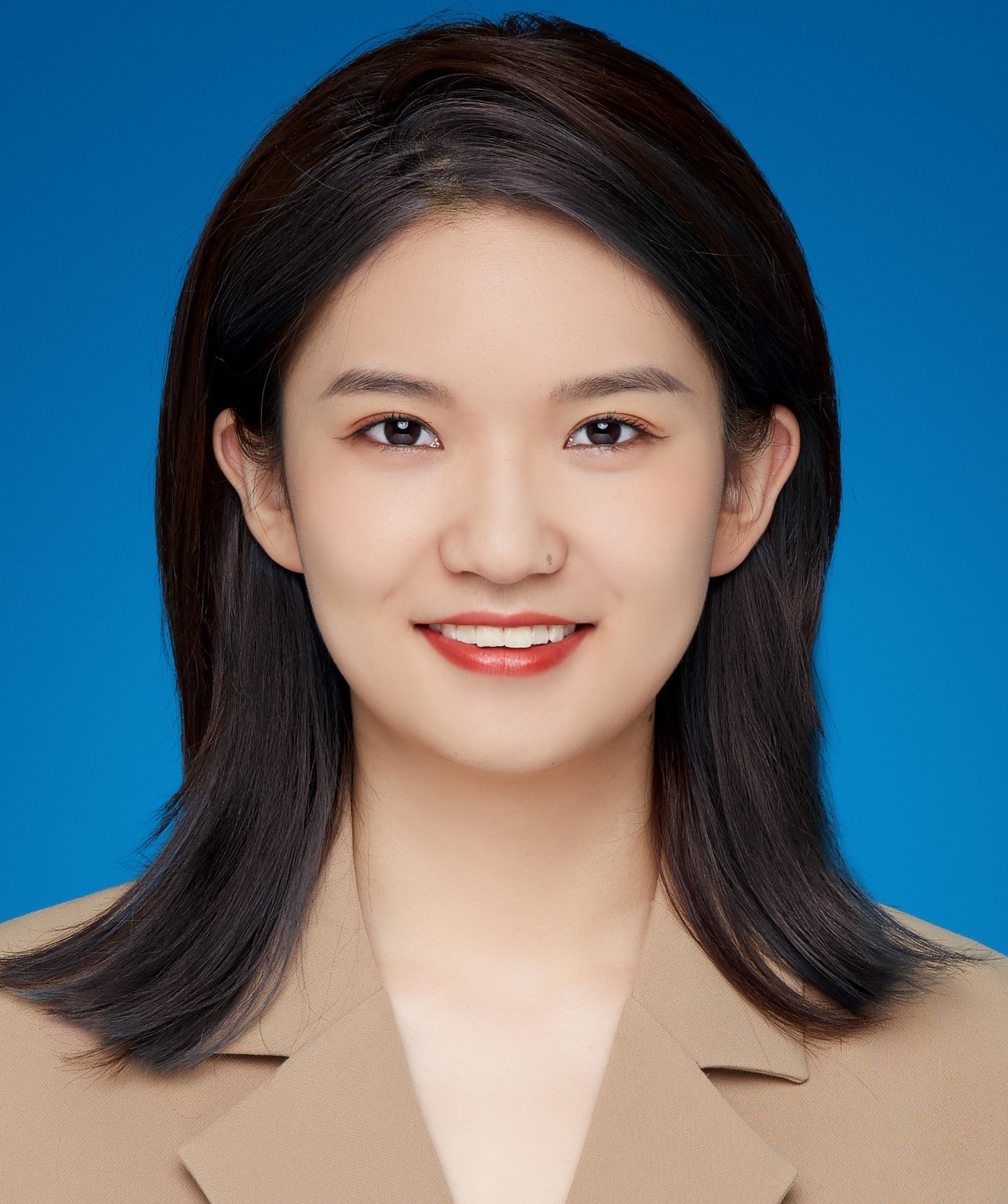}}]{Ruotong Zhao} (Graduate Student Member, IEEE) received the B.E. degree in telecommunications and the M.Phil. degree in electrical engineering from the University of New South Wales, Sydney, Australia, in 2022 and 2025, respectively, where she is currently pursuing the Ph.D. degree. She is a recipient of the Australian Government Research Training Program (RTP) Scholarship, the 2024 IEEE International Conference on Communications (ICC) Travel Grant, and the 2023 IEEE Global Communications Conference (GLOBECOM) Student Travel Grant. Her current research interests include resource allocation and flexible antenna-assisted wireless communications.
\end{IEEEbiography}
\begin{IEEEbiography}
[{\includegraphics[width=1in,height=1.3in]{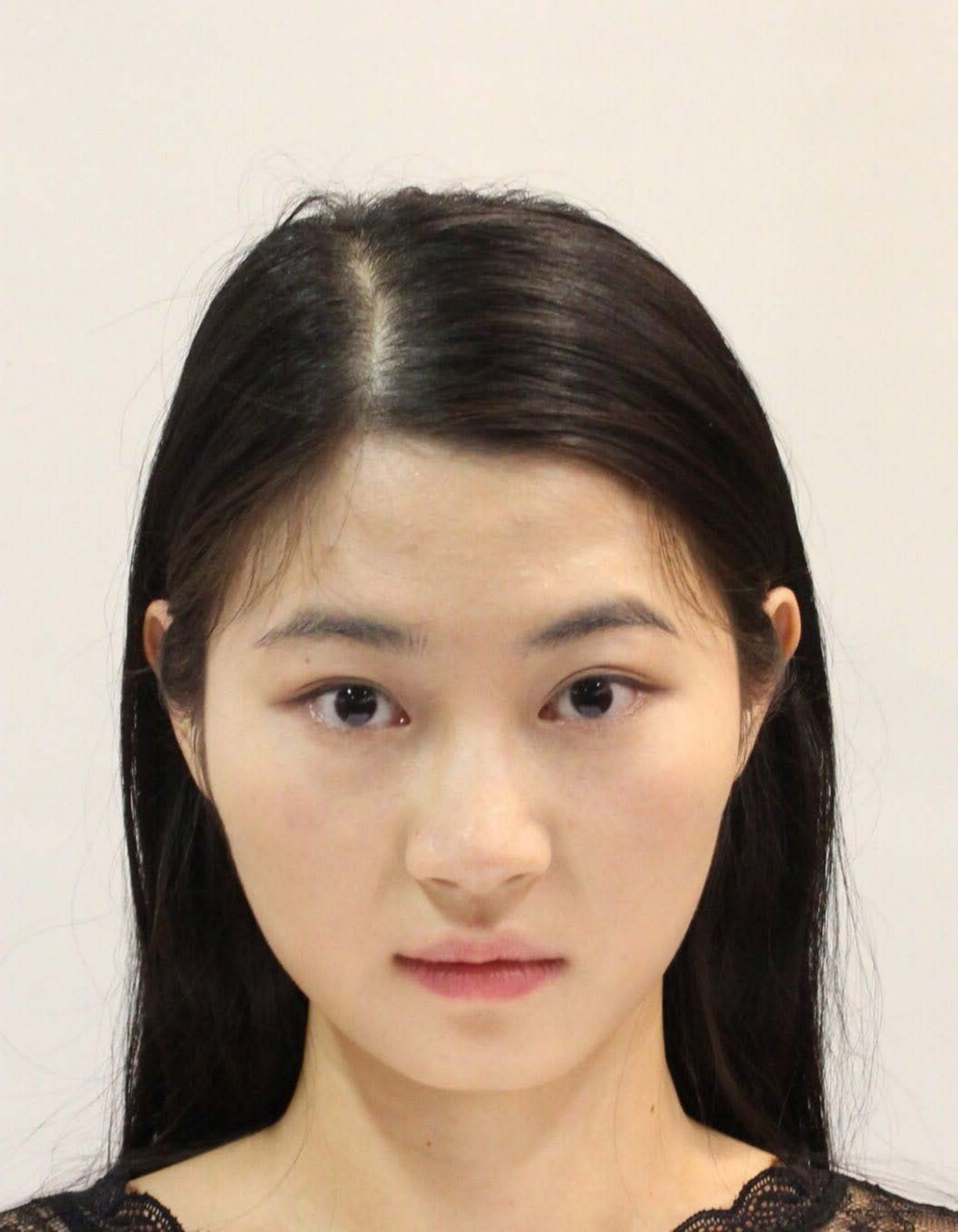}}]{Shaokang Hu} (Member, IEEE) received her B.S. degree in Engineering and Telecommunications, M.E. degree, and Ph.D. degree in Electrical Engineering and Telecommunication from the University of New South Wales (UNSW), Sydney, Australia, in 2017, 2018, and 2024, respectively. She is currently a postdoctoral researcher at UNSW's School of Electrical Engineering and Telecommunications. Her research interests include convex and non-convex optimization, IRS-assisted communication,  pinching antenna-assisted communication, UAV, resource allocation, physical-layer security, and energy-efficient (green) wireless communications.
\end{IEEEbiography}
\begin{IEEEbiography}
[{\includegraphics[width=1.1in,height=1.3in]{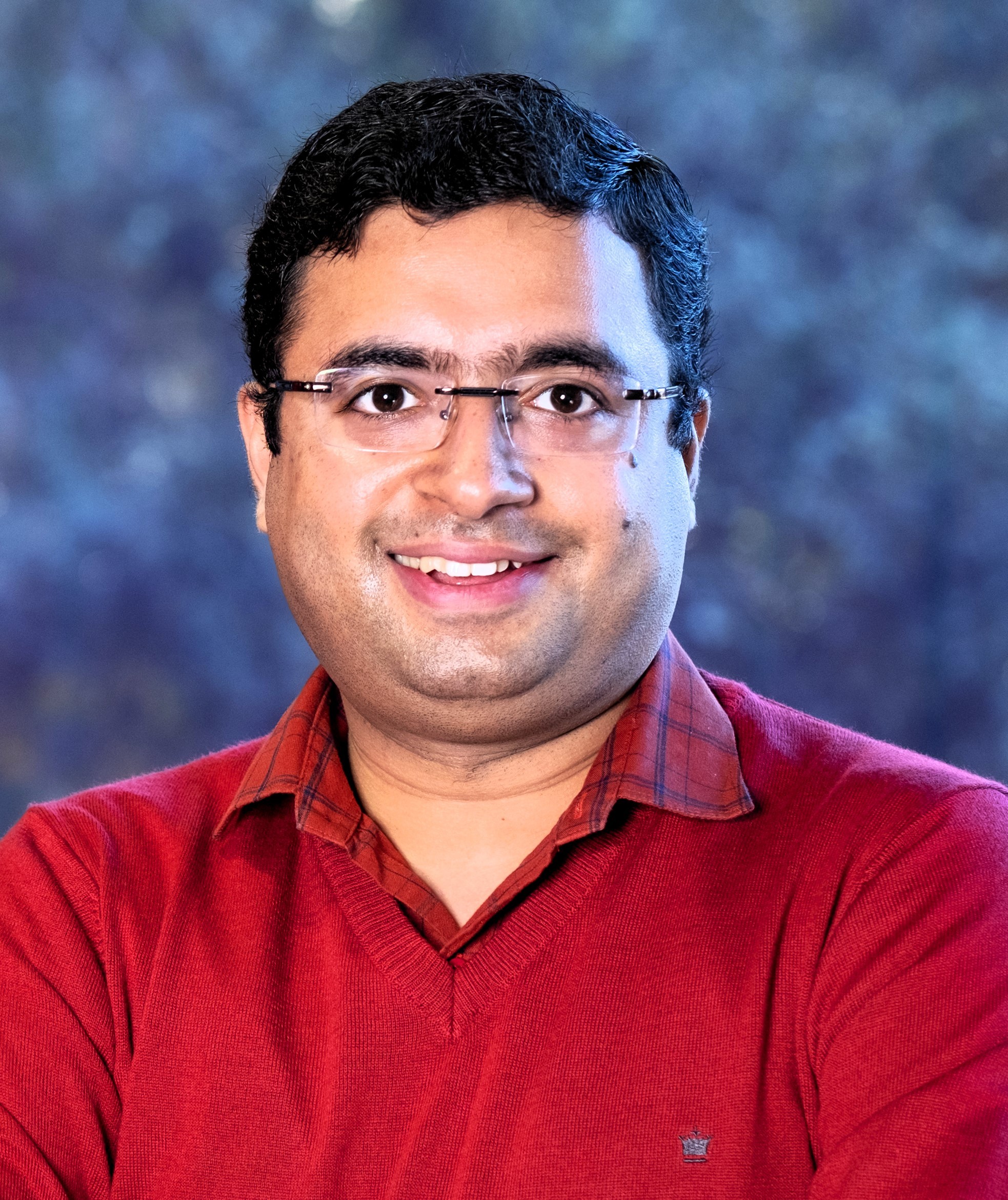}}]{Deepak Mishra} (Senior Member, IEEE) is a Senior Lecturer and Deputy Director of Postgraduate Academic Studies at the School of Electrical Engineering and Telecommunications at the University of New South Wales (UNSW), Sydney, Australia. He is also a co-founder and CTO of a UNSW-based startup company called GinigAI Pty Ltd, which utilises Embedded Radio Intelligence for safety purposes. He received his PhD degree in electrical engineering from the Indian Institute of Technology (IIT) in Delhi, New Delhi, in 2017. Before joining UNSW, he worked as a Postdoctoral Researcher at Linköping University in Sweden from August 2017 to July 2019. He has received the Australian Research Council (ARC) Discovery Early Career Researcher Award (DECRA) in 2022.
Deepak's research focus includes AI-enabled wireless sensing, energy harvesting cooperative communication networks, massive MIMO, privacy-aware backscattering, physical layer security, as well as signal processing and energy optimisation schemes for the uninterrupted operation of wireless networks. He is an Associate Editor of IEEE Transactions on Communications, IEEE Transactions on Green Communications and Networking, IEEE Transactions on Intelligent Vehicles, IEEE Communications Letters, IEEE Wireless Communications Letters, and IEEE Access, and has served as a Guest Editor for multiple Q1 publications. He was named a Best Editor for IEEE Wireless Communications Letters in 2023 and was selected as an Exemplary Reviewer of IEEE Transactions on Wireless Communications in 2017 and 2018, of IEEE Wireless Communications Letters in 2019, and of IEEE Transactions on Communications in 2019 and 2020.
\end{IEEEbiography}
\begin{IEEEbiography}
[{\includegraphics[width=0.95in,height=1.4in]{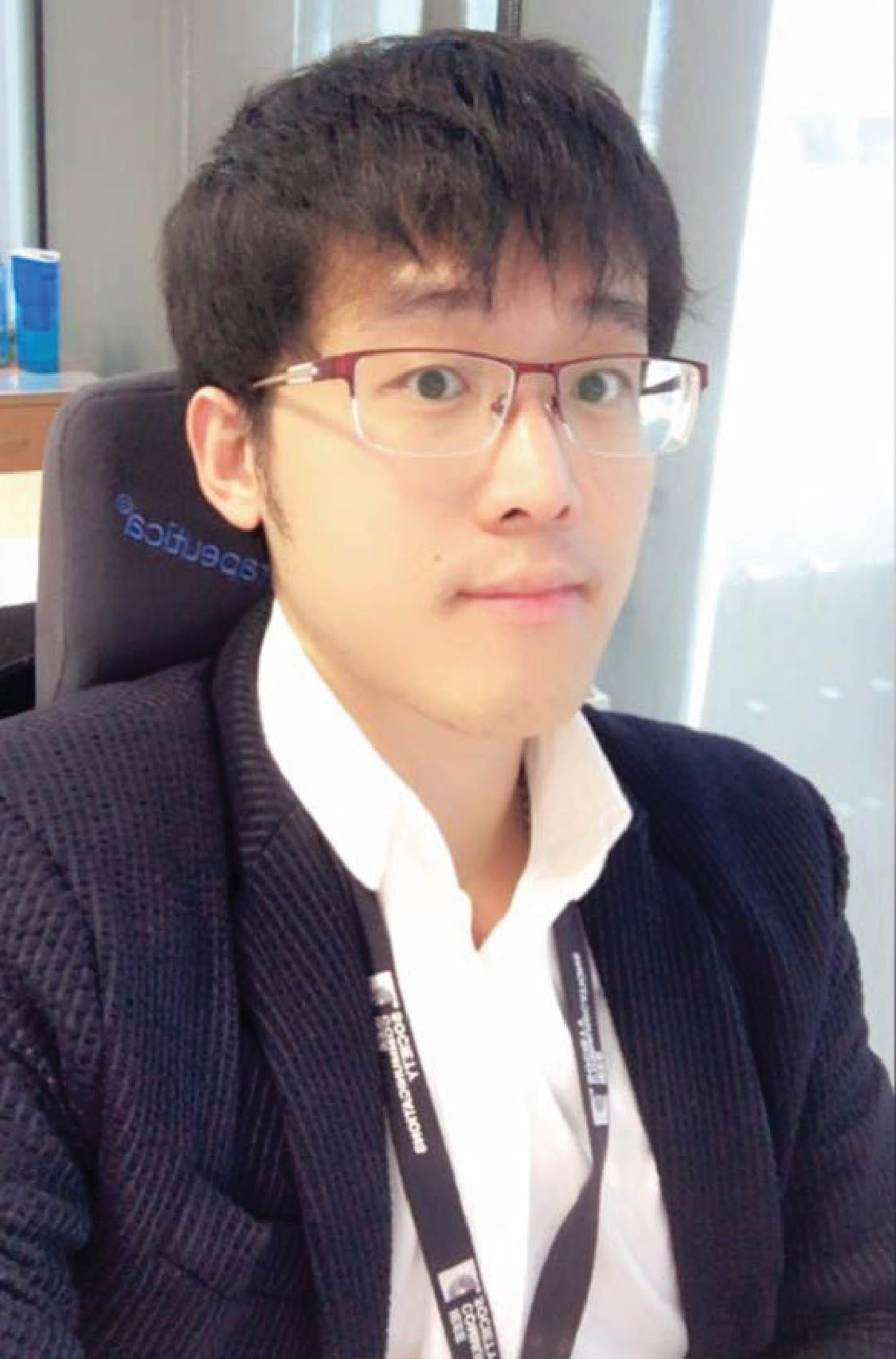}}]{Derrick Wing Kwan Ng} (S'06-M'12-SM'17-F'21) received his bachelor's degree (with first-class Honors) and the Master of Philosophy degree in electronic engineering from The Hong Kong University of Science and Technology (HKUST), Hong Kong, in 2006 and 2008, respectively, and his Ph.D. degree from The University of British Columbia, Vancouver, BC, Canada, in November 2012. Following his Ph.D., he was a senior postdoctoral fellow at the Institute for Digital Communications, Friedrich-Alexander-University Erlangen-N\"urnberg (FAU), Germany. He is currently a Scientia Associate Professor with the University of New South Wales, Sydney, NSW, Australia. His research interests include global optimization,  integrated sensing and communication (ISAC),  physical layer security, IRS-assisted communication, UAV-assisted communication, wireless information and power transfer, and green (energy-efficient) wireless communications.

He has been recognized as a Highly Cited Researcher by Clarivate Analytics (Web of Science) since 2018. He was the recipient of the Australian Research Council (ARC) Discovery Early Career Researcher Award 2017, IEEE Communications Society Leonard G. Abraham Prize 2023, IEEE Communications Society Stephen O. Rice Prize 2022, Best Paper Awards at the WCSP 2020, 2021, IEEE TCGCC Best Journal Paper Award 2018, INISCOM 2018, IEEE International Conference on Communications (ICC) 2018, 2021, 2023, 2024,  IEEE International Conference on Computing, Networking and Communications (ICNC) 2016, IEEE Wireless Communications and Networking Conference (WCNC) 2012, IEEE Global Telecommunication Conference (Globecom) 2011, 2021, 2023, 2025 and IEEE Third International Conference on Communications and Networking in China 2008. From January 2012 to December 2019, he served as an Editorial Assistant to the Editor-in-Chief of the IEEE Transactions on Communications. He is also an Area Editor of the IEEE Transactions on Communications,  an Associate Editor-in-Chief of the IEEE Open Journal of the Communications Society, and a member of the IEEE Transactions on Wireless Communications executive editorial committee.
\end{IEEEbiography}
\end{document}